\documentclass{aa}

\usepackage{natbib}
\bibpunct{(}{)}{;}{a}{}{,} 
\usepackage{graphicx}
\usepackage{txfonts,textcomp}
\usepackage[colorlinks=true,allcolors=blue]{hyperref}
\newcommand{\cii}{[C$\scriptstyle\rm~II$]~}     
\newcommand{\oiii}{[O$\scriptstyle\rm~III$]~}

\newcommand{\sigmag}{$\sigma_{\rm gas}$}
\newcommand{\vmax}{$V_{\rm max}$}

\newcommand{\msun}{M${_\odot}$}

\newcommand{\kms}{${\rm km~s^{-1}}$}

\begin{document}

   \title{ALMA hints at the presence of turbulent disk galaxies at $z>5$
   }


   \author{E. Parlanti\inst{1} \and S. Carniani \inst{1} \and A. Pallottini \inst{1}  \and M. Cignoni \inst{2,3,4} \and G. Cresci \inst{5} \and M. Kohandel \inst{1} \and F. Mannucci \inst{5} \and A. Marconi \inst{5,6} }

   \institute{Scuola Normale Superiore, Piazza dei Cavalieri 7, I-56126 Pisa, Italy  \and INAF -- Osservatorio di Astrofisica e Scienza dello Spazio, Via Gobetti 93/3, 40129 Bologna, Italy
   \and Physics Departement, University of Pisa, Largo Bruno Pontecorvo, 3, 56127 Pisa, Italy
   \and INAF -- Osservatorio Astronomico di Capodimonte, Via Moiariello 16, 80131, Napoli, Italy
   \and INAF -- Osservatorio Astrofisico di Arcetri, Largo E. Fermi 5, 50127, Firenze, Italy
   \and Dipartimento di Fisica e Astronomia, Università di Firenze, Via G. Sansone 1, 50019, Sesto Fiorentino (Firenze), Italy}

   \date{Received September 15, 1996; accepted March 16, 1997}

 
  \abstract
   {High-redshift galaxies are expected to be more turbulent than local galaxies because of their smaller size and higher star formation and thus stronger feedback from star formation, frequent mergers events, and gravitational instabilities. However, this scenario has recently been questioned by the observational evidence of a few galaxies at $z\sim4-5$ with a gas velocity dispersion similar to what  is observed in the local population.}
   {Our goal is to determine whether galaxies in the first billion years of the Universe have already formed a dynamically cold rotating disk similar to the  local counterparts.}
   {We studied the gas kinematic of 22 main-sequence star-forming galaxies at $z > 5$  and determined their dynamical state by estimating the ratio of the rotational velocity and of the gas velocity dispersion. We mined the ALMA public archive and exploited the \cii\ and \oiii\ observations to perform a kinematic analysis of the cold and warm gas of $z>5$ main-sequence galaxies. We compared our results with what was found in the local and distant Universe and investigated the evolution of the gas velocity dispersion with redshift. 
   We also compared the observations with theoretical expectations to assess the main driver of the gas turbulence at $z>5$.}
   {  The gas kinematics of the high-$z$ galaxy population observed with ALMA is consistent within the errors with rotating but turbulent disks. We indeed infer a velocity dispersion that is systematically higher by 4-5 times than the  local galaxy population and the $z\sim5$ dust-obscured galaxies reported in the literature. 
   The difference between our results and those reported at similar redshift can be ascribed to the systematic difference in the galaxy properties in the two samples: the disks of massive dusty galaxies are dynamically colder than the disks of dust-poor galaxies.  The comparison with the theoretical predictions suggests that the main driver of the velocity dispersion in high-redshift galaxies is the gravitational energy that is released by the transport of mass within the disk.
    Finally, we stress that future deeper ALMA high-angular resolution observations are crucial to constrain the kinematic properties of high-$z$ galaxies and to distinguish rotating disks from kiloparsec-scale mergers.
   }
   {}

   \keywords{Galaxies: high-redshift -- Galaxies: kinematics and dynamics
               }

   \maketitle
%
 
\section{Introduction}

The galaxy assembly and evolution in the first billion years of the Universe is still a hot topic of modern astrophysics.
Primeval galaxies are thought to be subjected to galaxy-galaxy interactions, merging processes, gravitational instabilities, and feedback from bursts of star formation and active galactic nuclei (AGN) \citep{Dekel:2009, Krumholz:2010, Green:2014, Somerville:2015, Dayal:2018, Krumholz:2018, Ginzburg:2022}, which may affect the formation of the galactic disks and their properties. It is therefore fundamental to explore the kinematical properties of galaxies to understand  the main processes that play a crucial role in the galaxy growth over cosmic time.

Several observing programs have focused on the gas kinematics of galaxies at $0 < z < 4$ and provided thousands of spatially resolved observations of ionized, neutral, and molecular gas \citep{Cresci:2009, Forster:2009, Epinat:2010, Gnerucci:2011, Ianjamasimanana:2012, Green:2014, Wisnioski:2015,  Mogotsi:2016, DiTeodoro:2016, Harrison:2017, Swinbank:2017, Turner:2017, Forster:2018, Johnson:2018, Ubler:2019, Girard:2021}. 
The kinematic analysis of these galaxies shows that the intrinsic velocity dispersion of gas (\sigmag) increases with redshift, suggesting that galaxies in the cosmic noon epoch are more turbulent than their local counterparts \citep{Wisnioski:2015, Johnson:2018, Wisnioski:2018, Ubler:2019}. 
These observations have differentiated between rotation-dominated and dispersion-dominated disks by comparing the ordered rotational velocities, \vmax (the maximum rotational velocity of the galaxy), with the \sigmag.
The distinction between the two kinematics is mainly based on the $V_{\rm max} / \sigma_{\rm gas}$ , with a threshold value that spans the range between 1 and 3. Studies of the high-$z$ Universe usually adopted the threshold $V_{\rm max} / \sigma_{\rm gas} =\sqrt{3.36}$ based on  the kinematics of the exponential disk (For an exponential disk, $V_{\rm c} = (V_{\rm rot}^2 +  3.36\sigma_{\rm gas}^2)^{0.5}$ Eq. 1 \cite{Forster:2018}). A ratio of $V_{\rm max} / \sigma_{\rm gas} <\sqrt{3.36}$  indicates that the dynamical support from random motions of gas is stronger than the rotational support from ordered motions \citep{Binney:2008, Wisnioski:2018, Forster:2020}.
In summary, high $V_{\rm max} / \sigma_{\rm gas}$ ratios suggest a dynamically cold disk, and low values are usually associated with a turbulent gas disk that is mainly supported by random motions.

At $0<z<1,$ the bulk of the galaxy population with a stellar mass $M_\star$ between $10^9$ \msun~ and $10^{11}$ \msun~ shows a $\sigma_{\rm gas}$ $\sim 10 - 40$ km s$^{-1}$, depending on the gas tracer, and a $V_{\rm max}/\sigma_{\rm gas}$ ratio of $\sim 10$ that indicates a rotationally supported gas disk similar to that of our Galaxy \citep{Epinat:2010, Ianjamasimanana:2012, DiTeodoro:2016, Mogotsi:2016, Wisnioski:2015, Harrison:2017, Swinbank:2017, Johnson:2018,  Ubler:2019}.
Observations in the distant Universe instead reveal a different scenario. At intermediate redshift ($1<z<3$),  different studies found galaxies with $\sigma_{\rm gas}$ up to 70~\kms and $V_{\rm max}/\sigma_{\rm gas}$ values down to 1-2, suggesting that distant galaxies can be described as rotating turbulent disks  \citep{Cresci:2009, Forster:2009,  Wisnioski:2015, Turner:2017, Forster:2018, Ubler:2019, Girard:2021}.

In the past decade, several models have been proposed to explain the observed evolution of the velocity dispersion with redshift \citep{Krumholz:2010,Forbes:2014, Krumholz:2018, Ginzburg:2022, Ostriker:2022}. A continuous energy injection into the system is necessary to maintain its high-velocity dispersion. 
This energy is thought to derive from three main mechanisms: feedback from star formation and AGN, gravitational instabilities within the disk, and gas accretion.
Supernova explosions and stellar winds are expected to inject energy and momentum into the surrounding environment. These processes increase the temperature and the velocity dispersion of the gas in the galaxy \citep{Hayward:2017, Orr:2020, Ostriker:2022}. Mergers and gravitational interactions between galaxies can also give rise to perturbations and generate gravitationally unstable regions in the galactic disk \citep{Kohandel:2020}. In these regions, the axisymmetry of the disk is broken, promoting the formation of clumps \citep{Zanella:2015, Kohandel:2019}. The gravitational field will exert a torque on the clumps that will transport the mass within the disk from a larger to a smaller radius, which increases the turbulence and the star formation rate (SFR). The transport of mass would release gravitational energy and restore the stability of the disk \citep{Krumholz:2010, Forbes:2014}.
Recently, even the accretion process of gas from the circumgalactic medium was taken into account as it increases the turbulence in galaxies. The gas accretes onto the galaxy via cold streams, and the streams fragment and turn into clumps. The energy of the accreting gas can turn into heat via shock or can remain kinetic energy that will turn into turbulence \citep{Dekel:2009, Forbes:2022, Ginzburg:2022}.

The increase in velocity dispersion over cosmic time and the decrease in $ V_{\rm max}/\sigma_{\rm gas} $ also seem to be supported by cosmological simulations. These simulations report an increase in the velocity dispersion at fixed stellar masses  \citep{Dekel:2014, Zolotov:2015, Pillepich:2019}.
The velocity dispersion of simulated high-redshift galaxies is higher by an order of magnitude than in their local counterparts.
The simulations are consistent with the observations up to $z=4,$ supporting the scenario that the contribution from random motion increases across cosmic time.

\cite{Wisnioski:2015} claimed that  the evolution of the gas kinematical properties of galaxies only depends on the evolution of the gas fraction ($f_{\rm gas} $) \citep{Tacconi:2020}. Star-forming galaxies are thought to be in steady equilibrium between outflow, star formation, and gas inflow \citep{Forster:2006, Mannucci:2010, Genzel:2011}, and the balance between the heating and the cooling of gas maintains the disk in a quasi-stable state with the Toomre parameter $Q \sim 1$ \citep{Toomre:1964}. This assumption leads to the relation
\begin{equation} \label{eq:Toomre_highz}
\frac{V}{\sigma} = \frac{a}{f_{\rm gas}}
,\end{equation}
where $1<a<2$, in particular, $a = \sqrt{2}$ for a flat rotation curve, $a = \sqrt{3}$ for a uniform disk, and $a = 2$ for a solid-body rotation.
With this model based on marginally stable disks, the decrease in $V_{\rm max}/\sigma_{\rm gas}$ and the increase in $\sigma_{\rm gas}$ at increasing redshift is due to the increment of the gas fraction at high redshift  \citep{Tacconi:2020}.  
This model would also explain the observed clumpy structure of high-redshift galaxies \citep{Forster:2006, Genzel:2008, Forster:2011, Carniani:2017b, Carniani:2018} because Toomre's stability criterion leads to the formation of   star-forming clumps in gas-rich disk when $Q \leq 1$.

With the advent of the Atacama Large Millimeter/ sub–Millimeter Array (ALMA), we can exploit the brightest rest-frame far-infrared (FIR) lines, which fall in the millimeter bands (0.5-2 mm), to trace the dynamics of galaxies at $z>4$. Among the rest-frame FIR lines, the \cii($157.7 \mu \rm m$) and \oiii ($88 \mu \rm m$) are the brightest lines \citep{Carilli:2013}. 
The \cii transition is one of the principal coolants of the interstellar medium (ISM) and one of the most luminous FIR lines \citep{Stacey:1991, Malhotra:1997, Luhman:1998, Luhman:2003}. The ionization potential of the carbon atom is $\sim 11.3$ eV, which is lower than that of  hydrogen  (i.e., 13.6 eV), and can be used to trace different phases (neutral, molecular, and mildly ionized gas) of the ISM of a galaxy \citep{Shibai:1991, Heiles:1994, Stacey:2010, Pineda:2013, Vallini:2015}.
On the other hand, the transition lines of \oiii\ are good tracers of HII regions because the oxygen atom has an ionization potential of 35.1 eV \citep{Carilli:2013}.

Through ALMA, a multitude of galaxies have been observed at $z>4$ targeting FIR emission lines \citep{Capak:2015, Willott:2015, Carniani:2017b, Carniani:2018, Smit:2018, Harikane:2020, Neeleman:2020, Rizzo:2020,Rizzo:2021, Lelli:2021,  Fraternali:2021,  Jones:2021, Herrera-Camus:2021, Herrera-Camus:2022, Schouws:2022, Bouwens:2022}. 
So far, only a few studies have focused on the gas kinematics of these high-redshift systems, and the results yield a contrasting picture of the kinematic properties of galaxies in the primeval Universe.
On the one hand, \cite{Jones:2021} and \cite{Herrera-Camus:2021} found a range of the velocity dispersion from 20 km s$^{-1}$ to 116 km s$^{-1}$ for nine galaxies at $4 < z < 6$, and $V_{\rm max}/\sigma_{\rm gas} \sim$ 3, which is consistent with what is observed at $z=1-3$.
On the other hand, other studies  have found a lower velocity dispersion and a higher rotational velocity by studying the \cii\ emission line for ten massive starburst dusty galaxies at $z \sim 4 - 5$ \citep{ Sharda:2019, Neeleman:2020, Rizzo:2020, Rizzo:2021, Fraternali:2021, Lelli:2021}.  The measured velocity dispersion is as low as 15-20 km s$^{-1}$ , with an average $V_{\rm max}/\sigma_{\rm gas}$ of 10 and an extreme value of 20 for one of the selected galaxies \citep{Fraternali:2021}. These estimates 
are comparable to the $\sigma_{\rm gas}$ and $V_{\rm max}/\sigma_{\rm gas}$ obtained for local spiral galaxies \citep{Epinat:2010, Ianjamasimanana:2012, DiTeodoro:2016, Mogotsi:2016}. %
This implies that the internal ISM turbulence of these galaxies is similar to that of local galaxies, and despite their young age, the gas is already settled into a dynamically cold rotating disk.
The existence of dynamically cold disks at $z>6$ is predicted in cosmological simulations \citep{Pallottini:2019, Kohandel:2020}, where $V_{max}/\sigma \sim 7$ was reported for a simulated Lyman-Break galaxy during its evolution when observed with the [CII] emission line. 

To summarize, despite the large progress in studying the gas kinematics in the first billion years of the Universe \citep{Sharda:2019, Neeleman:2020, Rizzo:2020, Rizzo:2021, Fraternali:2021, Jones:2021, Lelli:2021}, it is still unclear whether primeval galaxies are dominated by rotation or dispersion in the early phase of their evolution. 
Most of the studies in the literature  have focused on a limited number of targets that were preselected to be starburst ($SFR>100~{\rm M_\odot}$), massive ($M_{\star}>10^{10.5}~{\rm M_\odot}$), and luminous in [C$\scriptstyle\rm~II$]. As these galaxies might not represent the bulk of the galaxy population at high redshift, we need to analyze a larger sample of main-sequence star-forming galaxies  to  assess the dynamical state of early galaxies. Here we present a kinematic study of 22 galaxies observed with ALMA at $z>4$, which are expected to represent the bulk of the galaxy population in the first billion years of the Universe.
For the galaxies that were observed with both the \cii\ and \oiii\ lines, we also explore the possibility of a bias in the kinematics traced with one FIR tracer alone.

This paper is structured as follows. In Section \ref{sec:sampleselection} we describe the sample selection and data reduction processes.
In Section \ref{sec:dataanalysis} we present the data analysis process and the algorithm we adopted to recover the kinematic properties of the sample. 
In Sects. \ref{sec:results} and \ref{sec:discussion} we present our results, compare them with the other literature findings, and discuss the evolution of the velocity dispersion and $V_{\rm max}/\sigma_{\rm gas}$ as a function of redshift and other physical parameters of galaxies.
In section \ref{sec:conclusion} we summarize the findings of this work and draw our conclusions.
We adopt the cosmological parameters from \cite{Planck:2015}: $H_0$ =  67.8 \kms Mpc$^{-1}$
, $\Omega_m$ = 0.308, and $\Omega_\Lambda$ = 0.685.

\section{Sample selection and observations} \label{sec:sampleselection}
We selected our galaxy sample from the public ALMA data archive \footnote{\url{https://almascience.nrao.edu/aq}}. We queried the database (on 12 January 2022) to obtain only star-forming galaxies at $z > 4$ with observations of either \cii or \oiii or both emission lines in bands 5, 6, 7, or 8 and angular resolutions lower than 1.5\arcsec \footnote{1.5\arcsec at $z \sim 5$  corresponds to $\sim$ 7 kpc, which is $\sim 3-4$ times larger than the typical size of a galaxy at this redshift.}. 
We aimed to explore the kinematic properties of high-redshift star-forming galaxies and therefore excluded known submillimeter galaxies (SMGs) and quasars, which do not represent the bulk of the galaxy population \citep[e.g.,][]{Weiss:2013}.
We thus selected galaxies whose previous measurements of stellar masses and SFR are compatible main-sequence of star-forming galaxies (see details in Sec. \ref{sec:sampleproperties}). We also excluded from the selected targets galaxies with clear merger features. Finally, we removed observations in which the far-infrared lines are detected with a signal-to-noise ratio\footnote{We define the S/N as the median value across the map of the ratio of the integrated flux emission of the line (moment 0 map) and the pixel-by-pixel estimated error $\sigma^{\rm flux}$ defined in Eq. \ref{snr}} (S/N)$<7$, which is not sufficient to perform a kinematic analysis on the data.
Our final sample is composed of 22 galaxies at $4.2<z<7.6$, some of which were observed in the \cii and \oiii emission lines  (see Table \ref{tab:results} and \ref{tab:properties}).

We calibrated the visibilities with the  Common Astronomy Software Application (\verb'CASA') \citep{McMullin:2007} by using the pipeline scripts delivered with the data that are available in the archive. We used the appropriate \verb'CASA' version for each target as stated in the scripts. 
We then performed the `cleaning process on the calibrated visibilities by using the \verb'CASA' task \verb'tclean'. To obtain the best sensitivity from the ALMA data sets, we adopted a \verb'natural' weighting scale to generate the final datacubes, yielding an angular resolution  between 0.2\arcsec\ and 1.5\arcsec\ depending on the data sets. The spatial pixel size of each datacube was set up to be between $\sim$1/5 to $\sim$1/10 of the minor axis of the beam, which allowed us to both  sample the beam profile and reduce the number of correlated pixels in the datacube \citep{almahandbook}. The frequency channel spacing of the datacube was set up to exploit the maximum spectral resolution of each data set (10-20 km s$^{-1}$).

\section{Data analysis} \label{sec:dataanalysis}

In this section, we perform the kinematic analysis of the galaxies of our sample with the goal of determining their kinematic properties. We present two different methods for the kinematic fitting procedure that we adopted to determine the rotational velocity curves and velocity dispersion from the \cii\ and \oiii\ data on the basis of the angular resolution and sensitivity of the observations.

\subsection{Moment maps}\label{momentmaps}
Initially, we performed a pixel-by-pixel Gaussian fitting on the datacube to generate the moment maps: flux (zeroth moment), velocity (first moment), and velocity dispersion (square root of the second moment).
We also accounted for an additional constant to match any potential residual of continuum emission.

Before analyzing the moment and performing the kinematic fitting to the maps,  we removed the parts of the maps that were dominated by noise fluctuations.
We estimated the conservative pixel-by-pixel  error of the flux map as 
\begin{equation}\label{snr}
    \sigma^{\rm flux} = \sqrt{\sum_{k = v_{\rm min}}^{v_{\rm max}} (\sigma_k \cdot \Delta v)^2}~~{\rm Jy~beam^{-1}~km~s^{-1}}
,\end{equation}
where $\sigma_{k}$ is the ALMA noise level at the spectral channel $k$ determined in a free-target region  of the datacube, $v_{\rm min}$ and $v_{\rm max}$ are defined as the $16th^{}$ and $84th^{}$ percentiles in spectral channels of our best-fitting Gaussian model, respectively, and $\Delta v$ is the frequency channel spacing of the datacube in velocity units. We then excluded all the pixels in the kinematic maps with \cii integrated flux emission lower than 5$\sigma^{\rm flux}$. For \oiii observations, we adopted a lower threshold of 3$\sigma^{\rm flux}$ because the observed flux density of the oxygen line is usually fainter than that of the \cii line \citep{Carilli:2013, Inoue:2016, Carniani:2017b, Marrone:2018, Walter:2018, Hashimoto:2019, Harikane:2020, Witstok:2022}.

\subsection{Kinematic models}
\label{kinmodel}
To reproduce the kinematic maps of the FIR lines, we used a rotational velocity model, $V_{\rm rot}(r)$, defined by
\begin{equation}\label{circularvel}
    V_{\rm rot}(r)=\sqrt{V_d^2(r)+V_b^2(r)},
\end{equation}
where $r$ is the distance from the galaxy center, and $V_d(r)$ and $V_b(r)$ are the velocity profile due to the disk and bulge components, respectively.
We focus on high-redshift galaxies and used observations whose sensitivity is insufficient to probe the rotational velocity at the very large radii from the center. We therefore  neglected the contribution of dark matter to the circular motion and focused on the disk and bulge components, which are dominant within two scale radii from the galaxy center \citep{Sofue:2013}.
Our assumption is also supported by the results from \citet{Genzel:2017}, \citet{Fraternali:2021}, and \citet{Lelli:2021}, who have found that the rotation curves of $0.6 <z < 5$ galaxies do not show the classical flat rotation profile due to dark matter, but the gas kinematics is well modeled with a baryon-dominated disk with a negligible mass of the dark matter in the inner regions. 

Although most of the high-redshift studies \citep{Neeleman:2020, Fujimoto:2020, Genzel:2017, Ubler:2019, Gnerucci:2011} show that the gas kinematics of galaxies at $z= 1-4$ can be modeled only with an exponential gas disk neglecting the stellar bulge component due to the increasing gas fraction ($f_{\rm gas} = M_{\rm gas}/(M_{\rm gas}+M_{\rm \star})$) with  increasing redshift \citep{Tacconi:2020}, recent \cii observations \citep{Rizzo:2020, Lelli:2021} have revealed that massive ($M_\star > 10^{10}$~\msun ) $z>4$  galaxies might require a bulge component in addition to the exponential disk to reproduce their kinematics properties. 
In this work, we therefore fit the kinematics maps of massive galaxies with two models: one model that includes both components, and the other model with the exponential disk alone.
Then we select the best-fitting model that better reproduces the moment maps by minimizing the $\chi^2$.

In cases in which the sensitivity and resolution are not high enough to determine the presence of the bulge or $M_\star < 10^{10}$~\msun\ , we adopt the simple exponential disk model.

For the exponential disk, we assumed that the light is emitted by a thin disk  \citep{Freeman:1970,Forster:2006, Binney:2008} whose surface brightness radial profile is 
\begin{equation} \label{expdiskprofilegrightnes}
    I(r)=I_0 \exp{(-r/r_d)},
\end{equation}
where $r_d$ is the disk scale length.
Assuming a constant mass-to-light ratio, we  write the profile of the surface mass density as
\begin{equation}
    \Sigma(r)=\Sigma_0 \exp{(-r/r_d)}.
\end{equation}
The velocity contribution due to the exponential disk component can therefore be written as a function of the radius as

\begin{equation}\label{velocityexpdisk}
    V_d^2(r) =  2 y^2 \frac{G M_d(R_0)}{r_d} \frac{\mathcal{I}_0(y)\mathcal{K}_0(y) - \mathcal{I}_1(y)\mathcal{K}_1(y)}{1- \exp(-R_0/r_d)(1 + R_0/r_d)}
,\end{equation}
where $y \equiv \frac{r}{2r_d}$, while $\mathcal{I}_0$, $\mathcal{I}_1$, $\mathcal{K}_0$ , and $\mathcal{K}_1$ are modified Bessel functions.
As this works focuses on galaxies at $z>4$, which have a typical disk scale length of  $<2$ kpc \citep{Shibuya:2015}, we defined the mass of the disk, $M_d$, as the mass computed within a radius $R_0 = 5$ kpc, which corresponds to ~2-3 disk scale lengths: $M_d = M_d(R_0) = M_d(r = 5 {\rm kpc})$.

On the other hand, the light profile of the bulge component is  assumed to match a Sérsic profile \citep{Sersic:1963},
\begin{equation}\label{eq:sersicbrightness}
    I(r) = I_0 \exp{ \left[ - \tilde{b} \left( \frac{r}{r_e}\right)^{1/n}\right] }
,\end{equation}
where $I_0$ is the surface brightness at the center of the bulge, $r_e$ is the effective radius, and  $n$ is the Sérsic index. $\tilde{b}$ is a function of $n$ $\tilde{b}$ = 2n - 1/3 + 0.009876/n \citep{Prugniel:1997}.

We can then describe the contribution of the bulge component to the circular velocity assuming a constant mass-to-light ratio as 
\begin{equation}
V_b^2(r) = \frac{G M_b}{r} \frac{\gamma(n(3-p), b(r/r_e)^{1/n})}{\Gamma(n(3-p))}
\end{equation}
\citep{Terzic:2005}, where $M_b$ is the total stellar mass of the bulge, $p$ is a function of the Sérsic index ($p = 1 - 0.6097/n + 0.05563/n^2$; \citealt{Limaneto:1999}), and $\gamma$ and $\Gamma$ are the incomplete and complete gamma function, respectively.
In our analysis, we opted for a de Vaucouleurs profile by assuming $n=4$ for the bulge model \citep{deVaucouleurs:1956}. Because of the sensitivity and angular resolution of our observations, different values of the Sérsic index do not change the results of the velocity dispersion. Future studies based on high-sensitivity and high angular resolution observations should leave $n$ as a parameter in the kinematics fitting process.

A caveat of the present work is that we only considered rotating disk models. An alternative scenario would be that the observed velocity gradients are a consequence of two spatially unresolved sources at different systematic velocities that are located at a distance smaller than the half-beam size (i.e., $<$ 2 kpc). \citet{Rizzo:2022} showed that to perform a robust analysis and to distinguish disks from mergers in the early Universe, data with a resolution $\sim 0.2 \arcsec$  and a signal-to-noise > 10 are required. This means that the resolution of the ALMA observations is not sufficient to distinguish the two scenarios, and future high-angular resolutions will be fundamental to further this type of studies.

\subsection{Beam-smearing effect} \label{sec:beamsmearing}

The limited angular resolution strongly affects the maps of flux, velocity, and velocity dispersion, especially in high-redshift observations, where the angular resolution is comparable to the galaxy size.
From an analytical point of view, the beam-smearing effect introduces a convolution product in the equations of the moment maps. 
If the beam (or point spread function) of the telescope is greater than the characteristic size of the galaxies where its velocity profile changes steeply, the finite beam size smooths the velocity gradient in the velocity map and increases the velocity dispersion, especially in the central region, where the observed $\sigma_{\rm gas}$ is greater than the intrinsic velocity dispersion of the gas through the broadening of the emitting line. 
The broadening of the line can be mistaken as the result of random motions of the gas, yielding an overestimation of the gas velocity \citep{diTeodoro:2015, Kohandel:2020, Concas:2022}.
The net result of beam smearing might underestimate the rotation velocity
and overestimate the velocity dispersion. 

The ratio $V_{\rm max}/\sigma_{\rm gas}$ is then highly impacted by this effect and can be underestimated if no correction is applied. We thus take into account how the beam-smearing impacts the observations in the fitting process. In Appendix \ref{BeamSmearingAppendix} we test in detail the beam-smearing  effects on datacubes and our ability to recover the galaxy kinematics, even for low angular resolutions. We find that $V_{\rm max}/\sigma_{\rm gas}$ is underestimated if the beam size is more than three times the galaxy size.

In addition to these effects, the beam smearing introduces a strong correlation between neighboring spatial
pixels in the datacubes, and neglecting the pixel-pixel noise correlation might lead to an underestimation of the uncertainties of the kinematic properties.
In the next sections and in Appendix \ref{Correlation}, we describe how we corrected for these effects.

\subsection{3D datacube model} \label{sec:mockdatacube}

We used the Python routine \textsc{KinMS}  \citep{Davis:2013} to create the datacube models for the fitting process. 
\textsc{KinMS} initializes $5\times10^5$ line-emitting particles given by the user.
Each particle is assigned a position in the galaxy space ($x$,$y$,$z$) to reproduce the sky brightness  distribution given as input by the user (e.g., the exponential profile in Eq. \ref{expdiskprofilegrightnes}).
Each particle is also given a velocity (v$_x$, v$_y$, v$_z$) following the assigned velocity profile (e.g., rotational velocity in Eq. \ref{circularvel}) and a random velocity component $\sigma$ that represents the intrinsic velocity dispersion of the gas.

 The routine then projects the 6D space-velocity array of each line-emitting particle onto the sky plane based on a given inclination and rotated according to a position angle. After the rotation, every particle is assigned new coordinates onto the sky plane (x$_{\rm s}$, y$_{\rm s}$), and the velocity along the line of sight $v_{\rm los}$. 
Finally, \textsc{KinMS} creates a datacube: Each particle with a given position and projected velocity along the line of sight is assigned to a pixel ($i, j, k$) in the datacube.
The size of the spatial pixels is rebinned to match the size of the spatial pixels of the observation.
Each spectral slice of the datacube is convoluted with a point spread function (PSF) as large as the ALMA beam of the data sets analyzed in this work.
The result is a datacube with the same spatial and spectral resolution as the observations.
The moment maps derived from the mock datacube have the same effect of the beam smearing as the real data. In this way, the fitting process already takes into account the effect of beam smearing without the need for further corrections.

\subsection{Method I: 2D kinematic fitting} \label{sec:2Danalysis}

After we generated the moment maps for each target, we performed a kinematical fitting by using the mock datacube as the model to fit the data (see details in Appendix \ref{Correlation}). 
The parameter space was explored using a Bayesian approach based on the Monte Carlo Markov chains (MCMC) method, which allowed us  to infer the posterior probability distribution of the free parameters. In particular, we used the package EMCEE \citep{Foreman-Mackey:2013}.

The free parameters of our kinematic models were the scale radius of the exponential disk ($r_d$), the inclination ($inc$) and position angle of the disk ($PA$), the coordinates of the center ($x_0, y_0$), the rotational mass\footnote{$M_{\rm rot}$  is the mass of the disk (or disk + bulge) that is required to reproduce the velocity maps without taking into account the contributions from the turbulence support as well.}($M_{\rm rot}$),  the systemic velocity of the galaxy ($v_{\rm sys}$), and the intrinsic velocity dispersion ($\sigma_{\rm gas}$).
For the galaxies that were fit with both bulge and disk components, we did not add additional free parameters, but we assumed that the bulge effective radius is equal to half of the exponential disk scale radius and that the bulge mass is as high as the stellar mass, which has been reported in the literature for our galaxies (see Table \ref{tab:properties}). 

\begin{figure}[ht!]
   \centering
   \includegraphics[width=\hsize]{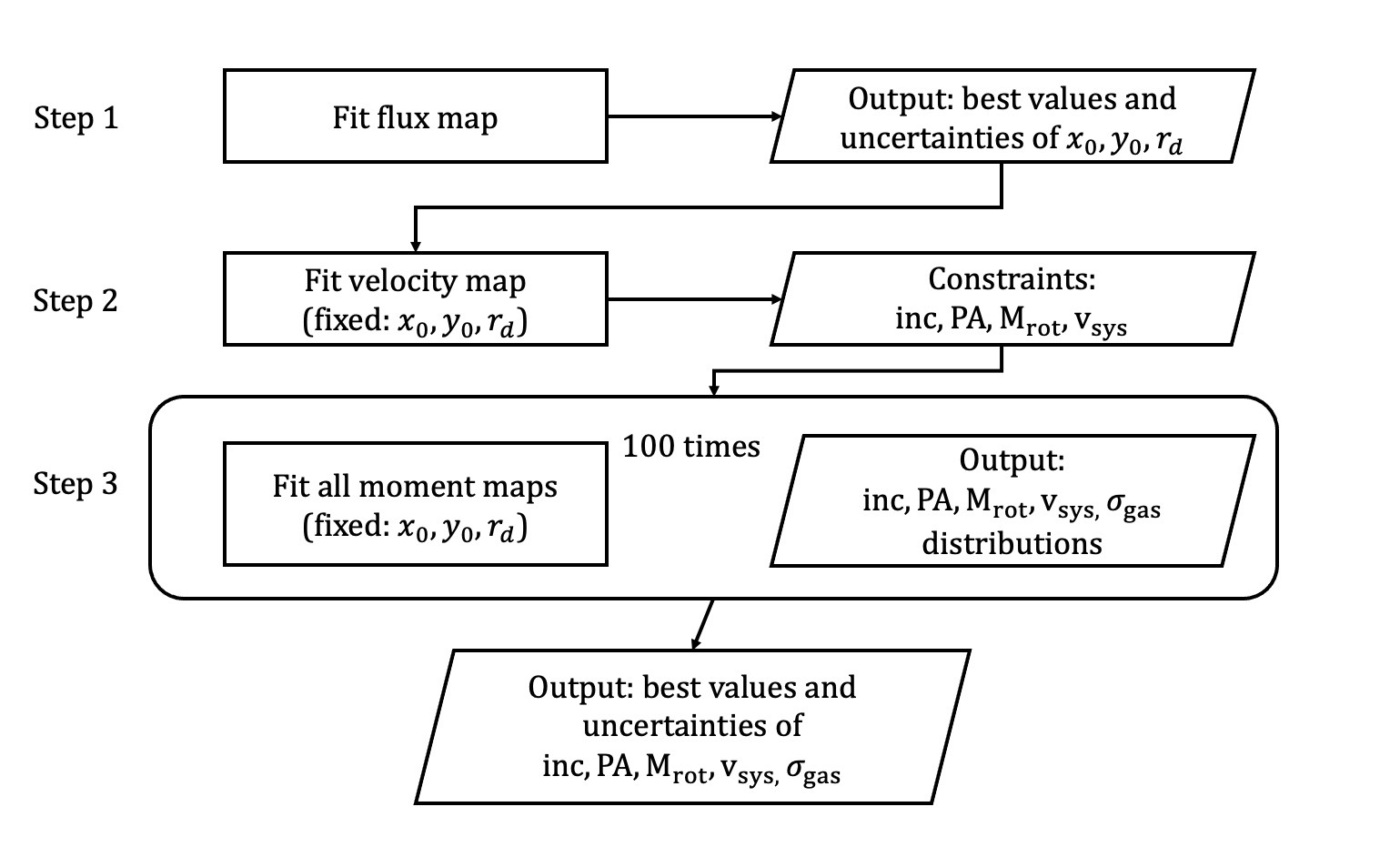}
    \caption{Summary of the method I fitting procedure to derive the model parameters.}
    \label{fig:flowchart}
   \end{figure}

The three main steps of the algorithm we adopted for the fitting procedure are shown in Figure \ref{fig:flowchart}. We initially fit the flux map and determined the center  and size of the galaxy. 
These best-fitting parameters were then kept fixed in the next steps.
In the second step, we fit the velocity map and inferred the best-fitting results for the $v_{\rm sys}$, $inc$, $PA,$ and $M_{\rm rot}$ , which were used as priors for the last steps, in which  all moment maps were fit simultaneously.
As the neighbor spatial pixels of the moment maps are correlated due to the beam-smearing effect (see Sec. \ref{sec:beamsmearing}), performing the standard kinematical fitting  on the maps might underestimate the parameter uncertainties because in the computation of the likelihood, all pixel are considered as independent values. 
Therefore, we opted to estimate the likelihood by  using only  independent pixels in the moment maps that were randomly extracted from the maps at a distance greater than the semi-major axis of the beam.
This procedure was performed 100 times for each target to ensure that the convergence of the free parameters did not depend on the pixel selection.
We then combined the walkers for every trial (after a 50\% burn-in phase) to derive the best parameters (median values) and error estimates (16th and 84th percentile) of the free parameter distribution for each target. We discuss this method in more detail in Appendix \ref{Correlation}.

\begin{figure*}[htb!]
   \resizebox{\hsize}{!}
    {  
    \includegraphics[width=\hsize]{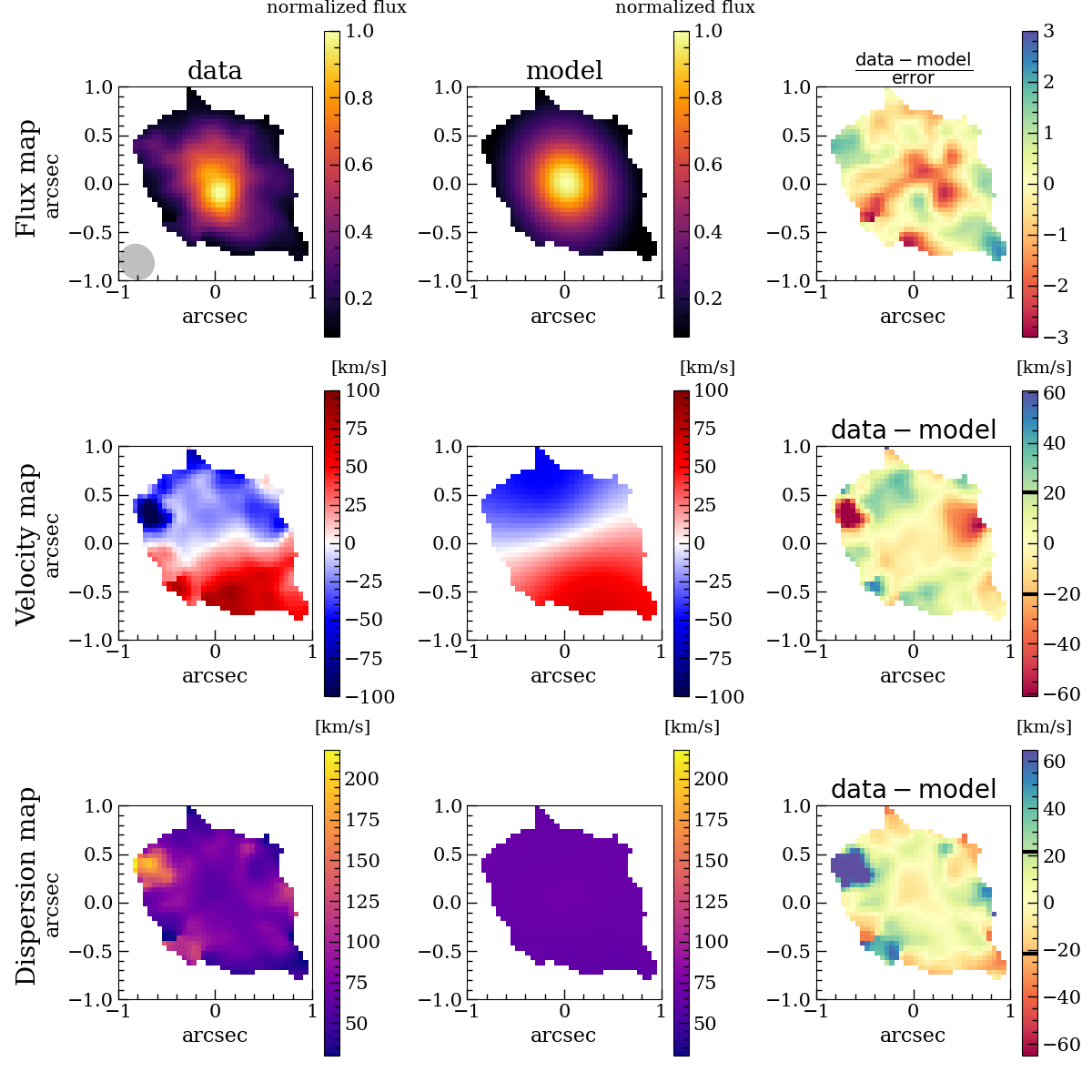}
    \includegraphics[width=\hsize]{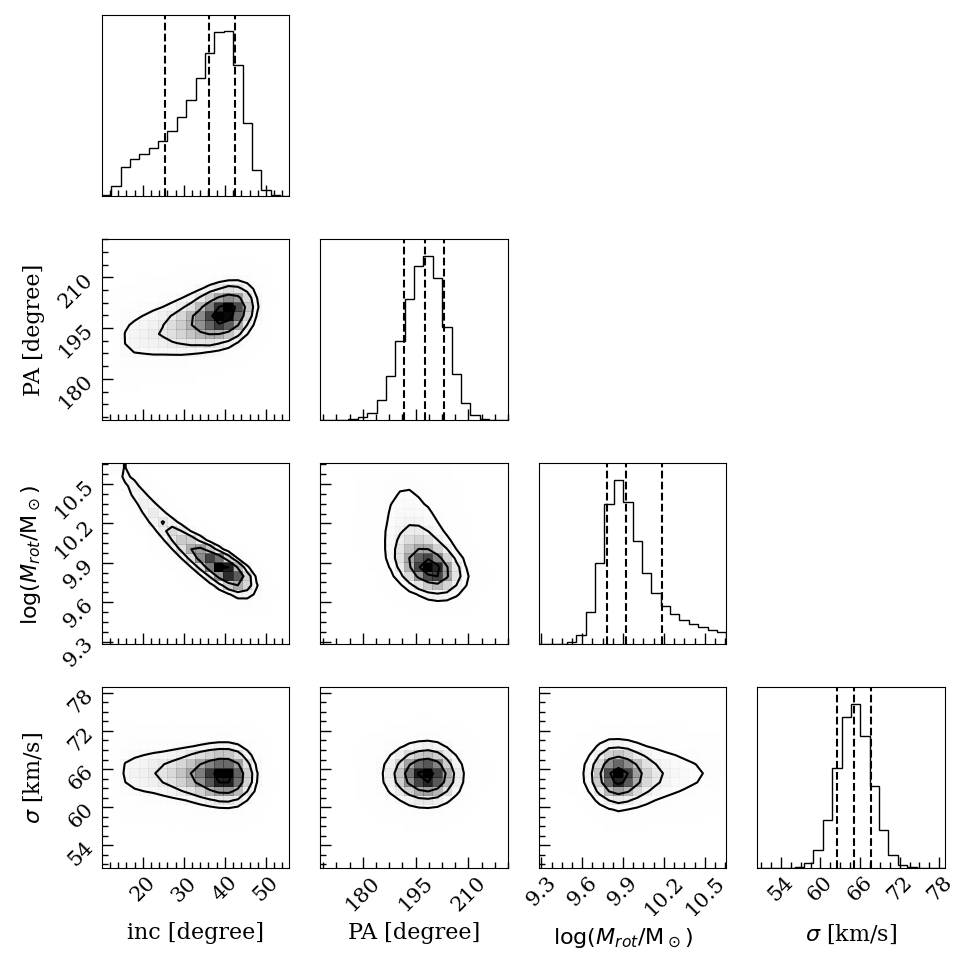}
        \includegraphics[width=\hsize]{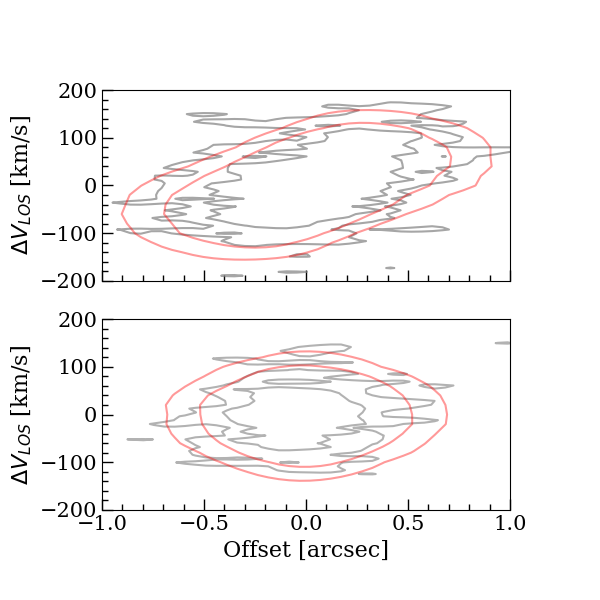}
    }
    \caption{Best-fitting results of HZ4 obtained with method I. 
    The left panels show the moment \cii\ maps: the normalized flux map (top), the velocity map (middle), and the velocity dispersion (bottom). From left to right, the panels illustrate the data, models, and residuals. The color bars of the residual range between $-3\sigma$ and $3\sigma$, and the black lines indicate $\pm1 \sigma$. The ALMA beam is shown as the gray ellipse in the flux map. 
    The multidimensional parameter space explored by step 3 of the method~I algorithm is shown in the corner plot on the right.}
    \label{Fig:HZ4_ciidistr}
   \end{figure*}

An example of the best-fitting results from method I is shown in Fig.~\ref{Fig:HZ4_ciidistr}.

\subsection{Method II: 1D kinematic fitting}\label{sec:method2}

Some of ALMA observations of our sample do not show a clear velocity gradient in the kinematic maps. 
The reasons why we could not see a clear velocity gradient could be:

\begin{itemize}
    \item low angular resolution of the observations; 
    
    \item low sensitivity; 
    
    \item the disk is not dominated by rotation, but it is instead dominated by random motion, hence we cannot see a velocity gradient. 
    
\end{itemize}
In these cases, the fitting process does not converge and thus returns a flat posterior distribution for the free parameters.
To overcome this problem and gain some information regarding the kinematics of these galaxies, we determined the kinematic properties from the spatially integrated spectrum. The spatially integrated spectrum is mainly used in radio observations of the HI emission line because it can show the kinematical properties of spatially unresolved galaxies \citep{Roberts:1978, yu:2020, Stewart:2014}. 
The basic idea is to determine the velocity dispersion by comparing the integrated spectrum of the observations with the integrated spectrum of the model datacube.
For this purpose, we need to limit the number of free parameters. In particular, we need to estimate the radius and mass of the galaxy a priori because these parameters have a strong impact on the shape of the line profile \citep{Roberts:1978, deBlok:2014} and are very  degenerate with the other disk inclination and intrinsic velocity dispersion parameters \citep{Kohandel:2019}.

We initially fit the flux map (step 1 of method I) and inferred both the scale length radius and a first constraint on the disk inclination. 
We then derive the mass as

 \begin{equation}
     M_{\rm rot} = M_{*} +M_{\rm gas}
 ,\end{equation}
where $M_*$ is the stellar mass, based on observations reported in the literature  \citep{Capak:2015, Willott:2015, Matthee:2017, Smit:2018, Harikane:2020, Herrera-Camus:2021, Fudamoto:2021, Bouwens:2022, Dayal:2022,  Schouws:2022}, and $M_{\rm gas}$ is the gas mass estimated by using the empirical relation by \cite{Zanella:2018},
\begin{equation}
    M_{\rm gas} = \alpha_{\rm [CII]} \cdot L_{\rm [CII]}
,\end{equation}
with $\alpha_{\rm [CII]}={\rm  30 M_\odot / L_\odot}$. The uncertainty on the conversion factor between \cii luminosity and  gas mass is about 0.2~dex.

After the scale radius and mass of the galaxy were estimated, the only free parameters of our datacube model are  the disk inclination $inc$ and the intrinsic velocity dispersion \sigmag.

In the fitting procedure, the model spectrum, $F^{\rm datacube}$, was extracted in the model datacube from the same region defined in the observations. In addition to the disk inclination and  \sigmag, we also added two other free parameters to match the normalization of the spectrum ($a$) and any offset due to the continuum emission ($b$). In conclusion, the spectrum of each galaxy was fit with a spectrum $S^{\rm model}$ defined as 
\begin{equation}
  S^{\rm model}(a,b,\sigma_{\rm gas}, inc) = b + a F^{\rm datacube}(\sigma_{\rm gas}, inc)
.\end{equation}

We fit the data spectrum by using an MCMC procedure and adopting a flat prior distribution for the velocity dispersion in the range 0 km s$^{-1}$ and 300 km s$^{-1}$. Since the inclination plays a major role in the global profile shape and the subsequent determination of velocity dispersion of the galaxy, we left as parameter the inclination assuming flat priors between $ \pm 3 \sigma$ of the best-fitting  inclination estimated from the flux map. 

We note that by applying method II to the galaxies that were fit using method I, the best-fitting velocity dispersion is systematically higher by 1.6 times than that obtained from method I (Appendix \ref{Confronto}). Hence, we applied a correction factor of 1.6 to the results inferred from method II.

\begin{figure*}[h]
   \resizebox{\hsize}{!}
    {   
        \includegraphics[width=\hsize]{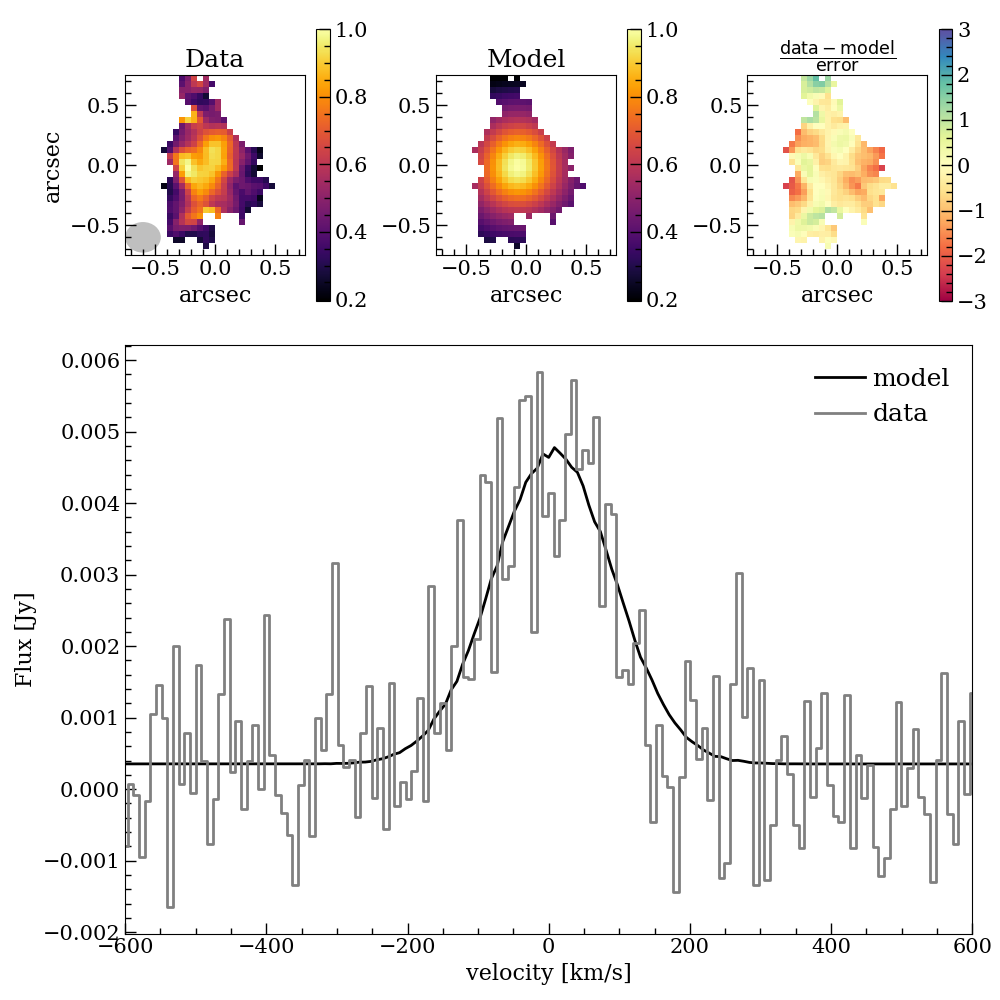}
        \includegraphics[width=\hsize, clip]{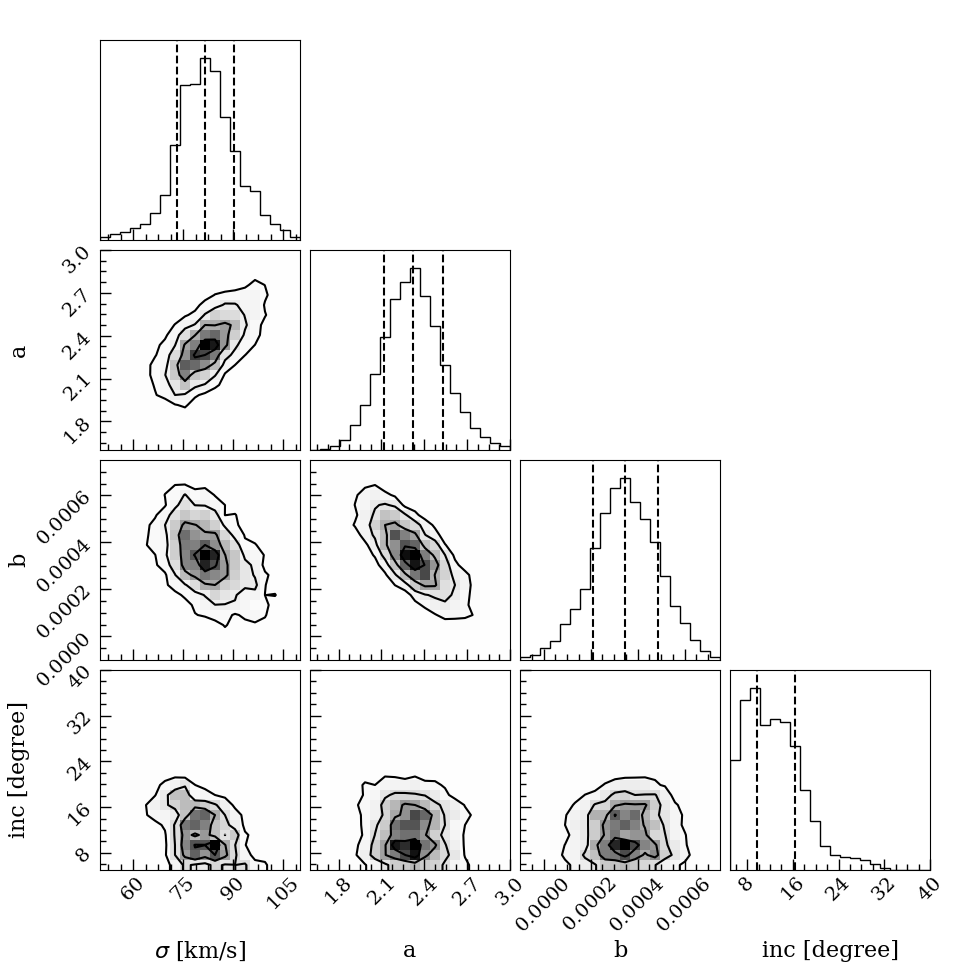}    
        }
    \caption{Best-fitting results from method II for the target HZ9. The top left panels illustrate the observed flux map, model, and residuals. The ALMA beam  is shown as  the gray ellipse in the bottom left corner of the observed flux map.  The bottom right panel shows the observed integrated spectrum and the best-fitting model. 
    The multidimensional parameter space explored by the method II algorithm is shown in the corner plot on the right.
          }  
    \label{Fig:hz9}
   \end{figure*}

Fig.~\ref{Fig:hz9} illustrates an example of the best-fitting results from method II.

\begin{table*}[ht!]
\centering
\caption{Results of the kinematical fitting for our sample of galaxies.}
\footnotesize   
\begin{tabular}{llllllllllll}
\hline
\hline
Name &
  Line &
  Beam &
  Method &
  $z$ &
  $r_D$ &
  inc &
  PA &
  $\log(\frac{M_{\rm rot}}{\rm M_{\odot}})$ &
  $V_{\rm max}$ &
  $\sigma_{\rm gas}$ &
  $V_{\rm max}/\sigma_{\rm gas}$ \\ \smallskip
  &
   &
  (arcsec) &
   &
   &
  (kpc) &
  (degree) &
  (degree) &
   &
  (km/s) &
  (km/s) &
   \\
\textit{\small{(1)}} &
  \textit{\small{(2)}} &
  \textit{\small{(3)}} &
  \textit{\small{(4)}} &
  \textit{\small{(5)}} &
  \textit{\small{(6)}} &
  \textit{\small{(7)}} &
  \textit{\small{(8)}} &
  \textit{\small{(9)}} &
  \textit{\small{(10)}} &
  \textit{\small{(11)}} &
  \textit{\small{(12)}} \\  \hline \smallskip
DLA0817g  \tablefootmark{b} &
  \cii &
  0.21$\times$0.15 &
  I &
  $4.249 $ &
  $1.13 ^{+0.03} _{-0.03}$ &
  $41_{-5}^{+5}$ &
  $108 \pm 2$ &
  $10.8_{-0.2}^{+0.2}$ &
  $358 ^{+94} _{-74}$ &
  $71^{+5}_{-5}$ &
  $5.04 ^{+1.81} _{-1.31}$ \\ \smallskip
ALESS 073.1 \tablefootmark{b} &
  \cii &
  0.17$\times$0.14 &
  I &
  $4.746 $ &
  $1.209 ^{+0.002} _{-0.002}$ &
  $32 ^{+3} _{-2}$ &
  $45 ^{+1} _{-1}$ &
  $10.7_{-0.1}^{+0.1}$ &
  $324 ^{+91} _{-34}$ &
  $54 ^{+1} _{-3}$ &
  $6.00 ^{+1.83} _{-0.91}$ \\ \smallskip
HZ7 &
  \cii &
  0.35$\times$0.32 &
  I &
  $5.245 $ &
  $1.6 ^{+0.1} _{-0.1}$ &
  $<47$ &
  $269^{+29}_{-29}$ &
  $9.7^{+0.6}_{-0.7}$ &
  $81 ^{+83} _{-45}$ &
  $85 ^{+12}_{-10}$ &
  $0.95 ^{+1.23} _{-0.58}$ \\ \smallskip
HZ9 &
  \cii &
  0.30$\times$0.25 &
  II &
  $5.529$ &
  $2.7^{+0.2}_{-0.2}$ &
  ... &
  ... &
  $10.8 \pm 0.4$ &
  $265^{+155}_{-97}$ &
  $51^{+10}_{-9}$ &
  $5.19^{+4.80}_{-2.44}$ \\ \smallskip
HZ4 &
  \cii &
  0.39$\times$0.36 &
  I &
  $5.534$ &
  $2.07^{+0.04}_{-0.04}$ &
  $37^{+6}_{-10}$ &
  $199^{+5}_{-6}$ &
  $10.0^{+0.2}_{-0.1}$ &
  $108 ^{+28} _{-12}$ &
  $66^{+3}_{-3}$ &
  $1.64 ^{+0.52} _{-0.25}$ \\ \smallskip
J1211 &
  \cii &
  0.80$\times$0.57 &
  I &
  $6.019$ &
  $3.1 ^{+0.1}_{-0.1}$ &
  $45_{-20}^{+16}$ &
  $272^{+16}_{-14}$ &
  $10.3_{-0.3}^{+0.5}$ &
  $149 ^{+116} _{-43}$ &
  $58_{-11}^{+9}$ &
  $2.57 ^{+3.07} _{-1.01}$ \\ \smallskip
 &
  \oiii &
  0.76$\times$0.57 &
  I &
  $6.029$ &
  $1.1 ^{+0.1} _{-0.1}$ &
  $44^{+22}_{-22}$ &
  $293^{+26}_{-23}$ &
  $10.2_{-0.5}^{+0.6}$ &
  $155 ^{+160}_{-70}$ &
  $54^{+18}_{-24}$ &
  $3.12 ^{+3.71} _{-1.27}$ \\ \smallskip
J0235 &
  \cii &
  0.88$\times$0.70 &
  II &
  $6.076 $ &
  $1.8^{+0.4}_{-0.5}$ &
  ... &
  ... &
  $10.6 \pm 0.3$ &
  $219^{+107}_{-69}$ &
  $72^{+101}_{-50}$ &
  $3.04^{+11.77}_{-2.16}$ \\ \smallskip
 &
  \oiii &
  0.69$\times$0.63 &
  II &
  $6.090 $ &
  $1.2^{+0.1}_{-0.1}$ &
  ... &
  ... &
  $10.6 \pm 0.3$ &
  $248^{+108}_{-75}$ &
  $126^{+66}_{-53}$ &
  $1.96^{+2.90}_{-1.06}$ \\ \smallskip
CLM1 &
  \cii &
  0.32$\times$0.29 &
  II &
  $6.153 $ &
  $0.9^{+0.1}_{-0.1}$ &
  ... &
  ... &
  $10.3 \pm 0.3$ &
  $195^{+92}_{-62}$ &
  $42^{+16}_{-14}$ &
  $4.64^{+5.60}_{-2.34}$ \\ \smallskip
J0217 &
  \cii &
  0.75$\times$0.66 &
  II &
  $6.191 $ &
  $4.1^{+0.8}_{-1.1}$ &
  ... &
  ... &
  $10.7 \pm 0.3$ &
  $136^{+63}_{-42}$ &
  $>83$ &
  $<1.63$ \\ \smallskip
 &
  \oiii &
  0.68$\times$0.56 &
  I &
  $6.203$ &
  $0.8 ^{+0.2} _{-0.1}$ &
  $<60$ &
  $80_{-51}^{+59}$ &
  $<10.4$ &
  $<227$ &
  $111_{-32}^{+27}$ &
  $<2.8$ \\ \smallskip
VR7 &
  \cii &
  0.56$\times$0.52 &
  II &
  \textit{$6.517$} &
  \textit{$2.3^{+0.1}_{-0.1}$} &
  ... &
  ... &
  $10.5 \pm 0.3$ &
  $189^{+79}_{-56}$ &
  $57^{+15}_{-15}$ &
  $3.31^{+3.06}_{-1.46}$ \\ \smallskip
UVISTA-Z-349 &
  \cii &
  1.65$\times$1.17 &
  I &
  $6.564 $ &
  $1.5^{+0.2}_{-0.2}$ &
  $36_{-18}^{+26}$ &
  $238^{+55}_{-36}$ &
  $9.7^{+0.8}_{-1.0}$ &
  $81 ^{+128} _{-56}$ &
  $77_{-21}^{+13}$ &
  $1.05 ^{+2.68} _{-0.77}$ \\ \smallskip
UVISTA-Z-004 &
  \cii &
  1.36$\times$1.15 &
  II &
  $6.669 $ &
  $1.7^{+0.2}_{-0.2}$ &
  ... &
  ... &
  $10.4 \pm 0.3$ &
  $172^{+78}_{-54}$ &
  $37^{+24}_{-19}$ &
  $4.64^{+9.24}_{-2.71}$ \\ \smallskip
UVISTA-Z-049 &
  \cii &
  1.45$\times$1.16 &
  II &
  $6.716$ &
  $1.0^{+0.1}_{-0.1}$ &
  ... &
  ... &
  $10.4 \pm 0.3$ &
  $208^{+94}_{-64}$ &
  $51^{+46}_{-32}$ &
  $3.27^{+4.30}_{-1.83}$ \\ \smallskip
UVISTA-Z-019 &
  \cii &
  1.39$\times$1.21 &
  II &
  $6.740$ &
  $2.6^{+0.1}_{-0.1}$ &
  ... &
  ... &
  $10.5 \pm 0.3$ &
  $188^{+78}_{-55}$ &
  $52^{+21}_{-17}$ &
  $3.61^{+3.98}_{-1.79}$ \\ \smallskip
COS-29 &
  \cii &
  0.45$\times$0.34 &
  I &
  $6.794$ &
  $1.5 ^{+0.2}_{-0.2}$ &
  $<60$ &
  $146^{+20}_{-33}$ &
  $9.5^{+0.6}_{-0.8}$ &
  $70 ^{+83}_{-44}$ &
  $37^{+10}_{-13}$ &
  $1.89 ^{+4.48} _{-1.34}$ \\ \smallskip
COS-30 &
  \cii &
  0.43$\times$0.34 &
  I &
  $6.840$ &
  $1.3 ^{+0.3}_{-0.3}$ &
  $52_{-9}^{+8}$ &
  $73\pm 5$ &
  $10.7_{-0.1}^{+0.2}$ &
  $264 ^{+34} _{-55}$ &
  $68_{-13}^{+14}$ &
  $3.88 ^{+1.54} _{-1.33}$ \\ \smallskip
 &
  \oiii &
  0.89$\times$0.60 &
  II &
  $6.851 $ &
  $1.5^{+0.1}_{-0.1}$ &
  ... &
  ... &
  $10.2 \pm 0.3$ &
  $145^{+63}_{-44}$ &
  $>125$ &
  $<1.16$ \\ \smallskip
UVISTA-Z-001 &
  \cii &
  1.40$\times$1.14 &
  I &
  $7.057 $ &
  $1.3 ^{+0.3} _{-0.3}$ &
  $<65$ &
  $137^{+25}_{-32}$ &
  $10.5^{+0.6}_{-0.7}$ &
  $201 ^{+234} _{-116}$ &
  $51^{+34}_{-32}$ &
  $3.94 ^{+18.95} _{-2.94}$ \\ \smallskip
UVISTA-Y-004 &
  \cii &
  1.37$\times$1.25 &
  I &
  $7.077$ &
  $1.7 ^{+0.2}_{-0.2}$ &
  $<63$ &
  $312^{+24}_{-19}$ &
  $10.1^{+0.7}_{-1.0}$ &
  $124 ^{+144} _{-85}$ &
  $61^{+16}_{-30}$ &
  $2.03 ^{+6.61} _{-1.53}$ \\ \smallskip
UVISTA-Y-003 &
  \cii &
  1.64$\times$1.31 &
  II &
  $7.293$ &
  $1.6^{+0.1}_{-0.1}$ &
  ... &
  ... &
  $10.7 \pm 0.3$ &
  $281^{+120}_{-84}$ &
  $63^{+23}_{-27}$ &
  $4.46^{+6.67}_{-2.16}$ \\ \smallskip
UVISTA-Y-879 &
  \cii &
  1.44$\times$1.25 &
  I &
  $7.357 $ &
  $3.3 ^{+0.4} _{-0.4}$ &
  $<65$ &
  $143^{+22}_{-37}$ &
  $10.6^{+0.6}_{-1.2}$ &
  $212 ^{+215} _{-159}$ &
  $119^{+36}_{-48}$ &
  $1.78 ^{+4.23} _{-1.44}$ \\ \smallskip
SUPER8 &
  \cii &
  1.44$\times$1.27 &
  II &
  $7.351 $ &
  $2.2^{+0.4}_{-0.5}$ &
  ... &
  ... &
  $10.2 \pm 0.3$ &
  $134^{+61}_{-40}$ &
  $74^{+100}_{-53}$ &
  $1.59^{+8.78}_{-1.22}$ \\ \smallskip
UVISTA-Y-001 &
  \cii &
  1.57$\times$1.28 &
  I &
  $7.660 $ &
  $1.5 ^{+0.1}_{-0.1}$ &
  $<60$ &
  $122^{+31}_{-42}$ &
  $9.8^{+0.8}_{-0.9}$ &
  $90 ^{+139} _{-59}$ &
  $79^{+11}_{-18}$ &
  $1.14 ^{+1.22} _{-0.79}$ \\ \hline
\end{tabular}

\tablefoot{{\it (1)} target name; {\it (2)} observed line; {\it (3)} beam FWHM; {\it (4)} best-fitting method; {\it (5)} redshift of the FIR line; {\it (6)} scale radius of the exponential disk; {\it (7)} disk inclination in degrees; {\it (8)} position angle of the galaxy; {\it (9)} logarithm of the dynamical mass in solar masses; {\it (10)} maximum rotational velocity as computed for an exponential disk with the dynamical mass and the scale radius obtained from the fitting; {\it (11)} best-fit velocity dispersion; {\it (12)} ratio of the maximum rotational velocity and the velocity dispersion.\\
\tablefootmark{b}{best-fit results with the model exponential disk + stellar bulge.\\}}
\label{tab:results}
\end{table*}

\section{Results}
\label{sec:results}

In this section, we report the results from the kinematic fitting by adopting the rotating models described in Sec. \ref{kinmodel}. Table  \ref{tab:results} summarizes the results provided by method I and II.  In addition to Fig. \ref{Fig:HZ4_ciidistr} and \ref{Fig:hz9}, the best-fitting result and posterior distribution of free parameter for each target are presented in Appendices \ref{Resultsmethod1} and  \ref{Resultsmethod2}.
For the galaxies that were fit by an exponential disk alone and by the model composed of bulge and exponential components, we present the model parameters for the model with the smaller $\chi^2$. 
Due to the poor angular resolution of some observations, we were not able to constrain the disk inclination, dynamical mass, and velocity dispersion of all galaxies. In these cases, we inferred  upper limits to the kinematics parameters. 

The values of the velocity dispersion obtained for galaxies at $z>5$ that were studied with method I range between 37~\kms and 132~\kms, with a median velocity dispersion of $67^{+15}_{-13}$~\kms.
For the rotational support, we find a median value of $V_{\rm max}/\sigma_{\rm gas} = 2$ that is compatible with turbulent but still rotationally supported disk galaxies.

Galaxies analyzed with method II show a velocity dispersion, corrected by a factor 1.6 (see Sec.~\ref{sec:method2}), between 37 and 126 km s$^{-1}$ and a median value of $60^{+19}_{-11}$~\kms,  which is compatible with the median value obtained with the other method. The median value of $V_{\rm max}/\sigma_{\rm gas}$ is 3.2, which is slightly higher than the result found with method I.
This can be due to the uncertainties on the mass inferred from the luminosity of the \cii emission and the rest-frame UV observations, to the uncertainty of the scale radius of the exponential disk obtained with the fitting of the flux map, or to the  uncertainties on the inclination estimates, which can broaden the line. 

Interestingly, we note that for the galaxies that are observed in the two FIR lines, the velocity dispersion of \cii\ and \oiii\ are similar, but they trace different gas phases of the interstellar medium. In only one out of four galaxies is the \oiii\ velocity twice higher than the velocity dispersion  obtained by fitting the \cii map. The \oiii and \cii observations in this sample predict similar values for the velocity dispersion of the galaxies. Simulations and observations showed that different gas tracers give rise to different velocity dispersions \citep{Kretschmer:2022, Ejdetjarn:2022}.

The  rotation-to-random motion ratios estimated in our sample are lower on average than the $V_{\rm max}/\sigma_{\rm gas}$ ratios observed in the local galaxies and in the dusty massive galaxies at $z>4$ reported in the literature.
To exclude any possible bias due to the different fitting algorithms, we also analyzed four high-redshift galaxies that were studied by \citet{Lelli:2021}, \citet{Neeleman:2020} \citet{Jones:2021}, \citet{Herrera-Camus:2022}, and \citet{Posses:2022}, and we compare their results with ours.

\citet{Lelli:2021} studied the kinematic properties of the massive dusty galaxy ALESS 073.1 using the software  \textsc{$^{3D}$Barolo} \citep{DiTeodoro:2016}, which directly takes the beam-smearing effect into account on the datacube and uses a 3D fitting algorithm with tilted-ring models to infer the kinematical properties of galaxies. 
Our best-fitting results based on the moment maps agrees with the results found by \citet{Lelli:2021}, who reported low values of $\sigma_{\rm gas}$ ranging from $\sim 60$ km s$^{-1}$ in the internal regions to $\sim 10$ km s$^{-1}$ in the outer region. Our results in Figure \ref{Fig:aless} show a bimodal distribution of the velocity dispersion that converges at $\sim 30$ km s$^{-1}$ when most of the selected pixels were in the outer regions, and to $\sim 56 $ km s$^{-1}$ when the central pixels were selected. 
Despite the two different methods of fitting, our results agree with the kinematical cold disk with values of $V_{\rm max}/\sigma_{\rm gas}$ comparable to those obtained for local galaxies.
\citet{DeBreuck:2014} analyzed the same galaxy using a 2D fitting algorithm of data with a poorer resolution and found values of $V_{\rm max}/\sigma_{\rm gas} \sim 3.1,$ but this is mostly due to the underestimated rotational velocity ($\sim$ 120 km s$^{-1}$) and not to the overestimated velocity dispersion ($\sigma_{\rm gas} = 40 \pm 10$ km s$^{-1}$).

Another galaxy that was observed with ALMA and was recently analyzed by \citet{Neeleman:2020} and \citet{Jones:2021} is  DLA0817g. \citet{Neeleman:2020} exploited the higher-resolution observations ($\sim 0.2''$ as used in this work) and used an approach similar to ours, in which the flux was fit assuming a rotating thin disk with an exponential brightness profile. They compared it with the results from \textsc{$^{3D}$Barolo}.
 \citet{Jones:2021} studied the kinematics by exploiting lower-resolution data ($\sim 1''$) using \textsc{$^{3D}$Barolo}.
Their results agree with those by \citet{Neeleman:2020}, who they found  $\sigma_{\rm gas}=$ 80 $^{+13}_{-11}$  km s$^{-1}$ and $V_{\rm max}/\sigma_{\rm gas} = $3.4$ ^{+1.1}_{-0.3}$. We note that we obtain a higher value for the maximum rotational velocity because we also considered the bulge component in our fitting because this model better reproduced the kinematic maps.  When we use the exponential disk alone, we obtain a $V_{\rm max}/\sigma_{\rm gas} = 4.01^{+0.21}_{-0.73}$ , while the velocity dispersion that we obtain is the same using the bulge+exponential disk or the exponential disk model.
Our results disagree with the results found by \citet{Jones:2021}, who reported a velocity dispersion that is lower by a factor of two. 

\citet{Herrera-Camus:2022} presented the kinematical study of the target HZ4 by modeling the galaxy as a turbulent disk plus a dark matter halo.
They found values of the velocity dispersion of $\sigma_{\rm gas} =$ 65.8 $^{+2.9}_{-3.3}$ \kms and a value of $V_{\rm max}/\sigma_{\rm gas}=2.2$. These results are comparable within the uncertainties to our best-fitting values.

Finally, the target COS-29 was analyzed by \citet{Posses:2022} using the algorithm \textsc{$^{3D}$Barolo}. They found that the average velocity dispersion has an upper limit of 30 \kms and a ratio of the velocity and the velocity dispersion greater than 1.4, which is comparable to the values found in this work.

Overall, we find that the parameters inferred from our algorithms agree with the results found by other high-redshift studies, and we found no evidence of a bias in the recovery of the velocity dispersion. In the next sections, we therefore compare our results to those of other studies from $z=0$ to $z=8$ to understand the evolution of turbulence with redshift and galaxy properties better.

   \begin{figure*}[ht!]
   \resizebox{\hsize}{!}
    {   
        \includegraphics{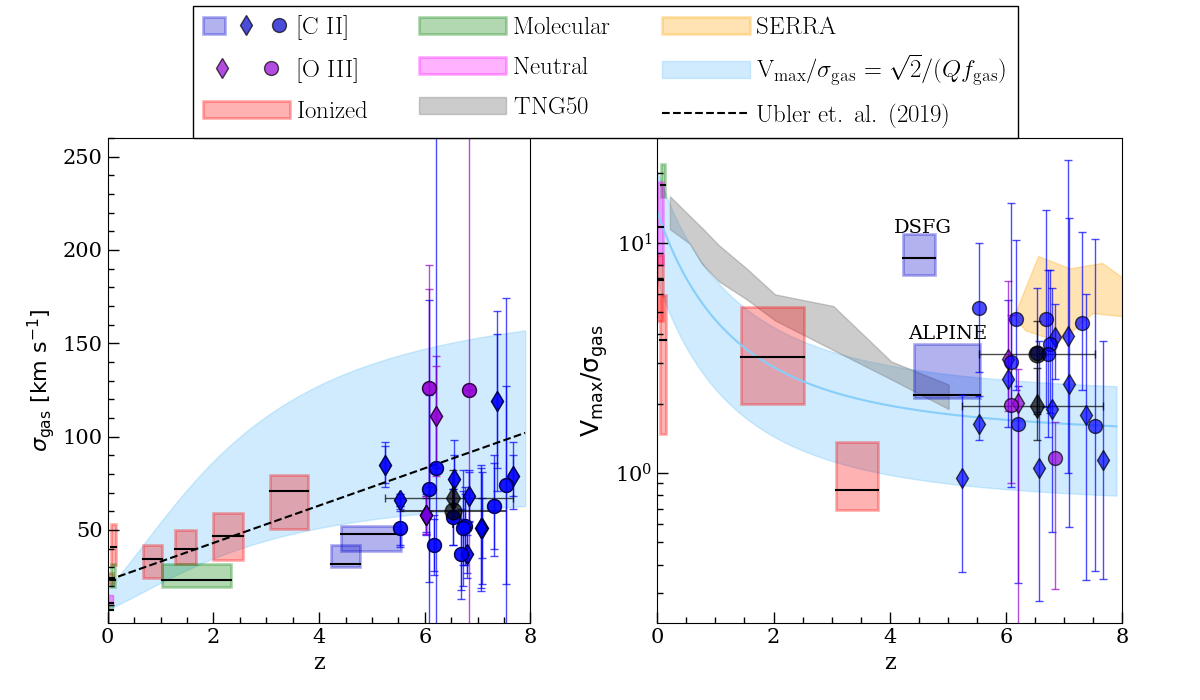}        
    }
    \caption{Evolution of the velocity dispersion (left) and the $V_{\rm max}/\sigma_{\rm gas}$ ratio (right) with redshift. Blue and purple marks are the values obtained from our sample by analyzing the \cii\ and \oiii\ observations, respectively. Diamonds are obtained by  fitting the three moment maps (method I), and  circles represent the results found for the marginally resolved galaxies (method II). Previous $\sigma$  and $V_{rot/\sigma}$ estimates reported in the literature are represented as color-coded rectangles: the observations with the \cii emission line at $4<z<5$ by \citet{Fraternali:2021}, \citet{Rizzo:2020},  \citet{Rizzo:2021}, \citet{Lelli:2021}, \citet{Sharda:2019}, and \citet{Neeleman:2020} at $4<z<6$ from \citet{Jones:2021} are plotted in blue. Red rectangles show the results of the studies of ionized gas (H$_\alpha$, [O III] $\lambda$ 5007) in $z<4$ galaxies \citep{Epinat:2010,Green:2014,Turner:2017,Ubler:2019}. Green rectangles represent the measurements derived from molecular gas \citep{Mogotsi:2016,Girard:2021} (CO). The pink symbols were derived with traced of neutral gas (H I) at $z = 0$  \citep{Mogotsi:2016}. 
    The boxes extend from the 25th to the 75th percentile of the data in the samples, and  solid black lines show the median of the value. The light blue area represents the predictions by \citet{Wisnioski:2015}, and the solid light blue line represents the $V_{\rm max}/\sigma_{\rm gas}$ values predicted for $Q=1$.
        The dashed black line represents the evolution of the best-fitting velocity dispersion  trend  by \citet{Ubler:2019}. The gray and orange areas show the results obtained by the TNG50 \citep{Pillepich:2019} and SERRA simulations (Kohandel et al. in prep), respectively. }  
    \label{Fig:evolution}
   \end{figure*}

\section{Discussion}
\label{sec:discussion}
\subsection{Velocity dispersion and $V_{\rm max}/\sigma_{\rm gas}$ evolution with redshift}
\label{Velocitydispersionevolution}
The left panel of Figure \ref{Fig:evolution} shows the evolution of the gas velocity dispersion as a function of the redshift from $z=0$ to $z=8$. The best-fitting velocity dispersion estimates of our sample are shown with blue ([C$\scriptstyle\rm~II$]) and purple ([O$\scriptstyle\rm~III$]) marks (diamonds for the results found with method I, and circles for the results found with method II). The median values inferred for the sample of galaxies analyzed with methods I and II are reported as black marks.
In the figure, we also report previous results from the literature that are shown as boxes that extend from the 25th to 75th percentile of the data in their samples and are color-coded as a function of the tracer adopted to study the gas kinematics.
Differently from the conclusion reported by \citet{Lelli:2021}, \citet{Rizzo:2020, Rizzo:2021}, and \citet{Fraternali:2021}, who reported that high-redshift  ($z>4$) galaxies have low values of the velocity dispersion (i.e., $\sigma_{\rm gas} \sim 30-60$ km s $^{-1}$; blue region at $4<z<4.5$ in figure \ref{Fig:evolution}) with respect to theoretical models, our sample at $z\sim 5-7$ seems to be consistent with the prediction by \citet{Wisnioski:2015}  of the velocity dispersion evolution with redshift,\begin{equation}
\label{eq:Wisnioski}
    \sigma = \frac{1}{\sqrt{2}} V_{\rm max} Q f_{\rm gas}(z).
\end{equation}
The trend of eq.~\ref{eq:Wisnioski} with $Q=1$ and $100$ \kms$<V_{\rm max}< 250$  \kms\  is shown with the light blue shaded curve in the left panel of Figure \ref{Fig:evolution}. As the gas fraction increases with redshift \citep{Tacconi:2020}, the velocity dispersion also increases, reaching values of $\sim40-100$ km s $^{-1}$ at $z\sim 6-8$, independently of the tracer that is used to map the gas kinematics, which seems to be confirmed by our results based on two different FIR lines, \cii\ and \oiii.

The evolution of the $V_{\rm max}$/$\sigma_{\rm gas}$ ratio as a function of the redshift from $z=0$ to $z=8$ is shown in the left panel of Figure \ref{Fig:evolution}.
 We also superpose the expected decrease in $V_{\rm max}/\sigma_{\rm gas}$ with redshift based on the Toomre instability criterion \citep{Wisnioski:2015, Genzel:2011},
\begin{equation} 
\frac{V}{\sigma} = \frac{a}{Q f_{\rm gas}(z)}
.\end{equation}
The upper and lower boundaries of the  shaded light blue region represent the predicted $V_{\rm max}/\sigma$ for $Q=0.67$ and $Q=2$, respectively, and the solid light blue line represents the values predicted for $Q=1$ \citep{Wisnioski:2015}.
The shaded gray region illustrates the results from the TNG50 simulations \citep{Pillepich:2019}, and the shaded orange region reports the results from the SERRA simulations (\citealt{Pallottini:2022}, Kohandel et al. in prep) for the kinematic studied with the \cii emission line. 
While our  $V_{\rm max}/\sigma_{\rm gas}$ ratios agreet with the estimates found for galaxies in the ALPINE survey \citep{Jones:2021}, our analysis differs from the  results determined in the sample of galaxies studied by \citet{Sharda:2019},\citet{Rizzo:2020}, \citet{Rizzo:2021}, \citet{Fraternali:2021} and \citet{Lelli:2021}, hereafter the dusty star-forming galaxy (DSFG) sample (blue box at $4<z<5$).
The  DSFG sample indeed shows a level of rotational support similar to what is observed in the local Universe (i.e., $V_{\rm max}/\sigma\sim10$). In contrast, the sample targeted in this work and in the ALPINE survey has values of $V_{\rm max}/\sigma_{\rm gas}$ between 1 and 4 that are consistent with the values expected by \cite{Wisnioski:2015} and the trend in the TNG50 simulations \citep{Pillepich:2019} and is slightly lower than what was observed in the SERRA simulations ($V_{\rm max}/\sigma_{\rm gas}\sim 8-4$, Kohandel et al. in prep)

The discrepancy in the velocity dispersion and in  $V_{\rm max}/\sigma_{\rm gas}$ between our sample and the DSFGs suggests that the gas kinematic properties do not only depend on the redshift alone, but also on the properties of the galaxies themselves. In the next sections, we therefore analyze the properties of the galaxies and verify whether the gas kinematics depend on these properties.

\subsection{Sample properties}
\label{sec:sampleproperties}
To understand whether the physical properties of the galaxies can impact the galaxy kinematics, we analyzed the difference between the three samples (i.e, ALPINE, DSFG, and our sample) in the SFR-stellar mass diagram and dust content. The values of stellar mass, SFR, and dust mass were obtained from the literature (see Table \ref{tab:properties} for details).

   \begin{figure}[htb!]
   \centering
   \includegraphics[width=\hsize]{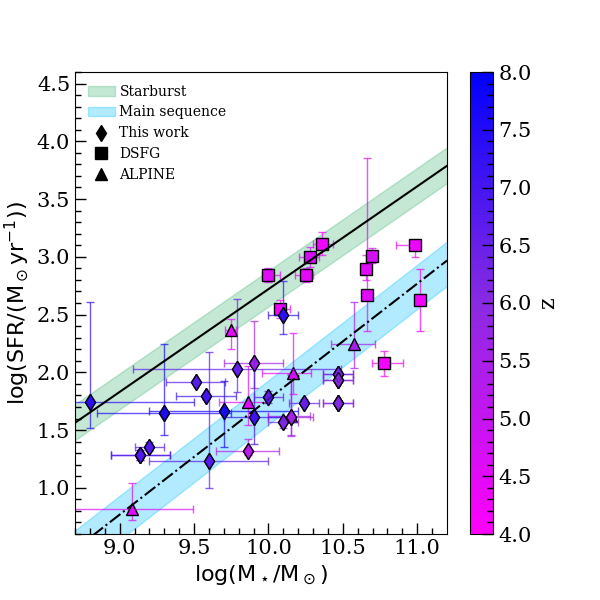}
      \caption{SFR as a function of the stellar mass for our sample of galaxies and the results found in the literature. The colors represent the redshift on the galaxy (see the color bar).
        The diamonds are the galaxies targeted in this work. The squares show the results found by \citet{Sharda:2019} \citet{Rizzo:2020}, \citet{Rizzo:2021}, \citet{Lelli:2021} and \citet{Fraternali:2021}. The triangles show the ALPINE sample \citep{Jones:2021}.
        In green we plot the starburst region at $z\sim 4-5$ as predicted by \citet{Caputi:2017}. In light blue we plot the main-sequence region extrapolated between $z \sim 4 - 8$ by \citet{Schreiber:2015}
              }
         \label{fig:mainsequence}
   \end{figure}
We analyzed the SFR-stellar mass diagram (Figure \ref{fig:mainsequence}), in which our galaxies, the ALPINE, and the DSFG sample are indicated with diamonds, triangles, and squares, respectively. 
We also plot the relation for main-sequence galaxies as parameterized by \cite{Schreiber:2015}, 
\begin{equation}
\log(SFR_{\rm MS}/({\rm M_\odot/yr})) = m - m_0 + a_0 r
-a_1 max(0, m - m_1 - a_2 r)^2,
\end{equation}
 extrapolated for $4<z<8$ \footnote{ $m = \log(M_{\star}/10^9{\rm M}_\odot)$, $r = log(1+z)$,  $m_0 = 0.5 \pm 0.07$, $a_0 = 1.5 \pm 0.15$, $a_1 = 0.3 \pm 0.08$,
$m_1 = 0.36 \pm 0.3$, and $a_2 = 2.5 \pm 0.6$},
and the empirical relation for starburst galaxies at $z \sim 4-5$ by \cite{Caputi:2017},
\begin{equation}
    \log(SFR_{\rm SB}/({\rm M_\odot/yr})) = (0.89 \pm 0.02)\log(M_*/{\rm M}_\odot )- 6.19^{+0.16}_{-0.15}.
\end{equation}

   \begin{figure}[htb!]
   \centering
   \includegraphics[width=\hsize]{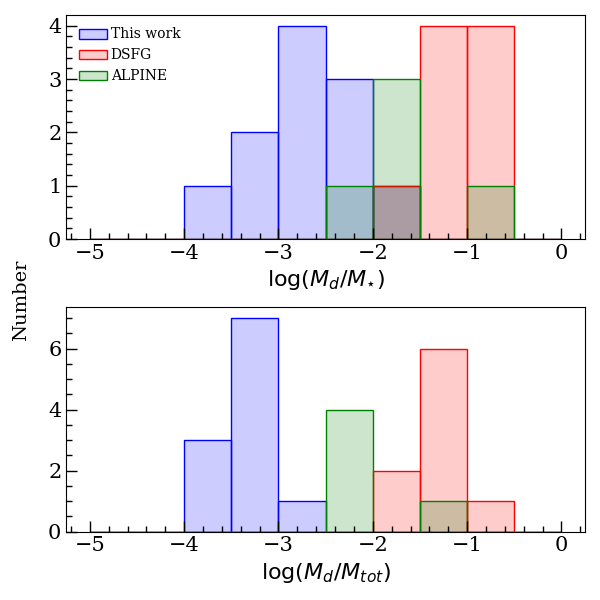}
      \caption{Dust mass content in the three galaxy samples targeted in this work. In the upper panel, we plot the distribution of the ratio of the dust mass and stellar mass for the sample targeted in this work (in blue) and the sample of DSFG targeted by \cite{Rizzo:2020, Rizzo:2021}, the starburst dusty galaxies \citep{Sharda:2019, Fraternali:2021} (in red), and the ALPINE sample \citep{Jones:2021} (in green).
      In the lower panel we present the distribution of the ratio of the dust mass and total baryonic mass ($M_{\rm gas} + M_*$) for the sample targeted in this work, the sample of DSFG targeted by other high-$z$ studies, and the ALPINE sample.}
         \label{fig:dust}
   \end{figure}
   
The ALPINE sample and our sample lie above the $z=5$ main-sequence relation, and the DSFGs are about 0.5 dex more massive and more consistent with the high-redshift starburst populations. 
Although at low redshift, starburst galaxies tend to higher velocity dispersion values than main-sequence galaxies \citep{Ubler:2019, Perna:2022}, our analysis reveals the opposite scenario in the distant Universe. The DSFG sample seems to be less turbulent than the high-$z$ main-sequence galaxies, suggesting that the gas is settled in a dynamically cold rotating disk despite the high star formation activity.    
The reason for the discrepancy in the velocity dispersion and $V_{\rm max}/\sigma$ between the two samples may therefore be ascribed to the different galaxy population properties of the selected galaxies.

In Figure \ref{fig:dust} we report the distributions for the three samples of $\log(M_{\rm dust}/M_*)$ and  $\log(M_{\rm dust}/M_{\rm tot})$ , where M$_{\rm tot}$ is the total baryonic mass computed as $M_{\rm gas} + M_*$ , and $M_{\rm dust}$ is the dust mass of the galaxy. The values of $M_{\rm dust}$ were obtained from \citet{Pozzi:2021} and \citet{Bethermin:2020} for the ALPINE sample, and from \citet{Aravena:2016} for the DSFG. While for the dust mass of the sample targeted in this work see Table \ref{tab:properties}.
For some galaxies, the dust temperature was assumed to be a fixed value. We therefore note that the correlation between the dust temperature and the dust mass may have overestimated some of the dust mass estimates.

The figure shows that while our sample and the ALPINE samples have a star-to-dust mass ratio of $10^{-3}$ on average, the DSFGs are characterized by $M_{\rm dust}/M_{\star}>10^{-2}$ , similar to what was observed in other high-$z$ DSFGs and in submillimeter galaxies \citep{Casey:2014}.

The low amount of dust and the high velocity dispersion in our sample with respect to DSFGs suggests that the mechanisms that cause the ISM to become turbulent might also have driven shocks that  affect the dust distribution.

\subsection{$V_{\rm max}/\sigma_{\rm gas}$ evolution with galaxy properties}
\label{sec:vsigmaevolution}

By combining the three samples studied at $z>4,$ we can also assess whether the $V_{\rm max}/\sigma_{\rm gas}$ ratio of our galaxies is correlated with stellar mass, dust mass, or gas fraction.  

Lower-redshift studies have shown a dependence on the $V_{\rm max}/\sigma_{\rm gas}$ and the stellar mass \citep{Kassin:2012, Wisnioski:2018, Forster:2020}, where more massive galaxies have higher values of $V_{\rm max}/\sigma_{\rm gas}$, while evidence of  rotating but turbulent disks are common in less massive galaxies. The cosmological simulation TNG50 \citep{Pillepich:2019} shows a slight but similar dependence on the stellar mass of the galaxy with greater $V_{\rm max}/\sigma$ with increasing stellar mass at all redshifts, while SERRA simulations predict a stronger dependence on the stellar mass, where $V_{\rm max}/\sigma$ at a fixed stellar mass remains almost constant for $6<z<8$.

\begin{figure}[htb!]
   \centering
   \includegraphics[width=\hsize]{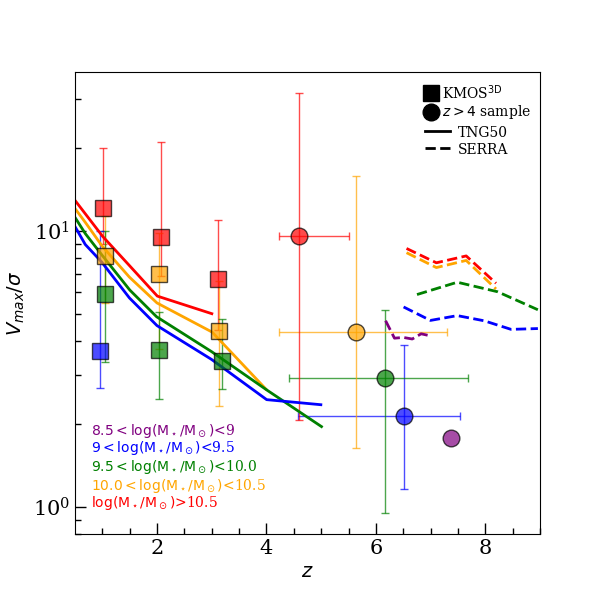}
      \caption{
      $V_{\rm max}/\sigma_{\rm gas}$ as a function of redshift and stellar mass. Squares represent the results from \textsc{KMOS$^{\rm 3D}$}  \citep{Wisnioski:2018}.
    Circles show the $V_{\rm max}/\sigma_{\rm gas}$ for the whole $z>4$ sample including the results of this work.
    Solid lines show the results obtained by TNG50 simulations \citep{Pillepich:2019}.  Dashed lines show the results obtained by SERRA simulations (Kohandel et. al. in prep). 
    Different stellar mass bins are represented with different colors.
              }
         \label{fig:stellar_massevolution}
   \end{figure}

Figure \ref{fig:stellar_massevolution}  shows the evolution of $V_{\rm max}/\sigma_{\rm gas}$ with redshift, color-coded with stellar mass bins.  The dependence of $V_{\rm max}/\sigma_{\rm gas}$ on stellar mass is still valid at $z>5,$ where galaxies with $M_{\star}>10^{10}$ have $V_{\rm max}/\sigma_{\rm gas}>3,$ while  $M_{\star}<10^{10}$ are characterized by lower values ($V_{\rm max}/\sigma_{\rm gas}\sim2$). As DSFGs and our sample cover two different ranges of stellar masses, the  discrepancy in $V_{\rm max}/\sigma_{\rm gas}$ values between the two populations might be due to a bias in the sample selection.

\begin{figure}[ht!]
   \centering
   \includegraphics[width=\hsize]{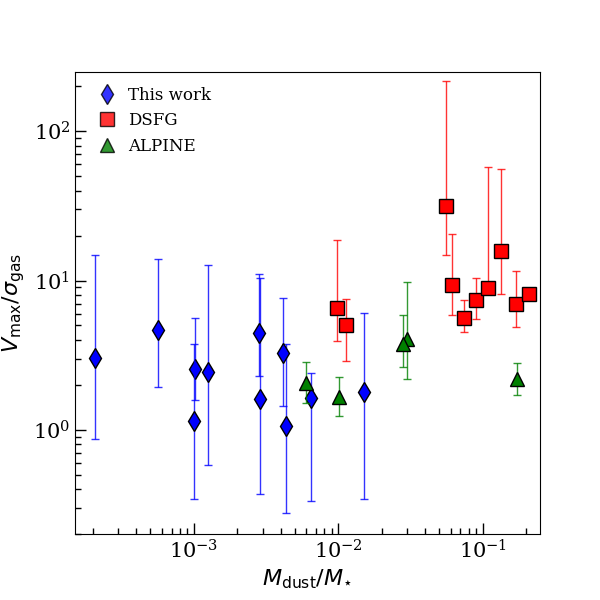}
      \caption{$V_{\rm max}/\sigma_{\rm gas}$ as a function of redshift, and the ratio of the dust mass and stellar mass.
      In red we show the DSFG sample \citep{Sharda:2019, Rizzo:2020, Rizzo:2021, Lelli:2021, Fraternali:2021}, in green we plot  the ALPINE sample \citep{Jones:2021}, and in blue we show the sample targeted in this work.}
         \label{fig:vsigmadust}
   \end{figure}

Analyzing the parameter space $V_{\rm max}/\sigma_{\rm gas}$  function of the ratio $ M_{\rm dust}/M_\star$ , we note a clear bimodality distribution between dusty galaxies and galaxies with low dust content. Figure \ref{fig:vsigmadust} shows that the kinematics of gas in high-$z$ galaxies is slightly related to the presence of dust in the interstellar medium.  The ISM of high-$z$ dusty galaxies seems to be less affected by star formation feedback and gravitational torques. This might imply that the processes that increase the gas velocity dispersion in the galaxies might also create shocks in the gas that destroy the dust.  

\begin{figure}[ht!]
   \centering
   \includegraphics[width=\hsize]{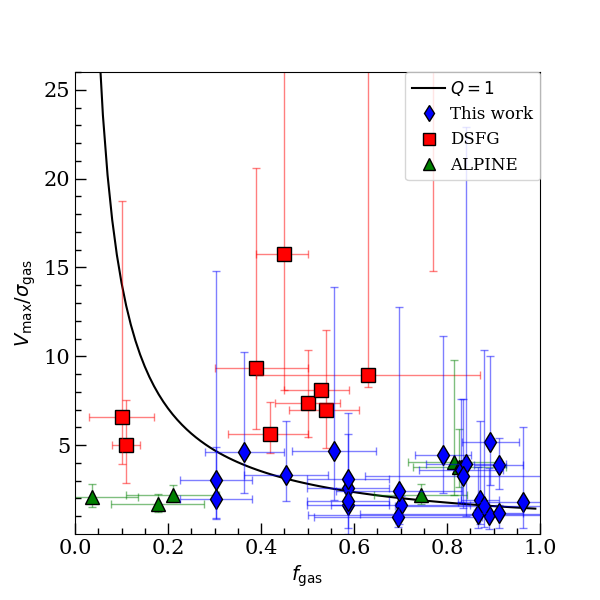}
      \caption{
      $V_{\rm max}/\sigma_{\rm gas}$ as a function of gas fraction. The black line represents the model by  \cite{Wisnioski:2015} with the assumption of Q=1 and a=$\sqrt{2}$.
    In dark blue, we show the data obtained from the sample of galaxies targeted in this work.}
         \label{fig:gasfraction}
   \end{figure}

Finally, we explored whether the $V_{\rm max}/\sigma$ ratio is driven by the gas fraction, as expected for a disk that is in equilibrium between gas heating and cooling, and assuming a Toomre parameter $Q=1$ (equation \ref{eq:Toomre_highz}; see also \citealt{Wisnioski:2015}). 
In Figure \ref{fig:gasfraction} we report the $V_{\rm max}/\sigma_{\rm gas}$ as a function of gas fraction for  $z>4$ galaxies.
Our data (blue marks) show a flat distribution between $f_{\rm gas}=0.3$ and  $f_{\rm gas}=1.0$, which is marginally consistent with the  relation by  \cite{Wisnioski:2015}. However, taking the distribution of $V_{\rm max}/\sigma_{\rm gas}$ values of all high-$z$ galaxies into account, we note that the dependence of $V_{\rm max}/\sigma_{\rm gas}$ on $f_{\rm gas}$ drops. Several galaxies show similar  $f_{\rm gas}$ but quite different $V_{\rm max}/\sigma_{\rm gas}$. This suggests that the state (i.e., $V_{\rm max}/\sigma_{\rm gas}$ ) of the rotating disk does not depend on the gas fraction alone.

Another possible explanation for the differences between the DSFG sample, this work, and the ALPINE sample is suggested by the theoretical work by \cite{Kretschmer:2022}. The zoom-in simulations developed by the authors show that when a galaxy is in a phase of constructive gas accretion (i.e., the cold gas streams are coplanar and corotate with the disk), the galaxy disk is then primarily supported by rotation with $V_{\rm max}/\sigma_{\rm gas} \sim 5$.
This phase only lasts for $ \text{about five}$ rotational periods ($\sim 400$ Myr), then the disk can be disrupted by merging episodes.
Therefore, we might see a difference because of the different merging rates for high- and low-mass galaxies. Massive galaxies undergo fewer major mergers than less massive galaxies \citep{Dekel:2020}, which means that the ordered disk can survive longer.
When this phase of low-velocity dispersion has a limited duration, however, this effect cannot explain why all of the galaxies in our reference sample are in this state while the majority of our galaxies are not.

\subsection{Drivers of the velocity dispersion}\label{section_drivers_of_velocity_dispersion}

   \begin{figure*}[ht!]
   \centering
   \resizebox{\hsize}{!}
    { \includegraphics{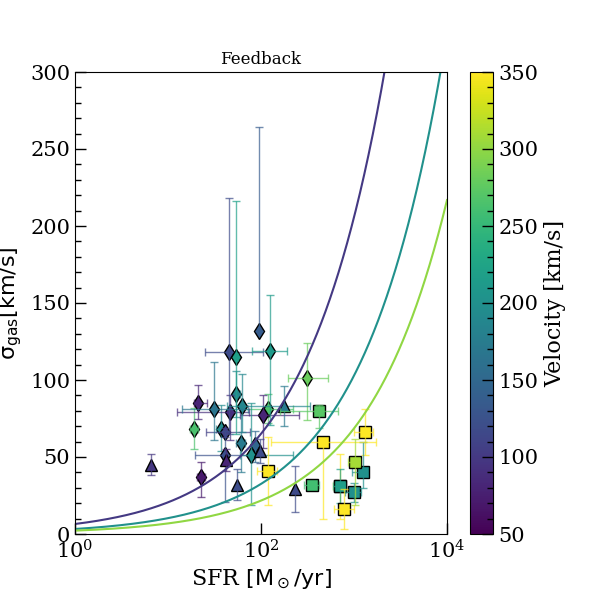}
      \includegraphics{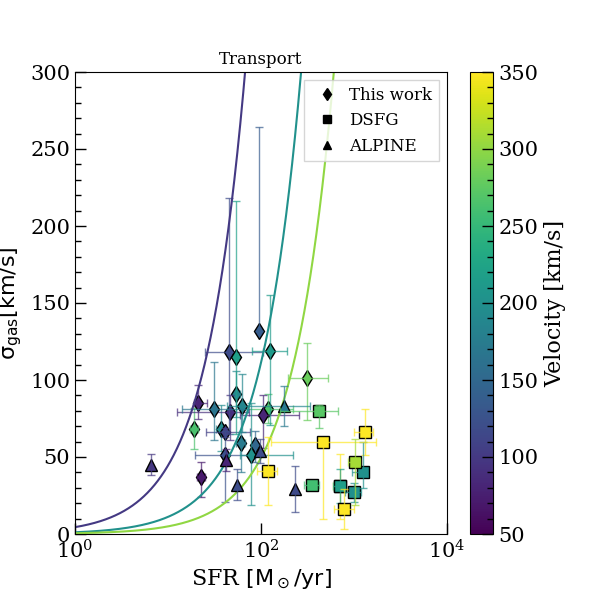}
      \includegraphics{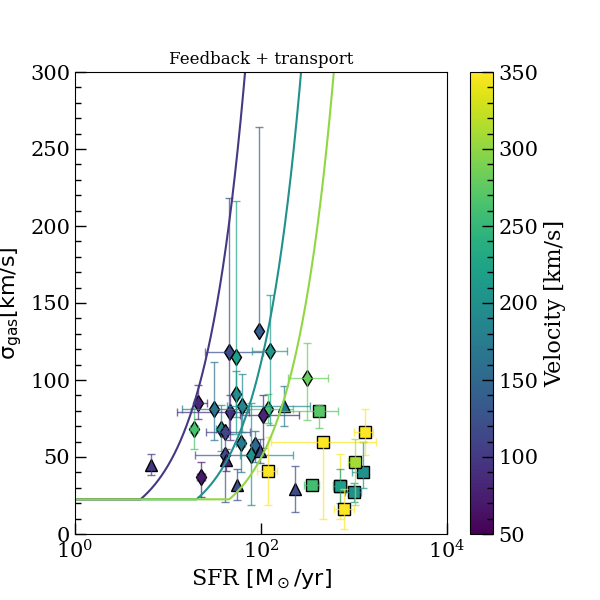}}
      \caption{
      Dependence of the velocity dispersion on the SFR and the rotational velocity. In the left panel, we show the model by \cite{Krumholz:2018} with the feedback as the only driver of the velocity dispersion. In the central panel, we show the model with the release of gravitational energy due to the transport of the gas across the disk as the only driver of the velocity dispersion. In the right panel, we show the model with the combination of feedback and transport as drivers of the velocity dispersion.}
    \label{fig:modelkrum}
   \end{figure*}

We analyzed the dependence of the velocity dispersion as a function of SFR.
The velocity dispersion is expected to increase with increasing SFR due to two main mechanisms: feedback, and gravitational instabilities.

In Figure \ref{fig:modelkrum} we report the velocity dispersion as a function of the SFR of the data as well as the empirical models derived by \cite{Krumholz:2018}, assuming the high-$z$ fiducial values for the model (see Table 3 from \citealt{Krumholz:2018}). These analytical models are based on the vertical hydrostatic equilibrium of the disk, on the marginal gravitational stability ($Q \sim 1$), and on the balance between the energy injected into the ISM by the star formation feedback and the gas transport from the outer to the inner region of the disk, and the dissipation of the energy from the velocity dispersion.

\begin{itemize}

\item  The left panel shows the prediction from the model called ``Feedback" here, which is based on the model that \cite{Krumholz:2018} called ``No-transport fixed Q". In this scenario, the authors assumed that galaxy-scale star-formation activity helps to maintain the Toomre parameter $Q$ close to the value of 1.
When the Toomre parameter drops below 1, star-forming clumps begin to form, which increases the SFR and therefore the feedback effects. Since the Toomre parameter is proportional to the velocity dispersion, $Q$ increases due to the turbulence induced by supernova explosions (i.e., feedback). When $Q>1$,  the star formation is instead not triggered, and when no other drivers of the velocity dispersion exist, the energy is dissipated, and the Toomre parameter decreases. 

\item The central panel depicts the model ``No-Feedback". In this model, the gravitational energy released by the mass transport within the disk is the only driver of the velocity dispersion. The assumption for this model is that the Toomre parameter $Q$ is always equal to 1, and all the gravitational energy released is dissipated by turbulence. 

\item The right panel describes the dependence of the velocity dispersion as a function of the SFR when the velocity dispersion is driven by both the star formation feedback and the mass transport within the disk (i.e., galaxy-scale gravitational instabilities).   At high SFR, the trend is similar to that of the model ``No-Feedback", while at low SFRs, there is a flat plateau in the velocity-dispersion-SFR trend that is based on the assumption that the star formation efficiency is constant for the entire disk, while the Toomre parameter is left free to vary.
These curves match the trend observed in local star-forming galaxies with a low SFR, which show a plateau in the velocity dispersion values independently of their SFR (\cite{Epinat:2010, Ianjamasimanana:2012, Mogotsi:2016}).
\end{itemize}

The gas velocity dispersion of our sample of high-redshift galaxies cannot be totally explained with the model in which only the feedback from star formation injects energy into the interstellar medium (left panel). To obtain high values of the velocity dispersion by using only the feedback effects, our galaxies should have an SFR higher by one order of magnitude than observed. Finally, models that include the gravitational processes for galaxy-scale turbulence driving (central and right panels) agree better with our data. On the other hand, the DSFG sample shows a velocity dispersion that agrees with the feedback-only driven turbulence \citep{Rizzo:2021}.
Unfortunately, our high-redshift observations are not able to explore the region at low SFR, so that we are not able to distinguish between the feedback+transport and the transport-only models.

   \begin{figure}[ht!]
   \centering
   \includegraphics[width=\hsize]{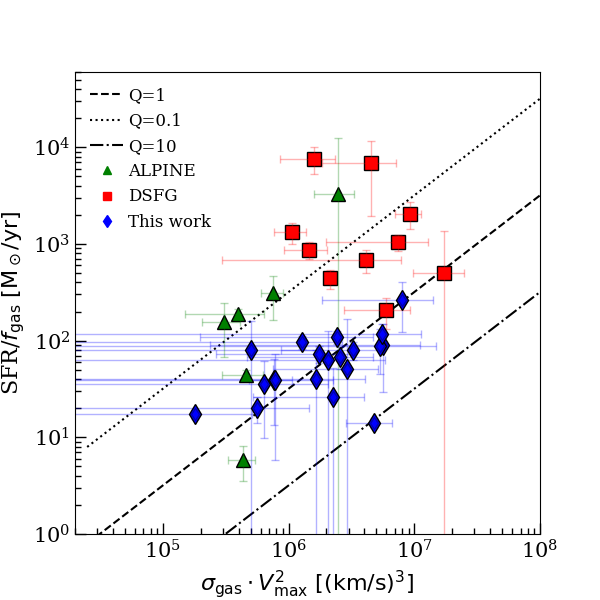}
    \caption{Relation of the kinematical properties and SF-related properties for our sample of galaxies following Eq. \ref{eq_gravity_vdisp}.
    The dashed line corresponds to the model with $Q=1$, the dotted line shows the model with $Q=0.1$ , and the dash-dotted line shows the model assuming $Q=10$}
    \label{fig:plotqsigma}
   \end{figure}

In Figure \ref{fig:modelkrum}  we study the dependence of $\sigma_{\rm gas}$ on SFR and rotational velocity. However, the \cite{Krumholz:2018} model expects that the gas velocity dispersion driven by the mass transport also depends on the gas fraction. In particular, for the "No Feedback" model, the dependence is
\begin{equation}\label{eq_gravity_vdisp}
    SFR = \frac{0.42}{\pi G} \frac{1}{Q} f_{\rm gas, p} \sigma_{\rm gas} v^2
,\end{equation}
where $f_{gas, p}$ is the gas fraction in the midplane of the galaxy ($f_{gas, p}$ = 1.5 $f_{\rm gas}$; \citealt{Ubler:2019}), Q is the Toomre parameter,  and $v$ and $\sigma$ are the rotational velocity and the velocity dispersion, respectively.
With this dependence, galaxies at fixed $\sigma_{\rm gas}$ and SFR can have a low rotational velocity but higher gas fraction, or conversely, a low gas fraction and high rotational velocity.
To take the different behavior at different gas fractions into account, we plot in Figure \ref{fig:plotqsigma} 
$SFR/f_{gas,p}$ as a function of $\sigma_{\rm gas} v^2$. Despite the large uncertainties and low statistics, our sample scatters around $Q = 1$ compatible with the lower redshift study \citep{Ubler:2019}, while the majority of the DSFG sample does not follow this trend because gravitational instabilities are not the dominant driver of turbulence in these galaxies. As even at lower redshift, the galaxies lie on the same relation, we conclude that main-sequence galaxies self-regulate over the cosmic time  (from $z=8$ to $z=0$) and evolve along roughly the lines of constant Q.

Figures \ref{fig:modelkrum} and \ref{fig:plotqsigma} show that our sample of galaxies is consistent with the models by \cite{Krumholz:2018}, who assumed a velocity dispersion that is mainly driven by gravitational instabilities.
The lower-redshift observations by \cite{Ubler:2019} and \cite{Girard:2021} also agree with the model of feedback+transport. 
\cite{Rizzo:2021} instead found that the velocity dispersion of their galaxy can be explained only by the energy injected by the stellar feedback due to their higher SFR. This scenario is also consistent with the fact that the majority of the galaxies analyzed by \cite{Rizzo:2021} are in a starburst phase (see Fig. \ref{fig:mainsequence}). 

\section{Conclusions}
\label{sec:conclusion}
Based on the kinematics analysis of the selected 22 galaxies observed with ALMA and studied through the \cii and \oiii emission lines, we found

\begin{itemize}
    \item The  median velocity dispersion of the gas in the sample is $\sim$ 65 km s$^{-1}$, even though we have large uncertainties due to the low signal-to-noise ratio and the low resolution of the observations. This value is about four to five times higher than what is observed in local galaxies ($z\sim0$) and two to three times higher than the velocity dispersion measured in galaxies at $z\sim1-2$. We therefore conclude that the velocity dispersion increases with redshift, which is consistent with the predictions of \cite{Wisnioski:2015} and the cosmological simulations of \cite{Pillepich:2019} and Kohandel et al. in prep. 
    
    \item The 2D kinematic analysis does not show a large bias in the recovery of the velocity dispersion and the value of $V_{\rm max}/\sigma_{\rm gas}$ due to beam smearing in comparison to the 3D fitting algorithm \textsc{$^{3D}$Barolo}.
    
    \item The velocity dispersion computed with both the gas tracer [C$\scriptstyle\rm~II$], which traces the multiphase medium, and [O$\scriptstyle\rm~III$], which only traces the ionized gas, does not show a significant difference for three out of four galaxies, while a galaxy shows a velocity dispersion traced with ionized gas greater than a factor of 2, as it does for lower-redshift galaxies \citep{Girard:2021}.

    \item The high-velocity dispersion of our sample of normal star-forming galaxies when compared with the \cite{Krumholz:2018} models can be explained by the release of gravitational energy from the transport of mass across the disk. This high-velocity dispersion cannot be sustained by similar models in which star formation feedback (i.e., supernova explosions) is the only source of the velocity dispersion. This is supported by the \cite{Kohandel:2020} simulation, in which they showed that 90\% of the observed \cii velocity dispersion is powered by bulk motions due to gravitational processes, such as merging and accretion events.

    \item The ratio $V_{\rm max}/\sigma_{\rm gas}$ does not show a clear dependence on the gas fraction, as was assumed by \cite{Wisnioski:2015}. However, we also found a dependence on the stellar mass and the dust mass. We therefore conclude that at fixed redshift,  galaxies with higher stellar mass have a more regularly rotating disk ($V_{\rm max}/\sigma_{\rm gas} \gg \sqrt{3.36}$) than those with lower stellar mass.

\end{itemize}

Current studies focusing on $z>5$ galaxies show a difference between rotation-dominated disks and dispersion-dominated disks.
Our analysis of a sample of normal star-forming galaxies at $z>4$ suggests that high-redshift galaxies are more chaotic and less rotationally supported than their local counterparts.  Our results contradict some previous studies \citep{Rizzo:2020, Rizzo:2021, Lelli:2021, Fraternali:2021, Sharda:2019} that analyzed $z\sim 4$ massive galaxies with low values of the velocity dispersion and higher values of the ratio $V_{\rm max}/\sigma_{\rm gas}$ comparable to local star-forming galaxies (see Sec. \ref{Velocitydispersionevolution}). 
However, we note that these massive ($M_* = 10^{11} M_\odot$) galaxies are highly star-forming galaxies ($SFR >100 M_\odot$ yr$^{-1}$), as shwon in figure \ref{fig:mainsequence}, and they appear to be more evolved than the galaxies in our sample, with the evidence of an already formed stellar bulge just 1.2 Gyr after the Big Bang \citep{Lelli:2021} and a greater dust content.

Because of the difference in the physical properties (SFR, stellar mass, and dust content) of the samples targeted (see Sec. \ref{sec:sampleproperties}), we can explain the contrast between their kinematical behavior, with these samples being part of two different galaxy populations that both exist at high redshift.
We showed in section \ref{sec:vsigmaevolution} that the gas fraction seems not the most important parameter for assessing the rotational support of a galaxy, as was shown by \cite{Wisnioski:2015}, and other physical quantities such as stellar mass and dust content may also affect the value of $V_{\rm max}/\sigma_{\rm gas}$.

The discrepancy between our results and those found in massive galaxies at similar redshift may also be explained by comparing the measured dispersion with the theoretical prediction by \cite{Krumholz:2018} to investigate the drivers of velocity dispersion. As discussed in Section \ref{section_drivers_of_velocity_dispersion}, while the velocity dispersion of massive
galaxies in the literature can be explained with star formation feedback alone, the high-velocity dispersion of the galaxies targeted in this work requires the contribution of gravitational instabilities as well, according to the models considered.

The challenge of inferring the kinematical properties of high-redshift data is particularly influenced by the angular resolution of the observations and the sample selection of high-redshift galaxies. The selection of high-redshift galaxies is biased toward DSFGs or UV-bright galaxies. To understand and explain the evolution of the velocity dispersion and the rotational support of galaxies with redshift and explain the presence of dynamically cold but extremely star-forming galaxies and dynamically hot main-sequence galaxies, we need more high-resolution observations for high-redshift galaxies and a higher S/N.

\begin{acknowledgements}
This paper makes use of the following ALMA data: ADS/JAO.ALMA \#2017.1.01052.S, \#2017.1.01471.S, $\#$2018.1.01359.S, \#2017.1.00508.S,  \#2017.1.01451.S, \#2018.1.01359.S, \#2017.1.00604.S, \#2018.1.00429.S, \#2018.1.00085.S, \#2019.1.01634.L. ALMA is a partnership of ESO (representing its member states), NSF (USA) and NINS (Japan), together with NRC (Canada), MOST and ASIAA (Taiwan), and KASI (Republic of Korea), in cooperation with the Republic of Chile. The Joint ALMA Observatory is operated by ESO, AUI/NRAO and NAOJ. EP
is grateful to F. Rizzo for useful comments and discussions. SC is supported by European Union’s HE ERC Starting Grant No. 101040227 - WINGS. MK and AP acknowledge support from the ERC Advanced
Grant INTERSTELLAR H2020/740120.
\end{acknowledgements}

\bibliographystyle{aa}
\bibliography{aa}

\begin{appendix}

\section{Galaxy properties} \label{galaxyproperties}

Table \ref{tab:properties} lists the properties collected from the literature of the galaxies analyzed in this work.

\begin{table*}[!h]
\caption{Properties of the analyzed sample of galaxies. Stellar mass, SFR, and dust mass.}
\centering
\begin{tabular}{lllllll}
\hline \hline
Target name &
  $\rm \log( M_\star/M_\odot)$ &
  Reference &
  SFR [$\rm M_\odot/yr$] &
  Reference &
  $\rm \log(M_{\rm dust}/M_\odot)$ &
  Reference \\ \smallskip
  
  \textit{\small{(1)}} &
  \textit{\small{(2)}} &
  \textit{\small{(3)}} &
  \textit{\small{(4)}} &
  \textit{\small{(5)}} &
  \textit{\small{(6)}} &
  \textit{\small{(7)}} \\

\hline  \smallskip
  
DLA0817g &
   ... & ...
   & 
  110 $\pm$ 10 &
  N17 &
   ... & ...
   \\\smallskip
ALESS 073.1 &
  10.7-11.1 &
  L21 &
  1050 $\pm$ 150 ~\tablefootmark{FIR}&
  G14 &
   $8.69 ^{+0.05}_{-0.06}$ & 
   G14
   \\\smallskip
HZ7 &
  9.86 $\pm$ 0.21 &
  C15 &
  $21^{+5}_{-2}$ &
  C15 &
   ... & ...
   \\\smallskip
HZ9 &
  9.86 $\pm$ 0.23 &
  C15 &
  $120^{+100}_{-60}$ &
  P19 &
   ... & ...
   \\\smallskip
HZ4 &
  $10.15^{+ 0.13}_{-0.15}$ &
  F20 &
  $40.7 ^{+35}_{-15}$ &
  F20 &
   ... & ...
   \\\smallskip
J1211 &
  10.47  &
  H20 &
  86 &
  H20 &
  $7.5^{+0.6}_{-0.6}$ &
  H20 \\\smallskip
J0235 &
  10.47 &
  H20 &
  54 &
  H20 &
   ... & ...
   \\\smallskip
CLM1 &
  10.11 &
  W15 &
  37 $\pm$ 4 ~\tablefootmark{SED} &
  W15 &
   ... & ...
   \\\smallskip
J0217 &
  10.47&
  H20 &
  96 &
  H20 &
  $8.3^{+1.5}_{-0.9}$ &
  H20 \\\smallskip
VR7 &
  10.23 &
  M17 &
  $54^{+5}_{-2}$ &
  M19 &
   ... & ...
   \\\smallskip
UVISTA-Z-349 &
  $9.79^{+0.74}_{-1.27}$ &
  D22 &
  $105.8^{+149.7}_{-47.0}$ &
  D22 &
  $7.43^{+0.29}_{-0.29}$ &
  So22 \\\smallskip
UVISTA-Z-004 &
  $9.83^{+0.19}_{-0.19}$ &
  D22 &
  $53.3^{+74.0}_{-23.8}$ ~ \tablefootmark{\cii} &
  F21 &
  $7.10^{+0.36}_{-0.32}$ &
  So22 \\\smallskip
UVISTA-Z-049 &
  $9.76^{+0.35}_{-0.37}$ &
  D22 &
  $31.2^{+42.6}_{-17.9}$ &
  D22 &
  $7.20^{+0.34}_{-0.32}$ &
  So22 \\\smallskip
UVISTA-Z-019 &
  $9.51^{+0.19}_{-0.18}$ &
  Sc22 &
   $63^{+270}_{-27}$ 
   & W22
   & $7.32^{+0.66}_{-0.53}$
   & W22
   \\\smallskip
COS-29 &
  $9.23^{+0.11}_{-0.6}$ &
  S18 &
  22.7$\pm$2.0~ \tablefootmark{UV} &
  S18 &
    ... & ...
   \\\smallskip
COS-30 &
  $9.14^{+0.18}_{-0.7}$ &
  S18 &
  19.2$\pm$1.6~ \tablefootmark{UV} &
  S18 &
  $7.84^{+0.57}_{-0.54}$
   & W22
   \\\smallskip
UVISTA-Z-001 &
  $9.58^{+0.09}_{-0.35}$ &
  Sc22 &
    $79^{+142}_{-20}$ 
   & W22
   & $6.6^{+0.4}_{-0.6}$ 
   & W22
   \\\smallskip
UVISTA-Y-004 &
  $9.90^{+0.25}_{-0.34}$ &
  D22 &
  $40.7^{+46.2}_{-21.2}$ &
  D22 &
  $7.13^{+0.36}_{-0.32}$ &
  So22 \\\smallskip
UVISTA-Y-003 &
  $10.1^{+0.15}_{-0.18}$ &
  D22 &
  $310.8^{+214.3}_{-115.1}$ &
  D22 &
  $7.55^{+0.30}_{-0.21}$ &
  So22 \\\smallskip
UVISTA-Y-879 &
  $9.00^{+0.69}_{-0.69}$ &
  D22 &
  $126.5^{+64.0}_{-45.3}$ &
  D22 &
  $6.98^{+0.13}_{-0.19}$ &
  So22 \\ \smallskip
SUPER8 &
  $9.69^{+0.45}_{-0.99}$ &
  D22 &
  $44.8^{+60.9}_{-20.0}$ &
  D22 &
  $7.06^{+0.35}_{-0.31}$ &
  So22 \\ \smallskip
UVISTA-Y-001 &
  $9.70^{+0.56}_{-0.73}$ &
  D22 &
  $46.5^{+27.7}_{-34.1}$ &
  D22 &
  $7.28^{+0.31}_{-0.31}$ &
  So22 \\\hline
\end{tabular}

\tablebib{G14~\citet{Gilli:2014}; C15~\citet{Capak:2015}; W15~\citet{Willott:2015}; M17~\citet{Matthee:2017};    N17~\citet{Neeleman:2017};   S18~\citet{Smit:2018}; M19~\citet{Matthee:2019};   P19~\citet{Pavesi:2019};      F20~\citet{Faisst:2020}; H20~\citet{Harikane:2020};   L21~\citet{Lelli:2021};      D22~\citet{Dayal:2022};  Sc22~\citet{Schouws:2022};  So22~\citet{Sommovigo:2022};  W22~\citet{Witstok:2022}.}

\tablefoot{
\textit{(1)}: target name; \textit{(2)} and \textit{(3)}: stellar masses computed from the UV luminosity and references; \textit{(4)} and \textit{(5)}:  SFR computed as SFR= SFR$_{\rm UV}$ + SFR$_{\rm IR}$ (if not indicated otherwise with a symbol) and  references; \textit{(6)} and \textit{(7)}: dust masses computed by SED fitting of the rest-frame FIR emission temperatures and references.\\
\tablefootmark{\cii}{Estimated from the \cii luminosity} \\
\tablefootmark{SED}{Estimated from SED fitting} \\
\tablefootmark{IR}{Estimated from the IR luminosity}\\
\tablefootmark{UV}{Estimated from the UV luminosity}}
\label{tab:properties}
\end{table*}

\section{Method I uncertainties and pixel correlation}\label{Correlation}

Our fitting process, similar to others used in the literature \citep{Neeleman:2020, Pensabene:2020, Neeleman:2021}, determined the best-fitting parameters by maximizing the likelihood function $\mathcal{L} = {\rm exp}(-0.5\chi^2), $ and  $\chi^2$ is defined as

\begin{equation}
 \chi^2 =\frac{\sum_{i,j}(D_{\rm ij}-M_{\rm ij})^2}{\sum_{i,j} \sigma_{\rm ij}^2},
\end{equation}
where the $D_{\rm ij}$  and  $M_{\rm ij}$ are the data and model moment maps, respectively, and $\sigma_{\rm ij}^2$ is the error associated with the data maps. We note that neighboring pixels within a beam are  strongly correlated and not independent because of the angular resolution of the data. This means that the assumption that  individual pixels in the moment maps are independent can lead to an underestimation of the free parameter uncertainties.

\begin{figure}[ht!]
    \centering
    \includegraphics[width=\hsize]{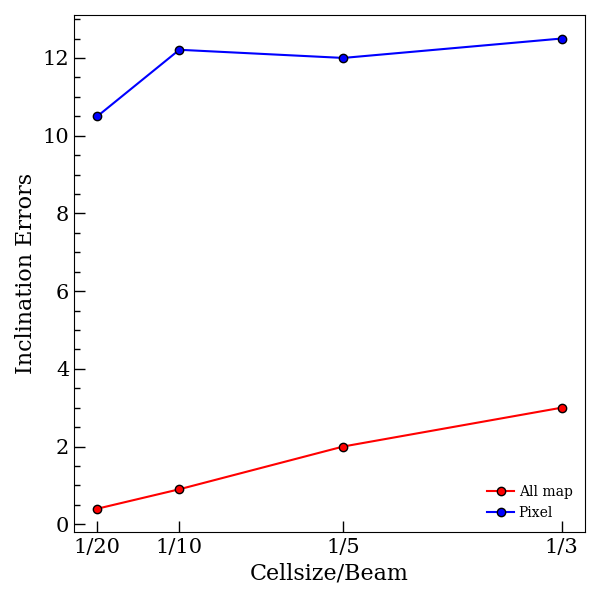}
    \caption{Error on the inclination parameter for different datacubes with different pixel sizes. The errors derived from our method in which we estimated the likelihood by using only independent pixels  is shown in blue, and the uncertainties obtained by assuming all pixel of a moment map as independent pixels are shown in red.}
    \label{fig:error_correlation}
\end{figure}

To better understand the impact of the pixel-pixel correlation on the kinematic analysis, we simulated an ALMA mock observation of a rotating gas disk with the \verb'CASA' task tool  \verb'simobserve'  and generated four different datacubes with the task \verb"tclean"  with the same angular resolution, but different pixel sizes: 1/20, 1/10, 1/5, and 1/3 of the beam.
Finally, we fit the 2D maps with the same algorithm as explained in Section~3.3 and left only the inclination as a free parameter, which is one of the parameters that is more affected by the beam smearing \citep{Pensabene:2020}. Figure \ref{fig:error_correlation} shows the  uncertainties on the disk inclination parameter obtained by assuming all the pixels in the map as independent pixels as the red line. Although the datacubes were generated from the same visibilities dataset, the uncertainties on the free parameter depend  on the  pixel size that is adopted to create the final image.

\begin{figure}[ht!]
    \centering
    \includegraphics[width=\hsize]{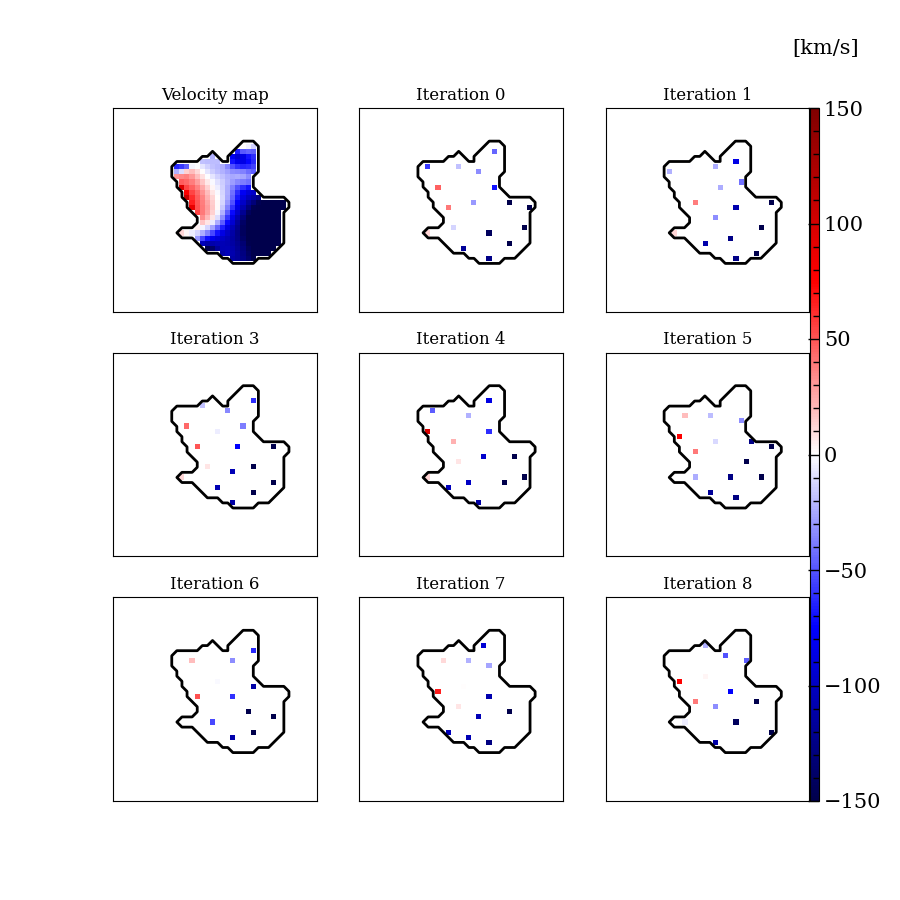}
    \caption{Example of a randomly selected pixel for each fitting iteration.}
    \label{fig:galaxyexample}
\end{figure}

To better constrain the uncertainties on our free parameter and avoid pixel correlation, we estimated the likelihood by using only uncorrelated pixels. In each map, we randomly selected pixels that were separated from each other by a distance greater than the semi-major axis of the beam (see an example of the randomly selected pixel for the target COS-30 in figure \ref{fig:galaxyexample}). We repeated this procedure 100 times to ensure that the likelihood estimate does not depend on the pixel selection. The blue line in Fig.~\ref{fig:error_correlation} shows the uncertainties obtained with this method for the four datacubes, and we note that the error does not depend on the pixel size. This confirms that the selected pixels are uncorrelated. 
This method therefore returns more conservative and physical uncertainties on the free parameters than the standard fitting procedure.

\section{Impact of beam smearing on the disk size estimate}
\label{BeamSmearingAppendix}
In this section, we test the reliability of method I when the kinematic properties of the galaxy sample are estimated even in cases of poor angular resolutions and a low signal-to-noise ratio.
As described in section \ref{sec:beamsmearing}, the beam size can strongly affect the recovery of the kinematical parameters, leading  to an underestimation of  $V_{\rm max}/\sigma_{\rm gas}$. This impact is more predominant when the radius of the galaxy is much smaller than the beam size.

\begin{figure}[ht!]
    \centering
    \includegraphics[width=\hsize]{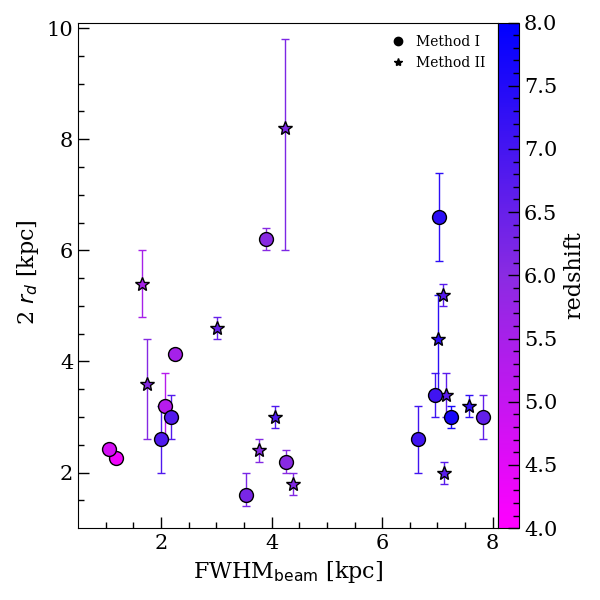}
    \caption{Recovered galaxy size (2$r_d$) as a function of the beam size. Circles are galaxies fit with method I, and stars are galaxies fit with method II. The colors represent the redshift of the galaxy (see the color bar).}
    \label{fig:radius_ratio}
\end{figure}

We  initially assessed whether the angular resolution of our observations might lead to a bias in the estimate of the galaxy disk sizes. Figure \ref{fig:radius_ratio} shows the  disk size (2$r_d$) from the fit of flux maps as a function of the full width at half maximum of the ALMA beam, ${\rm FWHM_{\rm beam}}$. Although the ratio of the beam and the disk size spans a wide range from 0.3 to 3.5,  we do not find any clear relation between the two parameters, suggesting that the galaxy size is well recovered within the uncertainties in all observations selected for this work.

\begin{figure*}[ht!]
    \centering
    \includegraphics[width=\hsize]{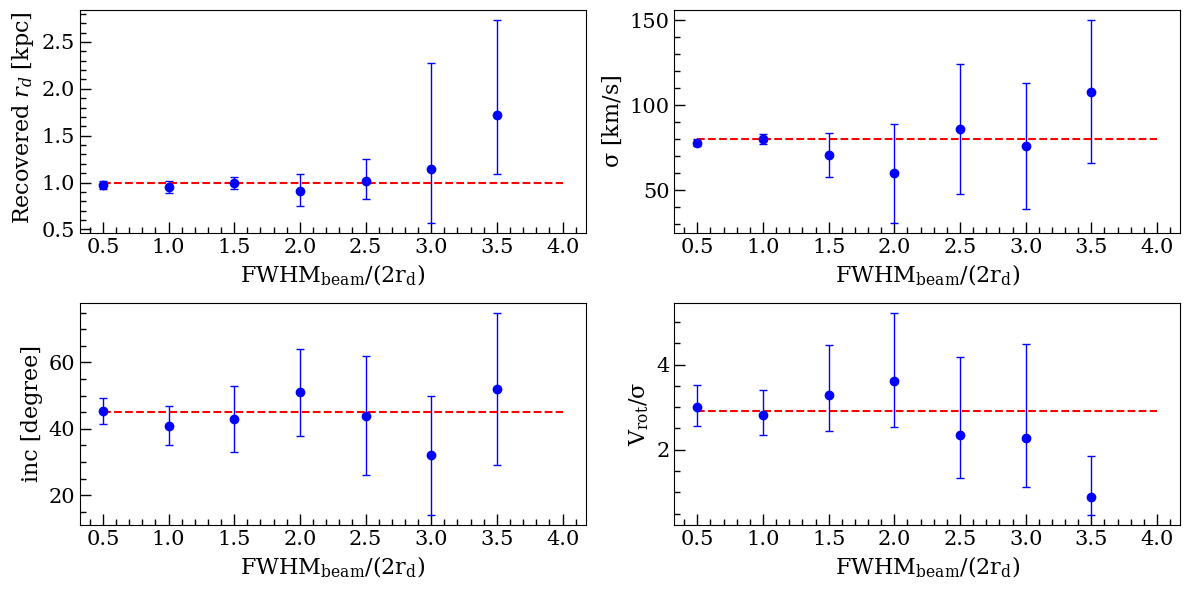}
    \caption{Best-fitting results of the scale radius (upper left), inclination (bottom left), \sigmag (top right), and $V_{\rm max}/\sigma_{\rm gas}$ (bottom right) for the mock datacubes with different $ \rm FWHM_{\rm beam}/ 2r_d$.
    The dashed red line represents the input parameters of the model we used to create the mock ALMA datasets. }
    \label{fig:sim_beam}
\end{figure*}

To verify the impact of the beam smearing on the inferred kinematic parameters, we simulated a set of ALMA observations, performed a kinematic analysis based on method I, and evaluated the discrepancy between the best-fits results and input parameters. In particular, we simulated ALMA observations with different angular resolutions of a disk galaxy at $z=6$ with a scale radius of 1 kpc. For each selected angular resolution, we produced a set of ten ALMA mock datacubes with a signal-to-noise ratio of $\sim10$ so that we can determine the uncertainties and biases associated with our inferred kinematic parameters.  In Figure \ref{fig:sim_beam} we show the results of the kinematical fitting of four important kinematical parameters (i.e, scale radius, velocity dispersion, inclination, and $V_{\rm max}/\sigma_{\rm gas}$) as a function of  the ${\rm FWHM_{\rm beam}}/2r_d$ ratio. The uncertainties on the kinematic parameters increase with decreasing angular resolution, but we do not find any clear bias or trend. However, we note that the  $V_{\rm max}/\sigma_{\rm gas}$ ratio is underestimated in all ten ALMA simulations with ${\rm FWHM_{\rm beam}}/2r_d=3.5,$ but this does not change the result of our work on the real data because we have only one dataset with a poor angular resolution like this.

We furthermore tested the reliability of this method in determining the velocity dispersion and the ratio $V_{\rm max}/\sigma_{\rm gas}$  even for data with low S/N. We created two different mock datacubes with infinite S/N and angular resolution, but different values of the kinematical parameters, such that one is dispersion dominated, and the other is rotation dominated. In particular, the dispersion-dominated model had a scale radius of 1.8 kpc, corresponding to 0.31\arcsec at redshift 6, an inclination of 40°, a mass of $10^{10}$ , and a velocity dispersion of 80 km/s, resulting in a $V_{\rm max}/\sigma_{\rm gas}$ = 1.4. The rotation-dominated model had the same values for the radius and the inclination, but a mass of $3\times 10^{10} \rm M_\odot$ and a velocity dispersion of 40 km/s, resulting in a $V_{\rm max}/\sigma_{\rm gas}$ = 5. We then simulated ALMA observations by changing both the angular resolution and the S/N. For the angular resolution, we tested two values of 0.8 and 1.5 times the mock galaxy radius that represents the bulk of our data, while for the S/N, we test four different levels of S/N that comprise the range of S/N of the data. Using method I, we then fit the mock cubes and recovered the kinematical parameters.

\begin{figure*}[ht!]
    \centering
    \includegraphics[width=\hsize]{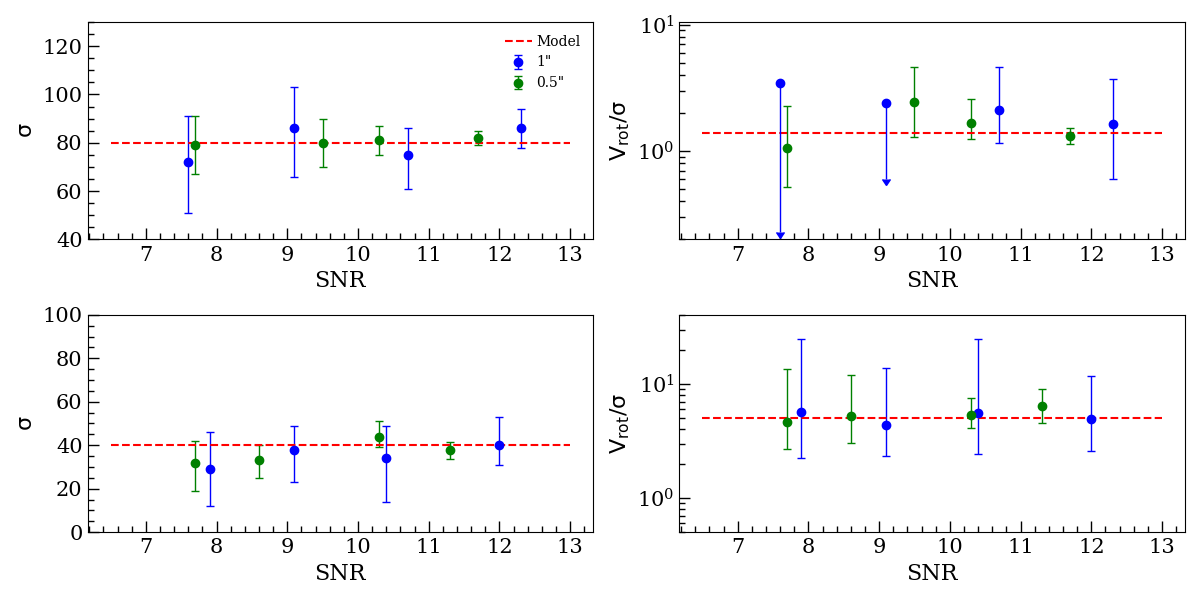}
    \caption{ Best-fit values of $\sigma_{\rm gas}$ (left) and $V_{\rm max}/\sigma_{\rm gas}$ (right) for the mock datacubes with different S/N and different beams for the dispersion-dominated model (upper panel) and rotation-dominated model (lower panel). In green we plot the results obtained from the model with 0.5\arcsec beam, and in blue we plot the results obtained from the model smeared with a beam of 1\arcsec. The dashed red line represents the values of the mock datacube with infinite angular resolution. }
    \label{fig:simulations}
\end{figure*}

In Fig. \ref{fig:simulations} we report the results of the velocity dispersion and the ratio $V_{\rm max}/\sigma_{\rm gas}$ for the dispersion-dominated model and the rotation-dominated model, respectively.
We do not find any significant bias in the recovery of these parameters. The median value of the velocity dispersion  is always within the limit of the velocity resolution of the datacube (10 km s$^{-1}$).
For the models with low S/N and 1\arcsec beam, especially with the dispersion-dominated datacube, we find that the degeneration between the rotational mass and the inclination of the disk does not allow us to constrain the value of the velocity, but we only find the upper limits, similar to what happens for our data (see Fig. \ref{Fig:J0217_oiiidistr}).

\newpage

\section{Method II tests and correction factor}\label{Confronto}

\begin{figure}[ht!]
    \centering
    \includegraphics[width=\hsize]{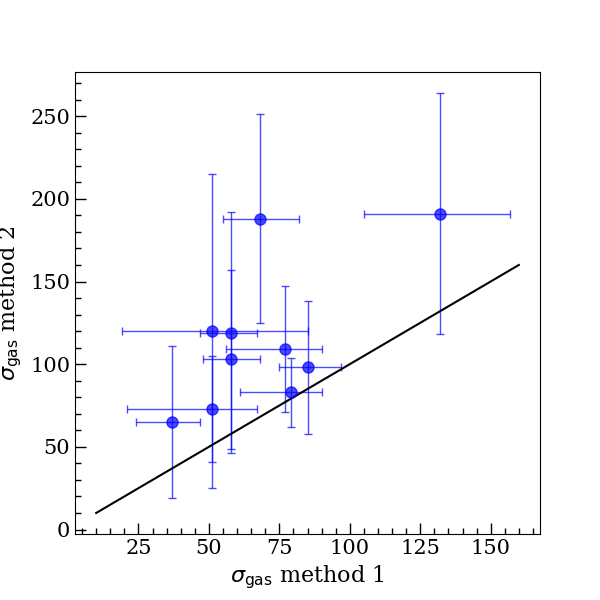}
    \caption{Comparison between the velocity dispersion of the same galaxies obtained with the two different methods.}
    \label{fig:sigma_comparison}
\end{figure}

In this section, we test the results of method II, first by using the results of method I to understand the correctness in the estimation of \sigmag, and then we try to understand the limits of this method, which is principally based on an a priori determination of some kinematical parameters.
To understand the correctness of the value of velocity dispersion extrapolated with method II, we exploited the galaxies analyzed with method I, and we also analyzed them with method II.
The results of the comparison are shown in Figure \ref{fig:sigma_comparison}. This analysis shows an overestimation of $\sigma$ obtained from method II in comparison with the value of $\sigma$ obtained with method I for the same galaxy. 
Analyzing the difference between the two different values, we found that the velocity dispersion estimated with method II is overestimated in the median by a factor of 1.6. Therefore, we applied this correction factor to the velocity dispersion obtained from galaxies analyzed with the second method.

\begin{figure}[ht!]
    \centering
    \includegraphics[width=\hsize]{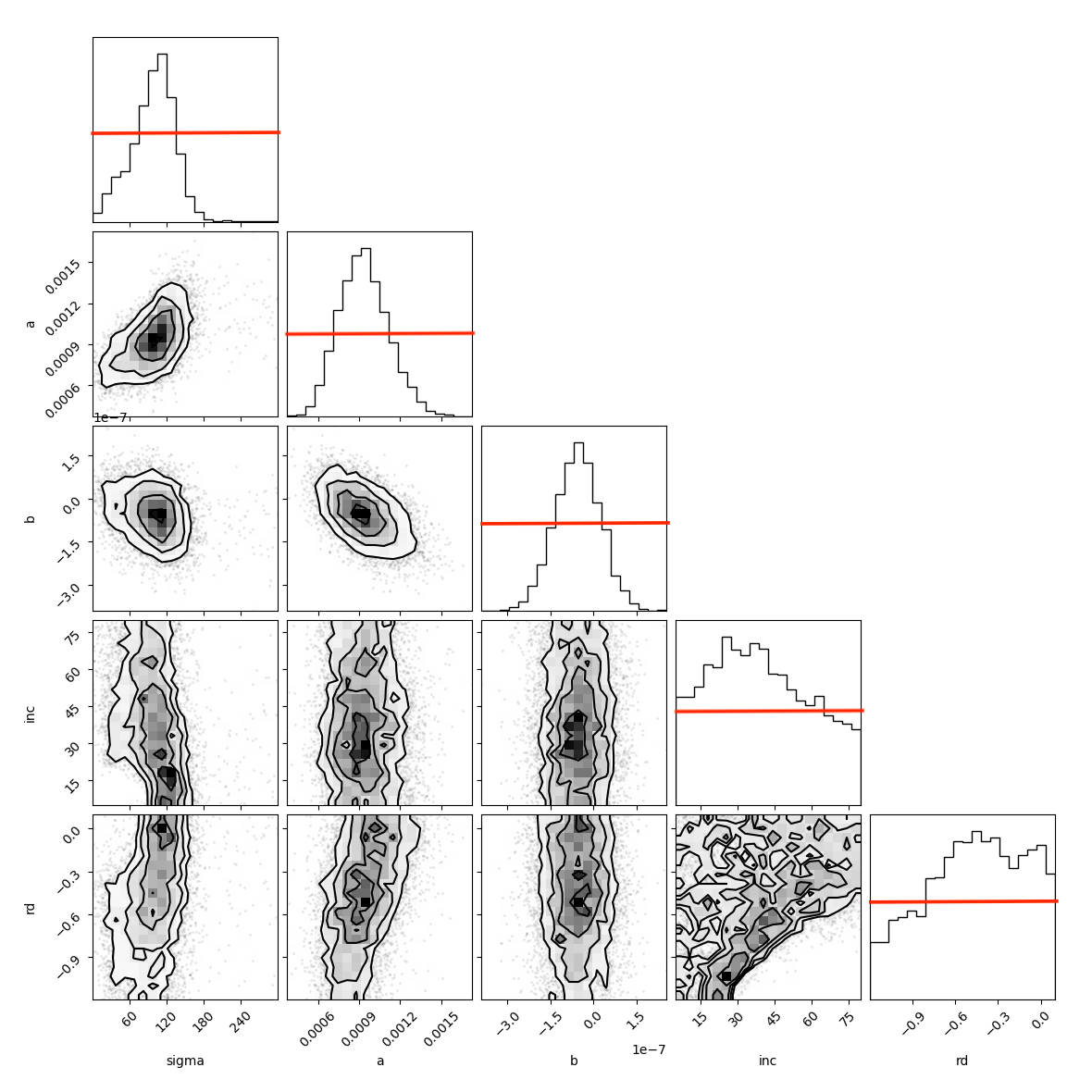}
    \caption{Corner plot distribution of the free parameters for the mock datacube analyzed with method II, in which the radius was left free to vary. In red we show the flat priors. }
    \label{fig:corner_plot_method2_sim}
\end{figure}

As already mentioned in Sec. \ref{sec:method2}, for this method, we had to determine the radius and the galaxy mass a priori while we constrained the inclination between 3$\sigma$ to the results of the flux map fitting. To understand the effect of the radius and inclination in the determination of the velocity dispersion, we performed a kinematic fitting on simulated ALMA observations with method II and also left the radius and the inclination as free parameters
The mock datacube simulated the ALMA observations of a disk galaxy with a radius of 1 kpc, an inclination of 45°, a mass of $3\times 10^{10} \rm M_\odot$ , and a \sigmag = 80 \kms. The S/N of the datacube was about 10.
In the kinematic fitting, all free parameters had flat priors, and in particular, the prior within [5 degrees, 85 degrees], [0.3~kpc, 5~kpc], and [0 \kms, 300 \kms ] was set for disk inclination, scale radius, and velocity dispersion, respectively.
In Figure \ref{fig:corner_plot_method2_sim} we show the corner plot of the posterior distribution from the kinematic fitting on the mock datacube. 
 We note that the posterior distributions of the disk inclination and scale radius are almost flat, while the posterior of the velocity dispersion parameters clearly peaks around 100 km$^{-1}$ and has a width of ~80 km/s. On the other hand, the posterior distribution that we obtain with a fixed scale radius peaks at 87 km$^{-1}$, which is more consistent with the input value, and has a width of~70km$^{-1}$. The combination of the two results indicates that the shape of the posterior distribution of the velocity dispersion parameter is only slightly modulated by the disk inclination and scale radius parameter.

\section{2D kinematical fitting results}\label{Resultsmethod1}
In this section, we show the moment maps derived from the pixel-by-pixel Gaussian fitting, the best model, and the 1D distribution of the free parameter recovered with the 2D fitting procedure explained in Sec. \ref{sec:2Danalysis}. 
To recover the best values of the free parameters, we combined the results of 100 trials, in which each time, we randomly selected pixels at a distance greater than half of the beam size to take the correlation between spatial pixels into account. To create the final distributions of the free parameters, we combined the results of each distribution for every fitting.

\begin{figure*}[p]
\resizebox{\hsize}{!}
    { 
    \includegraphics[width=\hsize]{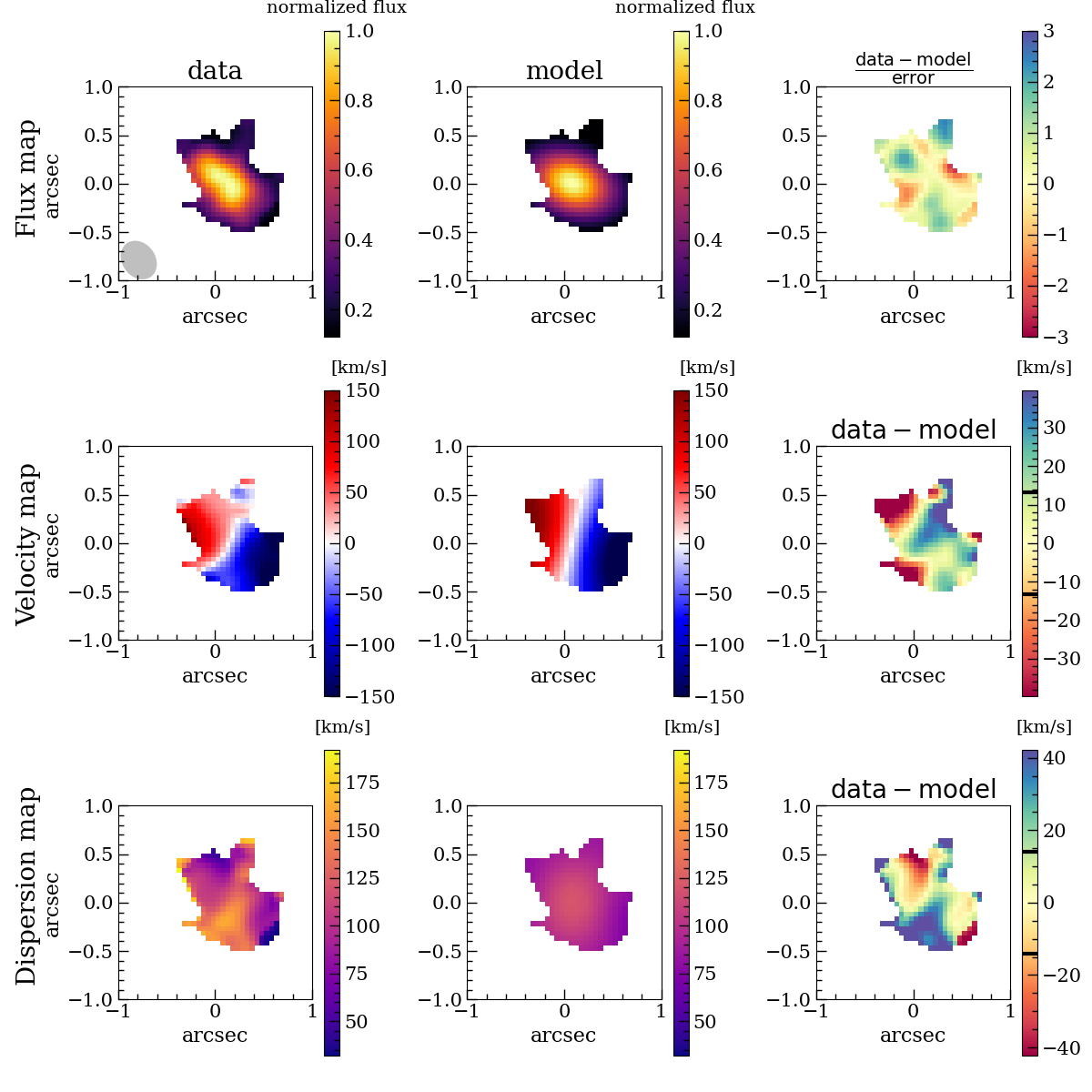}
    \includegraphics[width=\hsize]{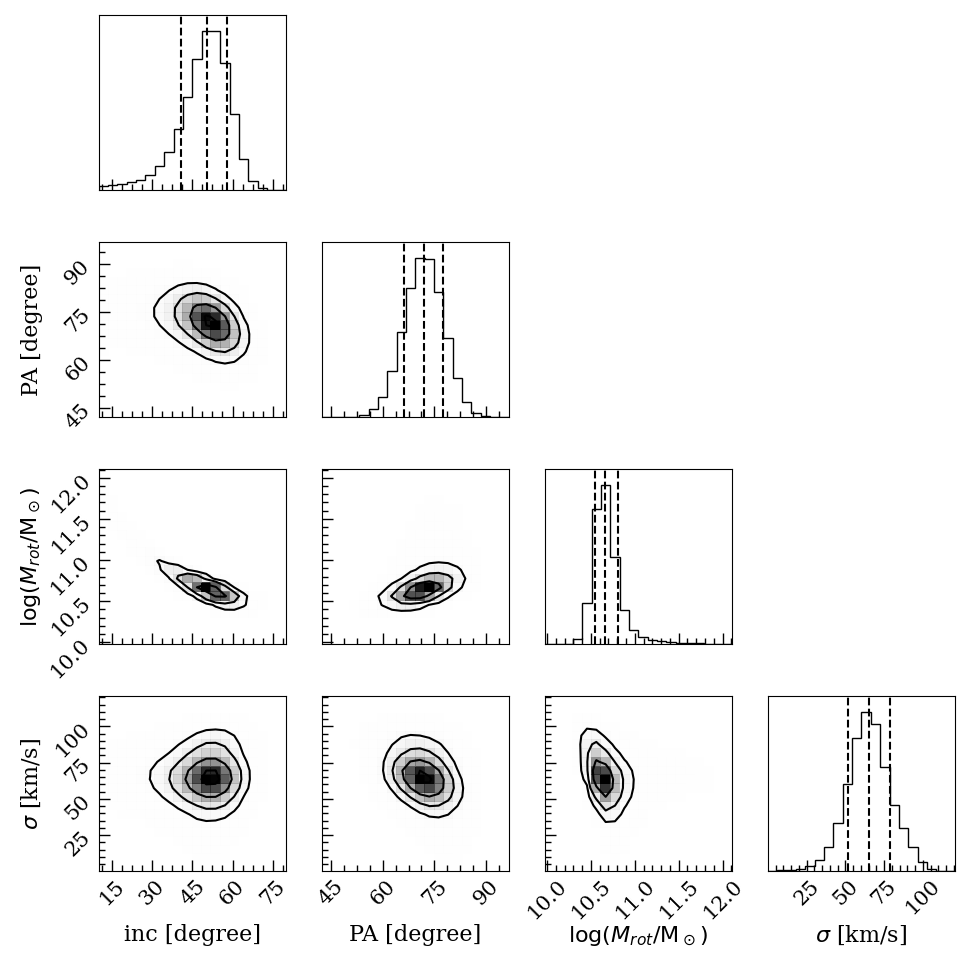}
    \includegraphics[width=\hsize]{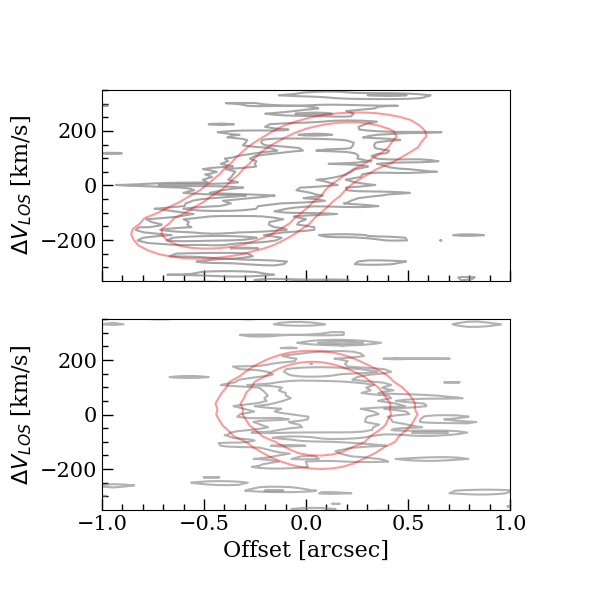}}
    \caption{Kinematical fitting using method I. On the left, we show moment maps for COS-30 observed with the \cii emission line.
      In the left column, we present the flux, velocity, and velocity dispersion map derived with the Gaussian fitting. The beam is shown as the gray ellipse in the flux map.
     In the central panels, we present from top to bottom the best-result maps of the zeroth, first, and second moment.
     In the right panels, we show the residuals of the fitting, with color bars ranging between $\pm 3\sigma$, the black line on the color bar indicates $1 \sigma$.
    The flux maps are normalized for a pixel with a maximum flux = 1.
    The x and y values of the kinematical maps are the measure in arcseconds from the central pixel of the image.
    In the center, we show the corner plot distribution of the free parameters.
    In the right panel, we show the position-velocity diagram along the major and minor axes in the upper and lower panel, respectively. In black we plot the data, in red we plot the model, and the contours are computed at 3 and 6 $\sigma$, respectively.
    }
    \label{Fig:cos30}
   \end{figure*}

\begin{figure*}[p]
\resizebox{\hsize}{!}
    { 
    \includegraphics[width=\hsize]{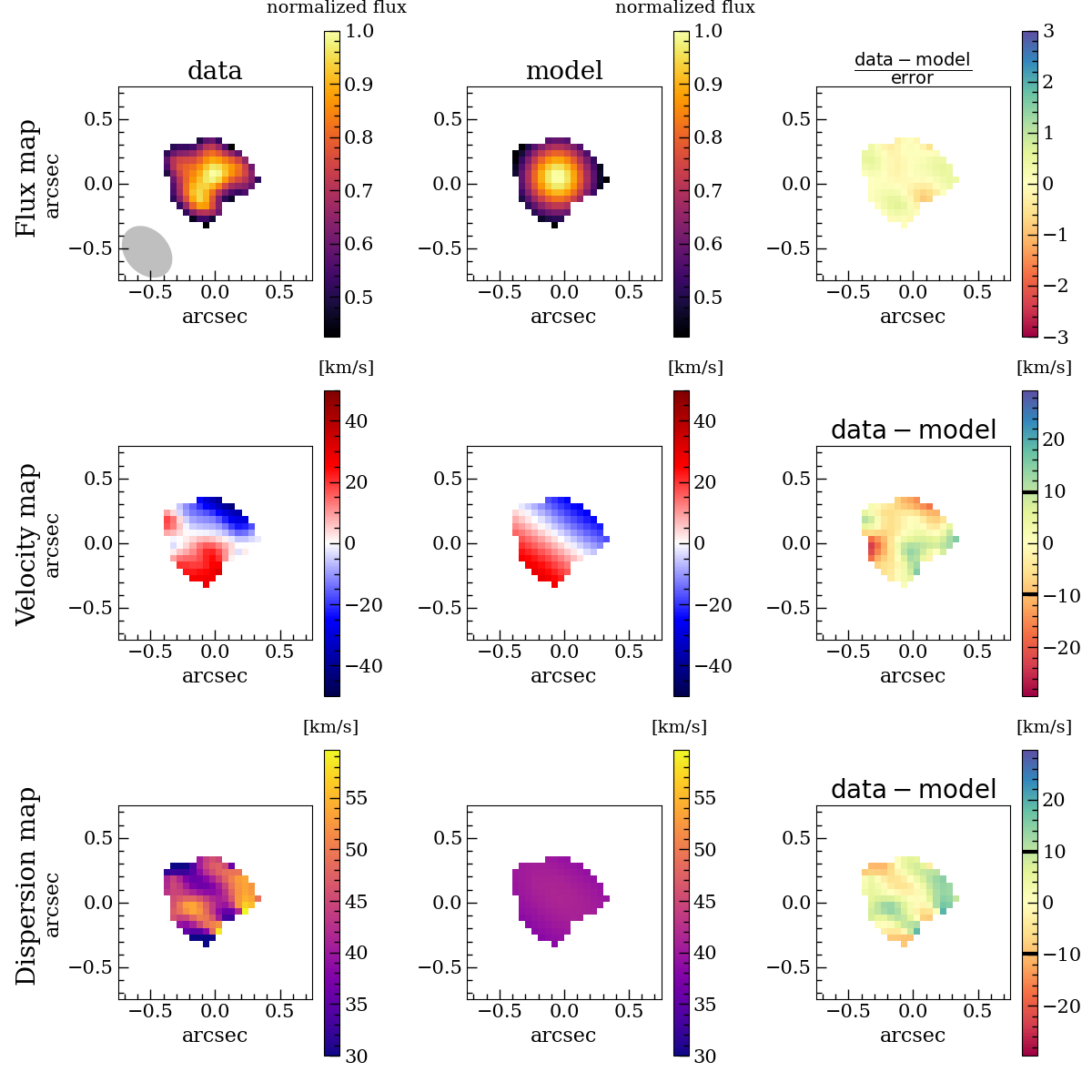}
    \includegraphics[width=\hsize]{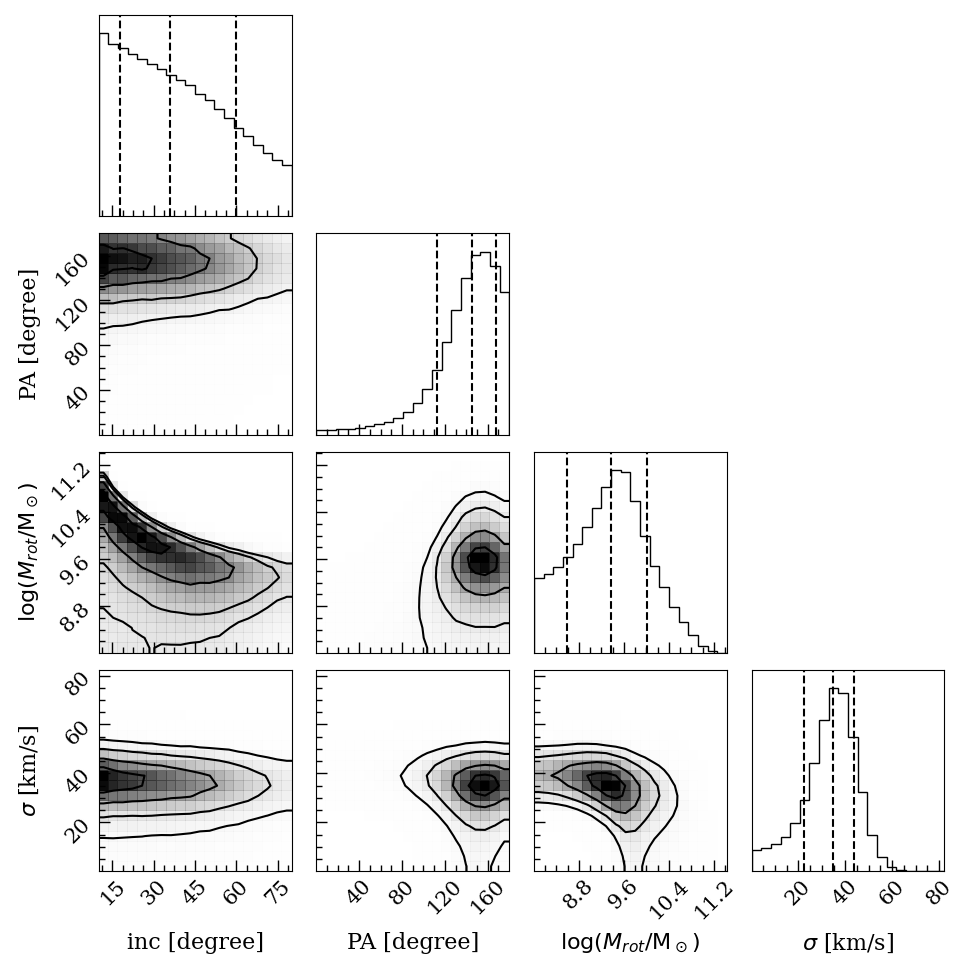}
    \includegraphics[width=\hsize]{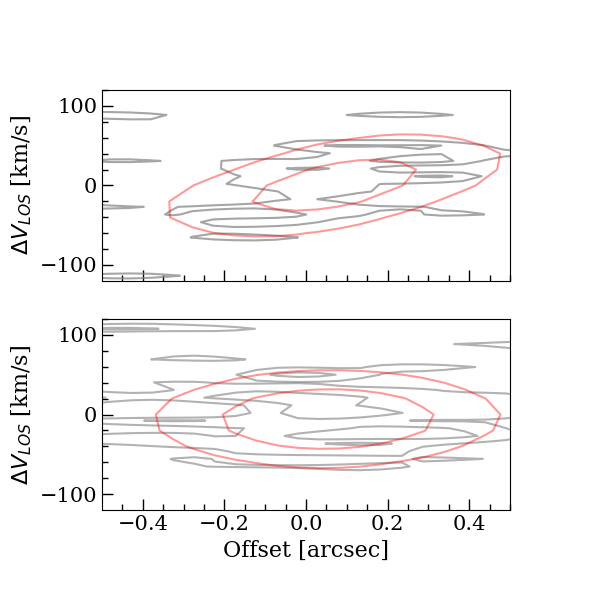}
    }
    \caption{Target COS-29 observed with the \cii emission line. See the caption of Fig. \ref{Fig:cos30}.}
    \label{Fig:cos29}
   \end{figure*}

\begin{figure*}[p]
   \resizebox{\hsize}{!}
    {  
    \includegraphics[width=\hsize]{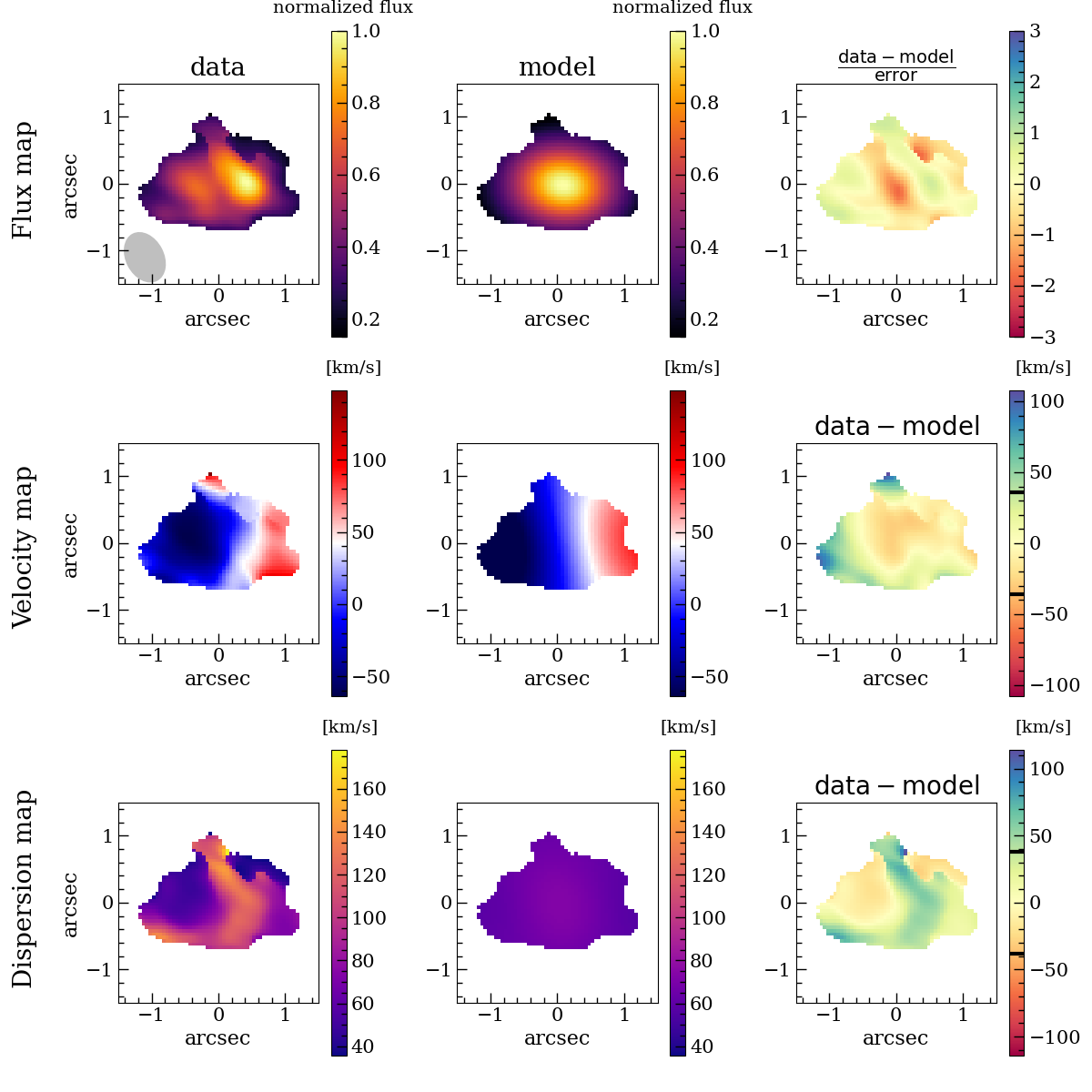}
    \includegraphics[width=\hsize]{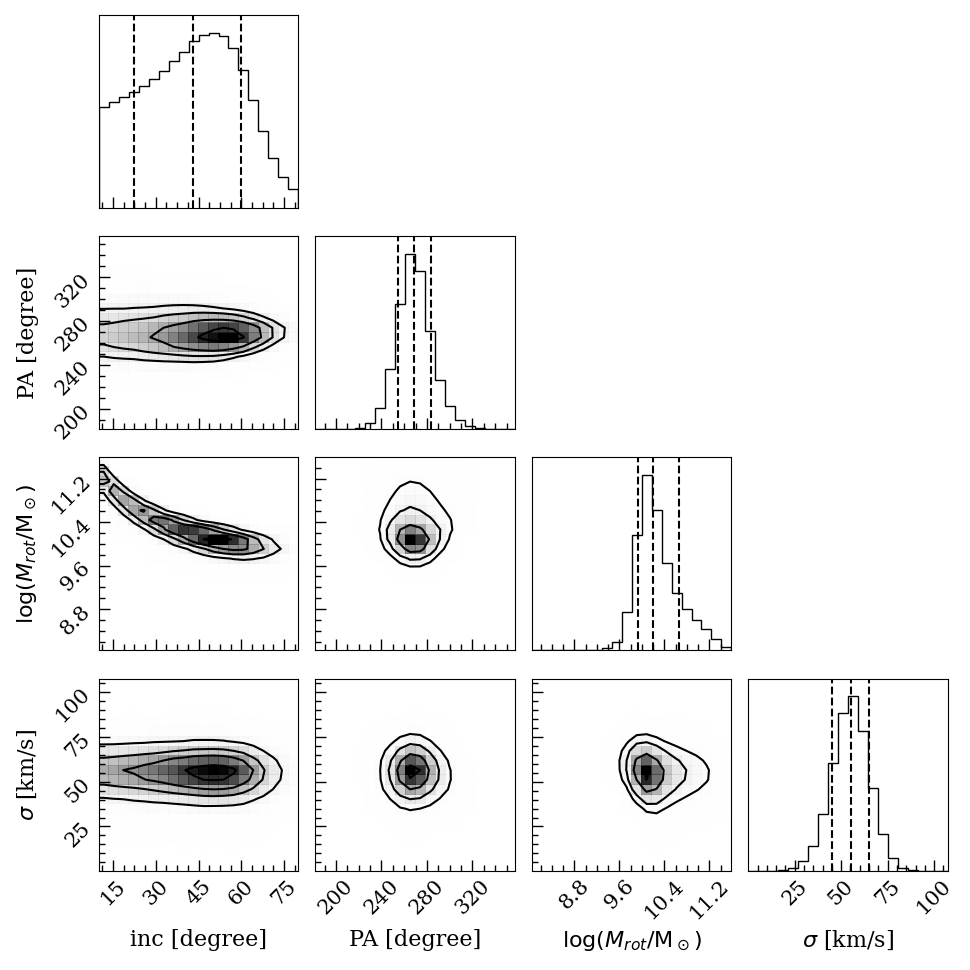}
        \includegraphics[width=\hsize]{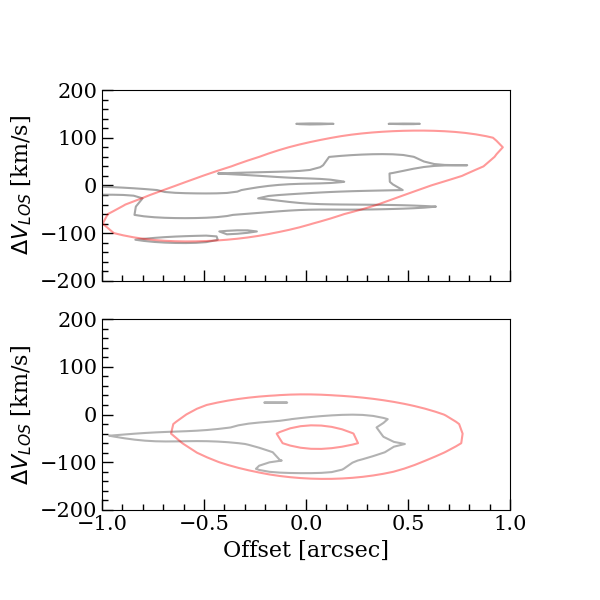}
    }
    \caption{Target J1211 observed with the \cii emission line. See the caption of Fig. \ref{Fig:cos30}.}
    \label{Fig:J1211_ciidistr}
   \end{figure*} 

\begin{figure*}[p]
   \resizebox{\hsize}{!}
    {  
    \includegraphics[width=\hsize]{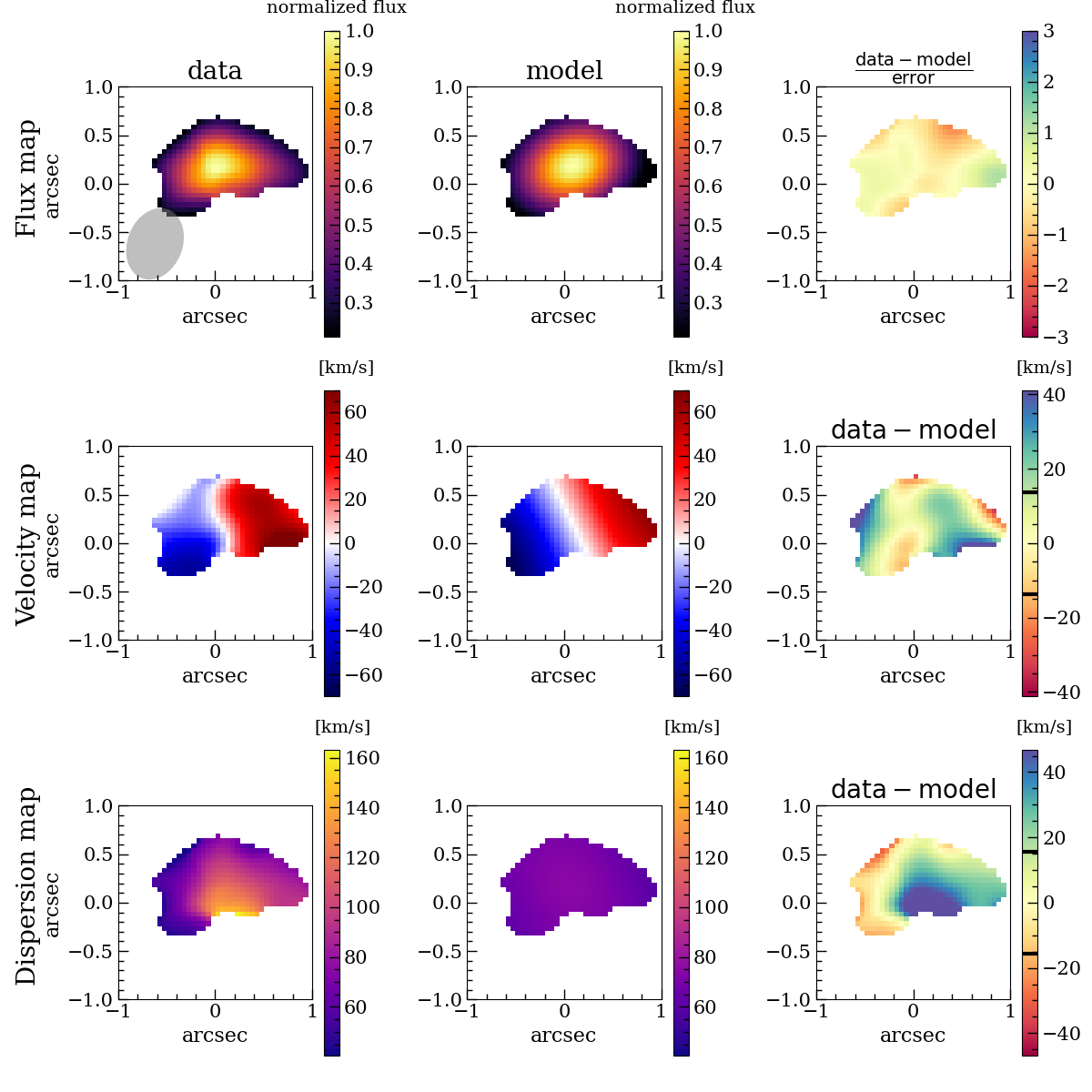}
    \includegraphics[width=\hsize]{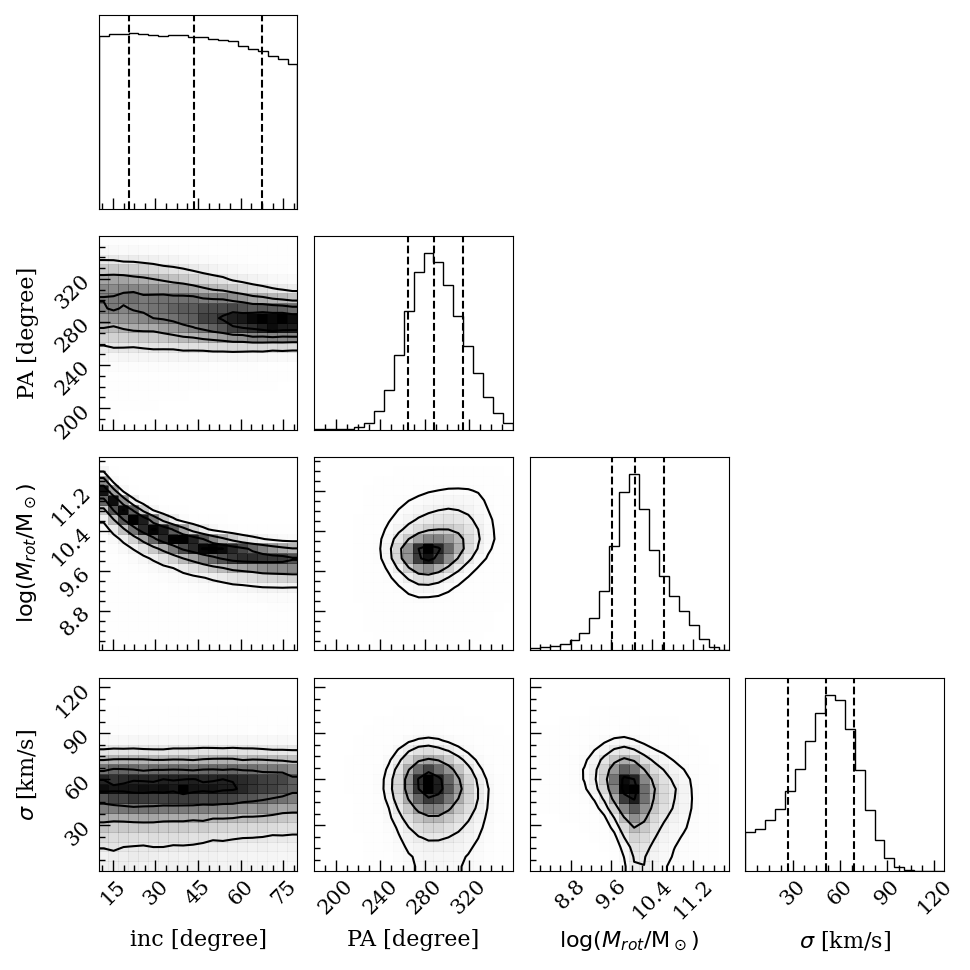}
    \includegraphics[width=\hsize]{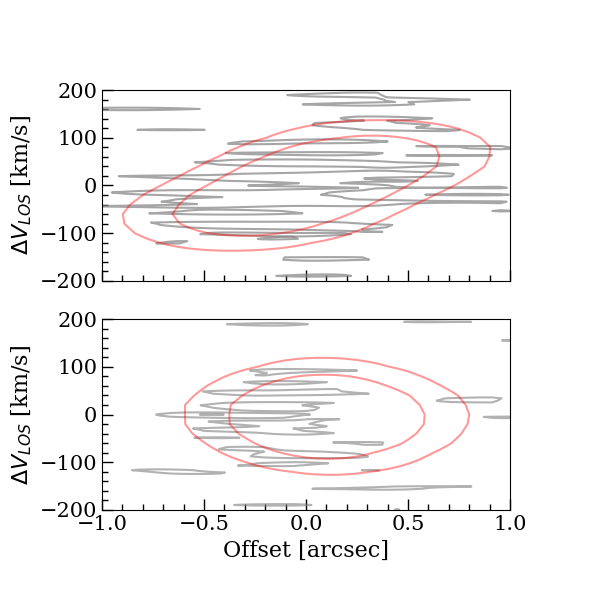}
    }
    \caption{Target J1211 observed with the \oiii emission line. See the caption of Fig. \ref{Fig:cos30}.}
    \label{Fig:J1211_oiiidistr}
   \end{figure*} 

\begin{figure*}[p]
   \resizebox{\hsize}{!}
    {  
    \includegraphics[width=\hsize]{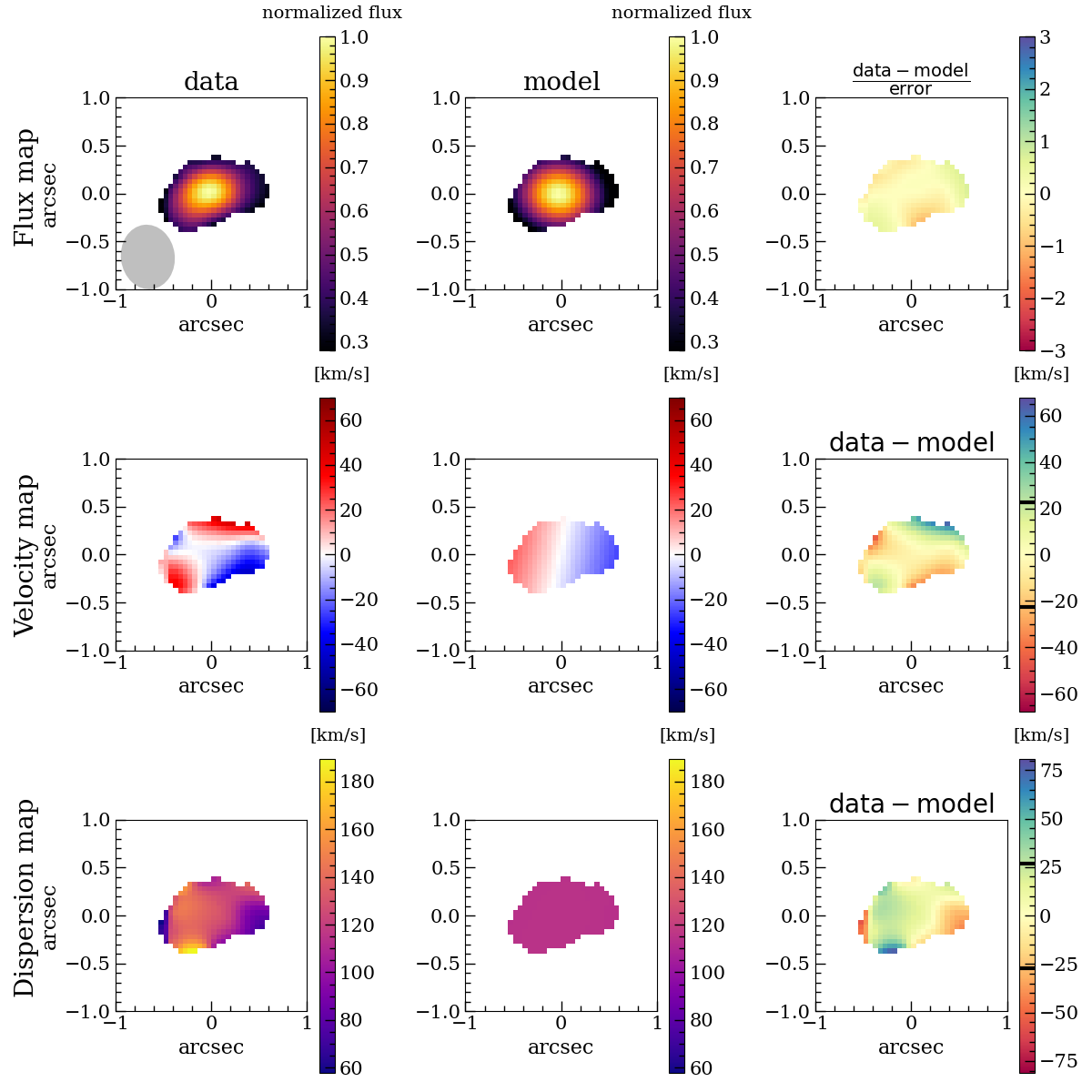}
    \includegraphics[width=\hsize]{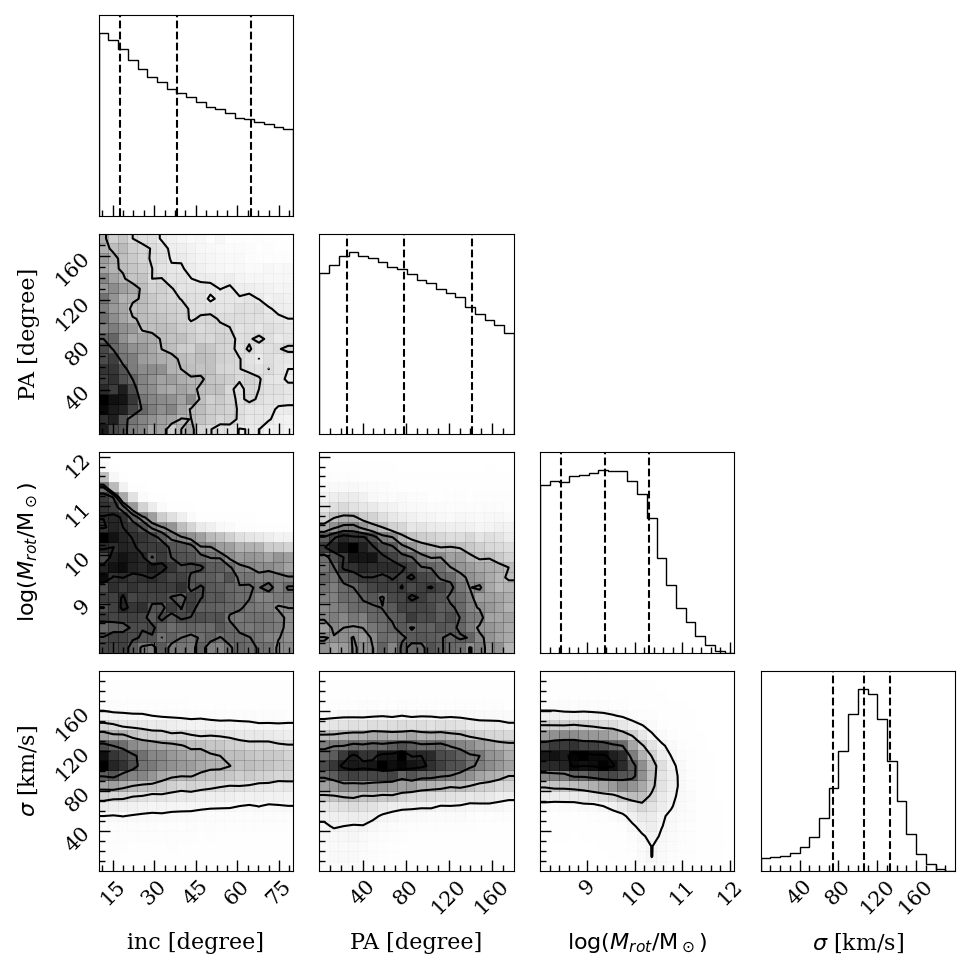}
     \includegraphics[width=\hsize]{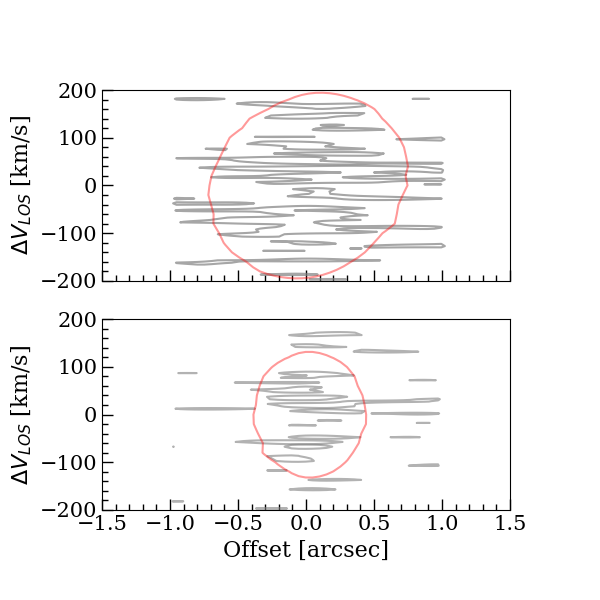}
    }
    \caption{Target J0217 observed with the \oiii emission line. See the caption of Fig. \ref{Fig:cos30}.}
    \label{Fig:J0217_oiiidistr}
   \end{figure*}

\begin{figure*}[p]
   \resizebox{\hsize}{!}
    {  
    \includegraphics[width=\hsize]{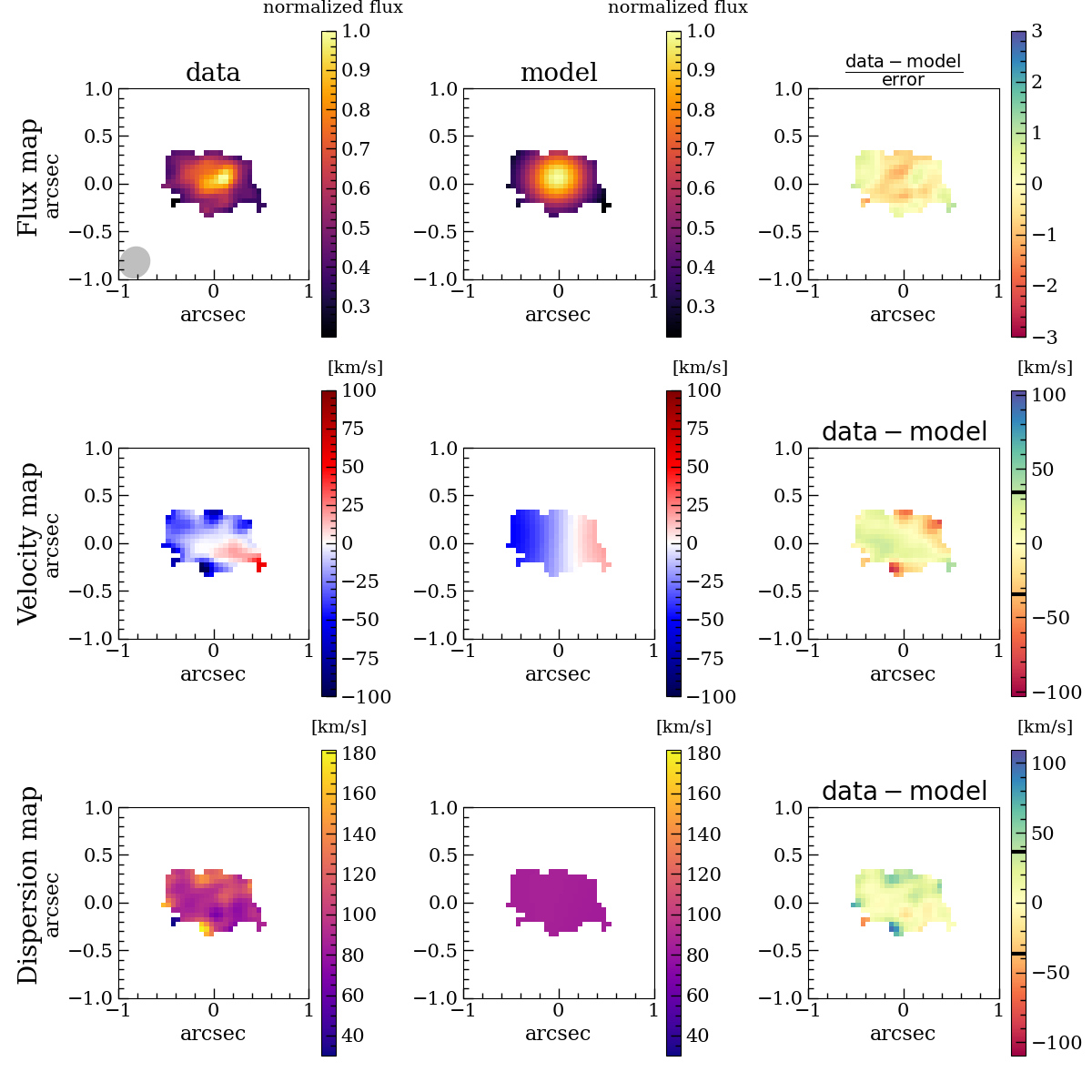}
    \includegraphics[width=\hsize]{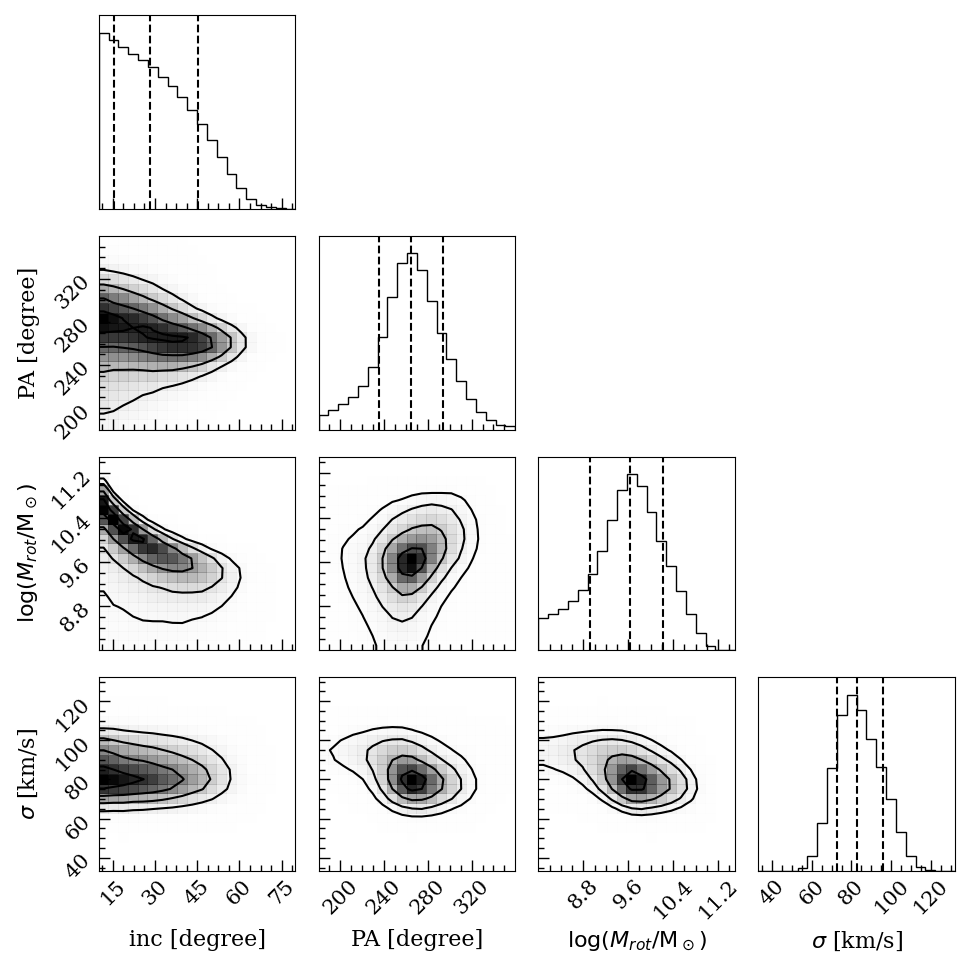}
     \includegraphics[width=\hsize]{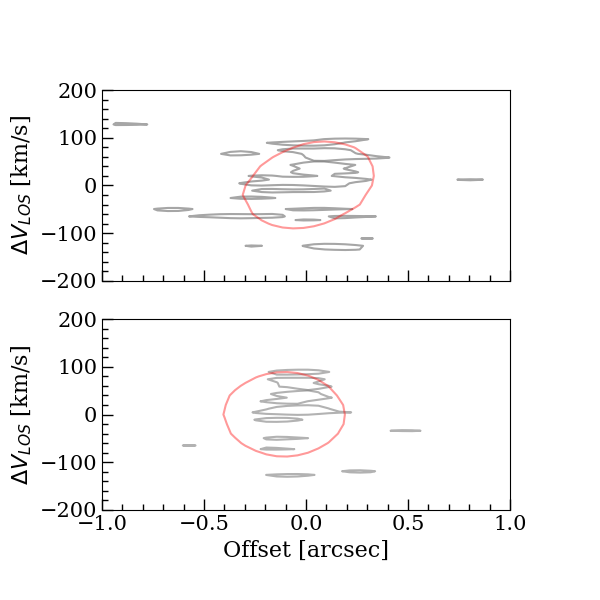}
    }
    \caption{Target HZ7 observed with the \cii emission line. See the caption of Fig. \ref{Fig:cos30}.}
    \label{Fig:HZ7_ciidistr}
   \end{figure*}

\begin{figure*}[p]
   \resizebox{\hsize}{!}
    {  
    \includegraphics[width=\hsize]{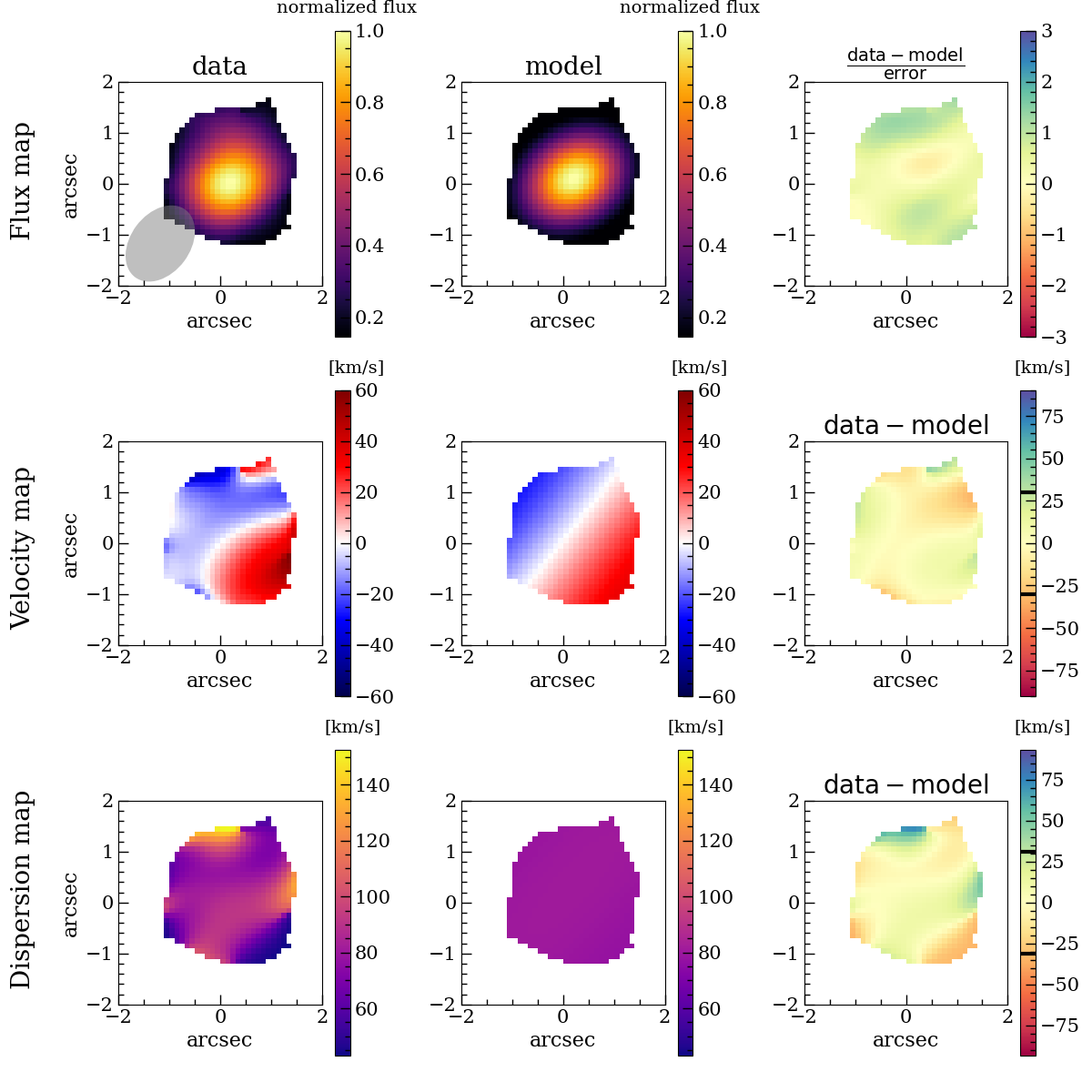}
    \includegraphics[width=\hsize]{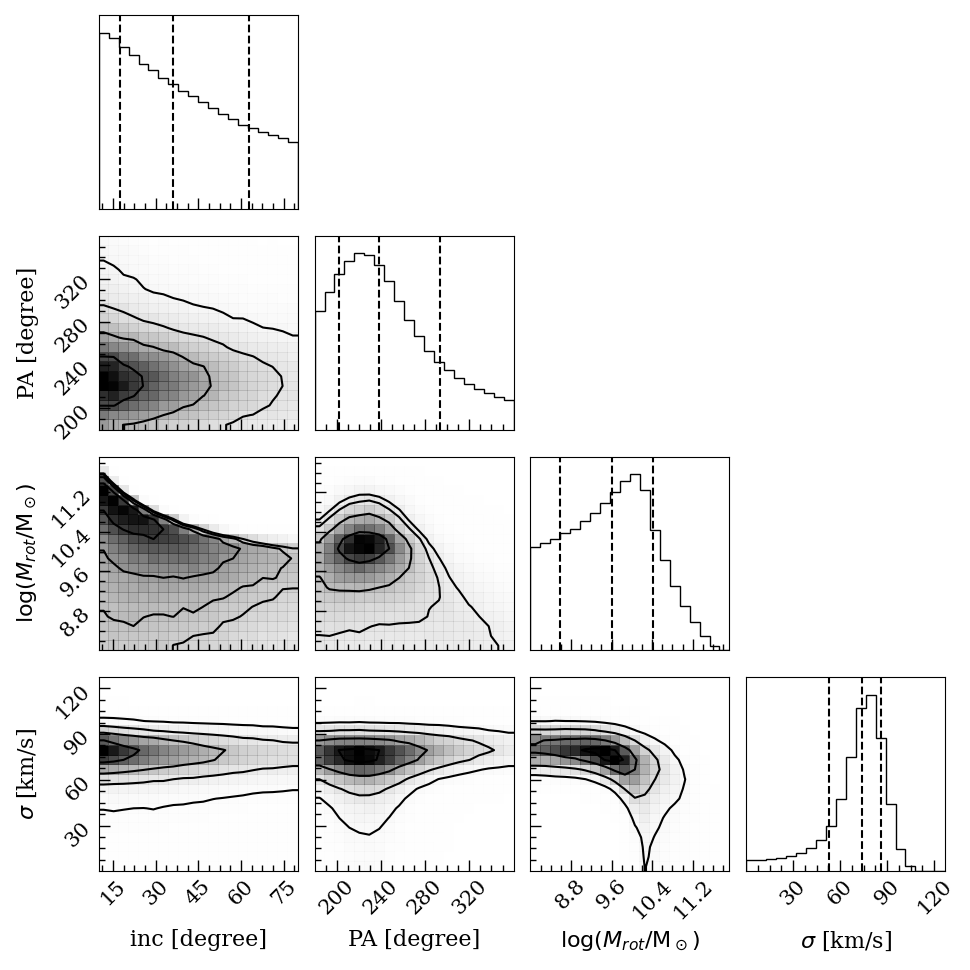}
         \includegraphics[width=\hsize]{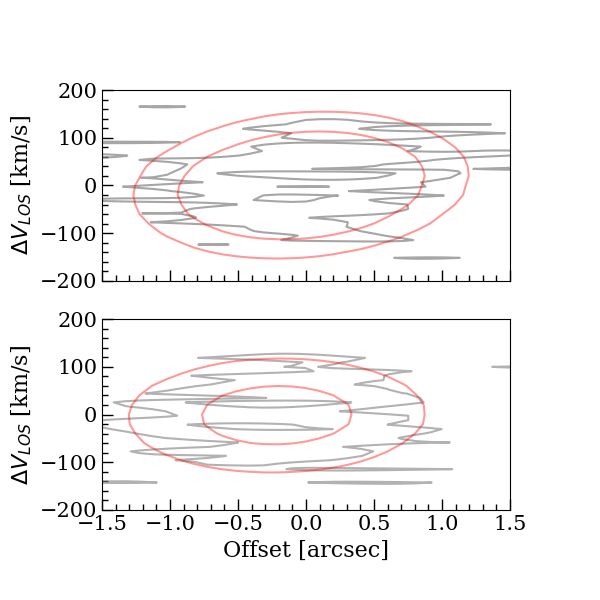}
    
    }
    \caption{Target UVISTA-Z-349 observed with the \cii emission line. See the caption of Fig. \ref{Fig:cos30}.}
    \label{Fig:z349distr}
   \end{figure*}

\begin{figure*}[p]
   \resizebox{\hsize}{!}
    {  
    \includegraphics[width=\hsize]{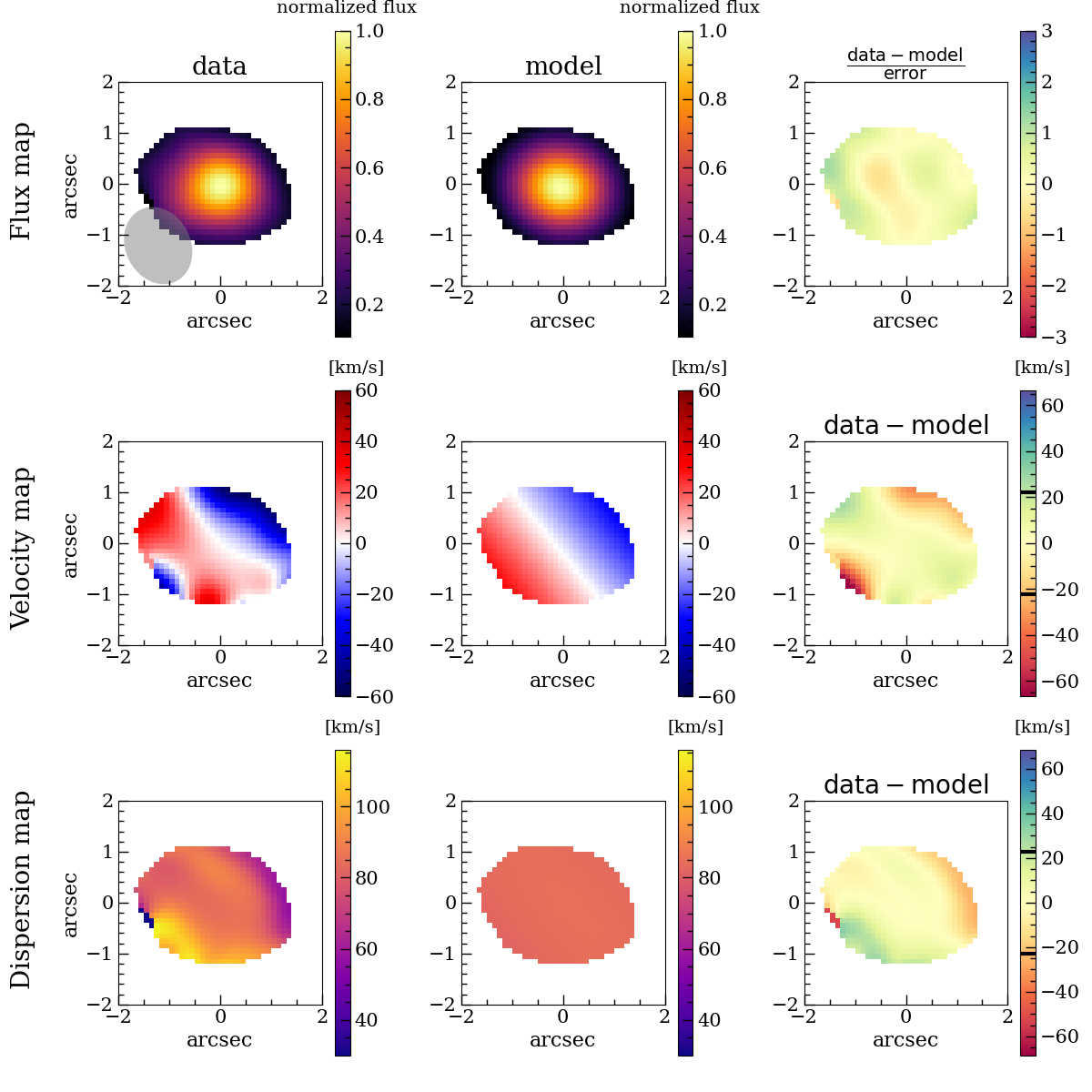}
    \includegraphics[width=\hsize]{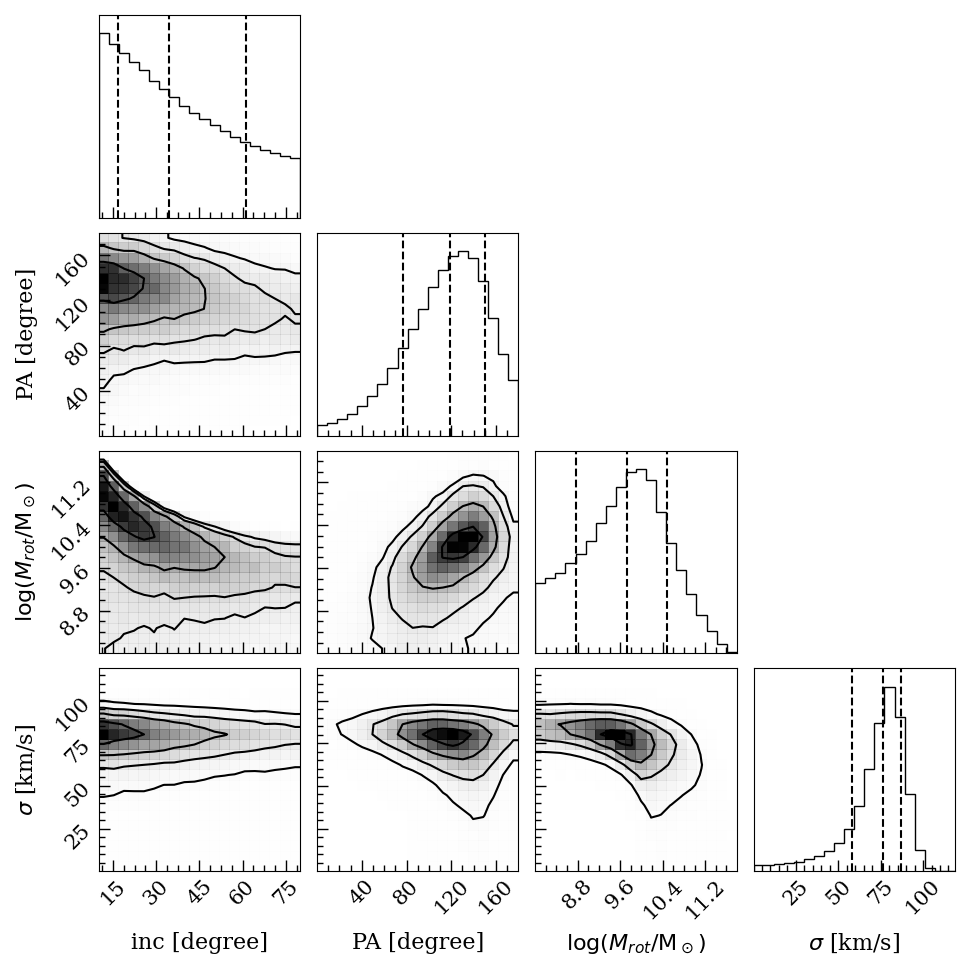}
    
        \includegraphics[width=\hsize]{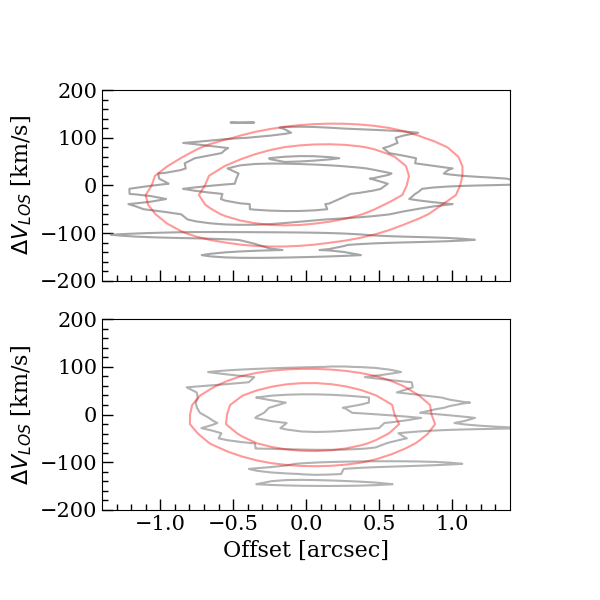}
    }    
    
    \caption{Target UVISTA-Y-001 observed with the \cii emission line. See the caption of Fig. \ref{Fig:cos30}.}
    \label{Fig:y001distr}
   \end{figure*}

\begin{figure*}[p]
   \resizebox{\hsize}{!}
    {  
    \includegraphics[width=\hsize]{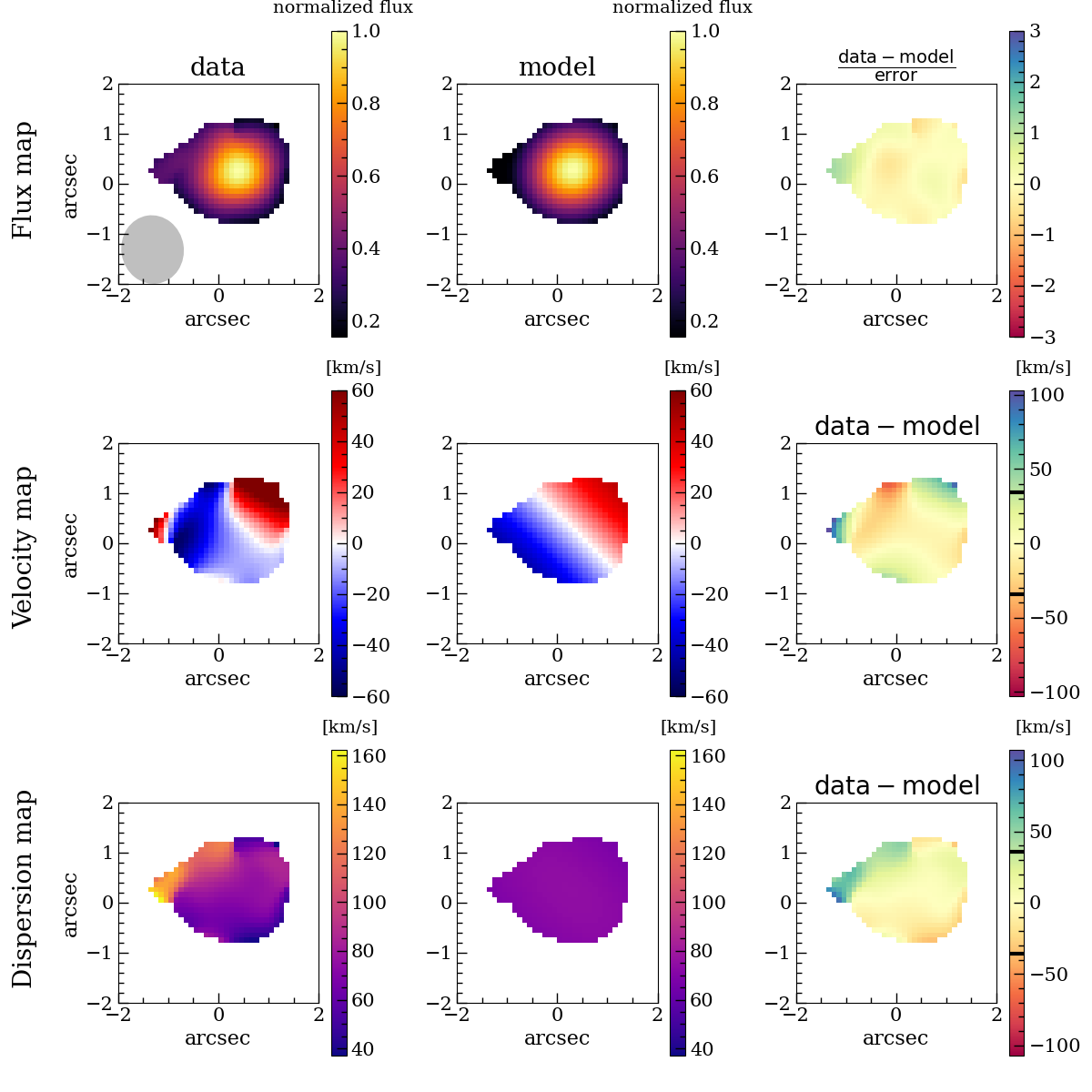}
    \includegraphics[width=\hsize]{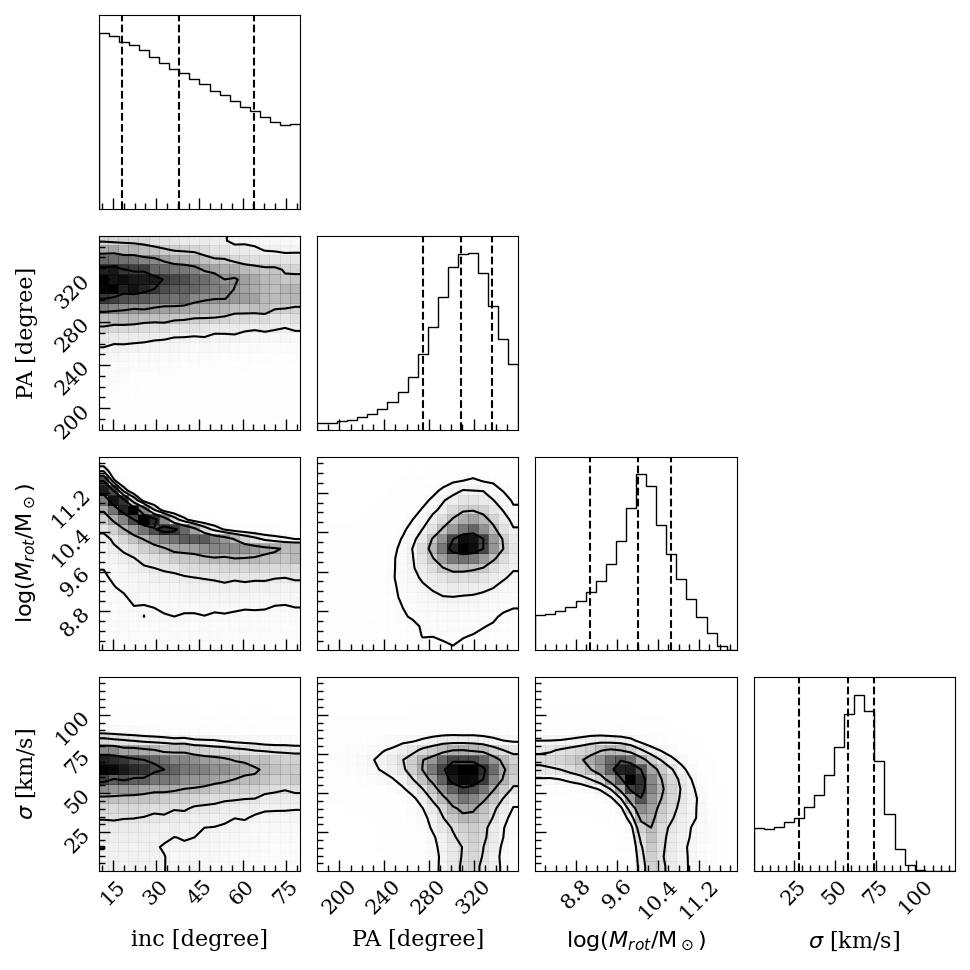}
            \includegraphics[width=\hsize]{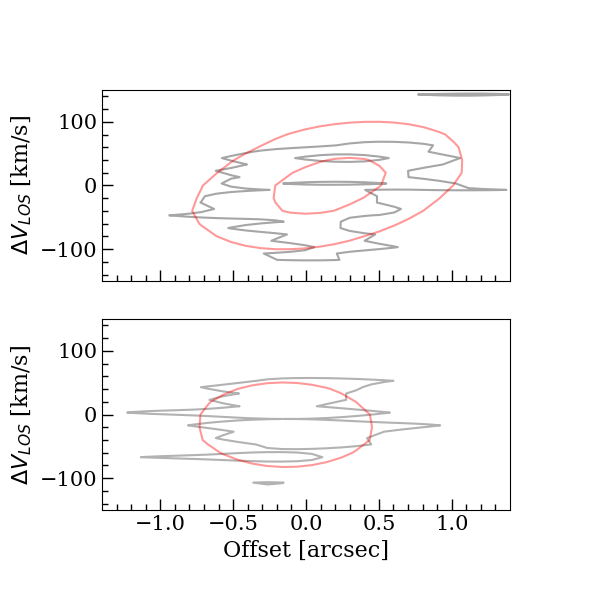}
    }
    \caption{Target UVISTA-Y-004 observed with the \cii emission line. See the caption of Fig. \ref{Fig:cos30}.}
    \label{Fig:y004distr}
   \end{figure*} 

\begin{figure*}[p]
   \resizebox{\hsize}{!}
    {  
    \includegraphics[width=\hsize]{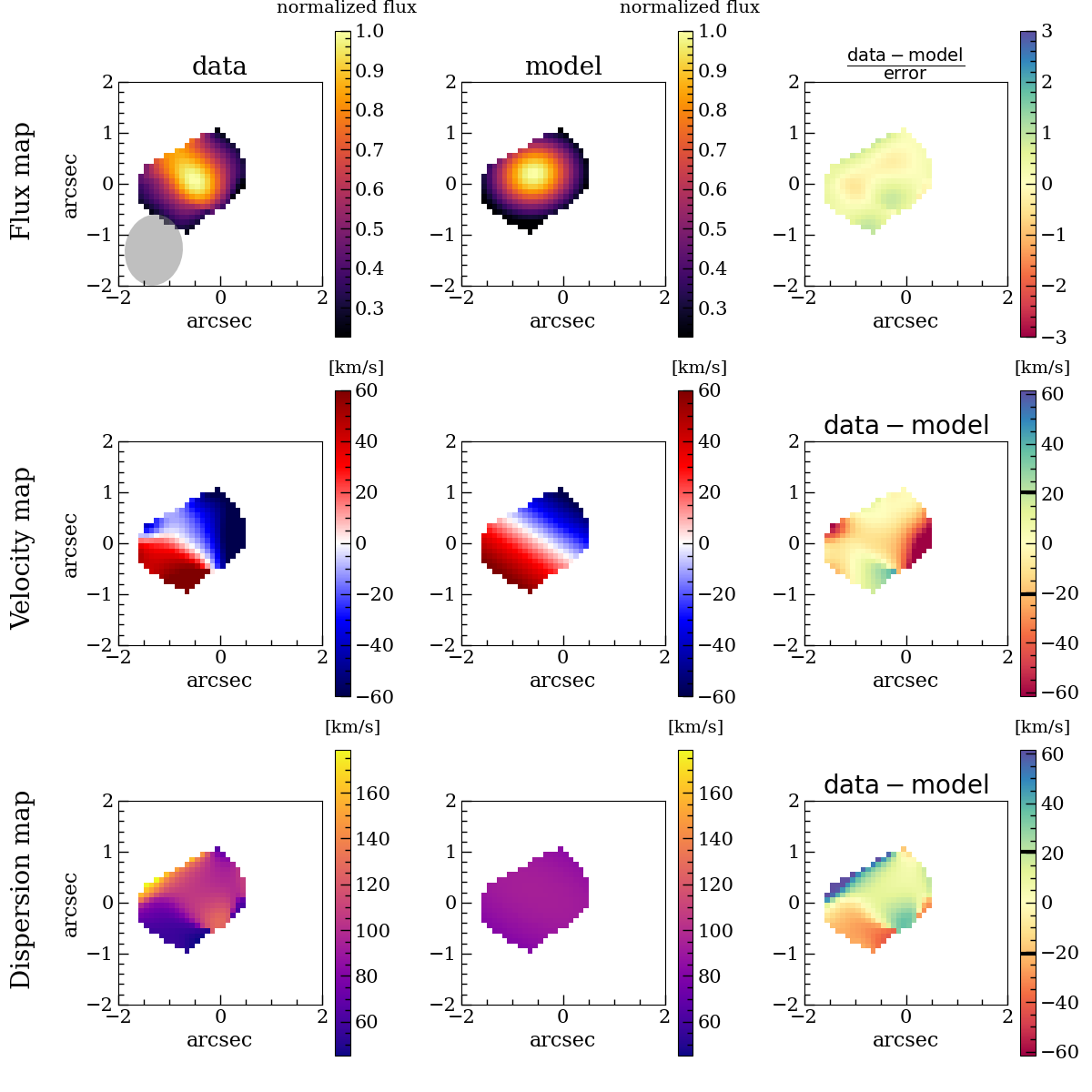}
    \includegraphics[width=\hsize]{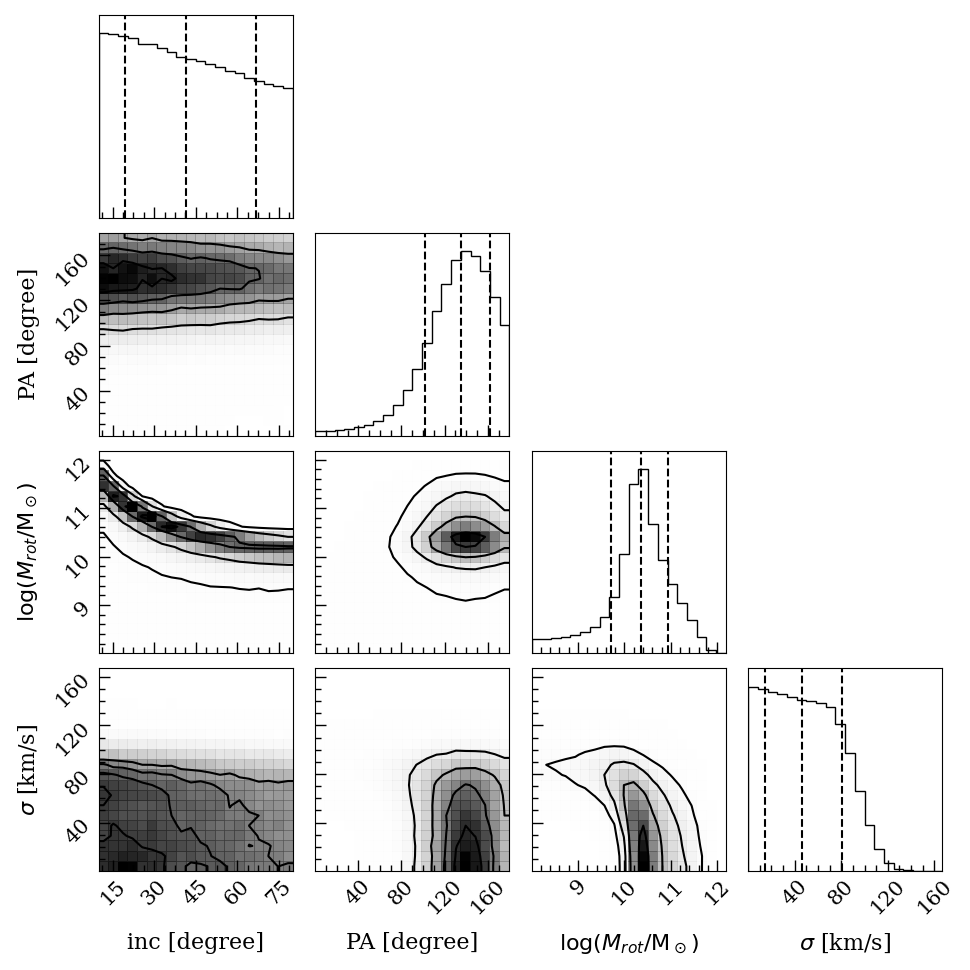}
        \includegraphics[width=\hsize]{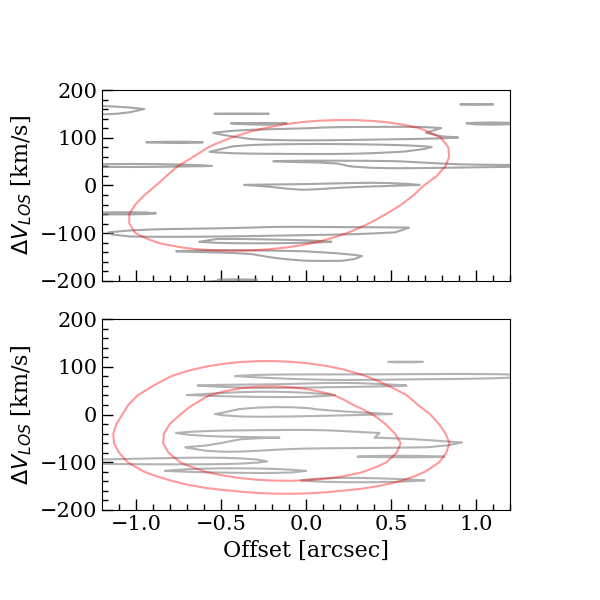}
    
    }
    \caption{Target UVISTA-Z-001 observed with the \cii emission line. See the caption of Fig. \ref{Fig:cos30}.}
    \label{Fig:z001distr}
   \end{figure*} 
   
\begin{figure*}[p]
   \resizebox{\hsize}{!}
    {  
    \includegraphics[width=\hsize]{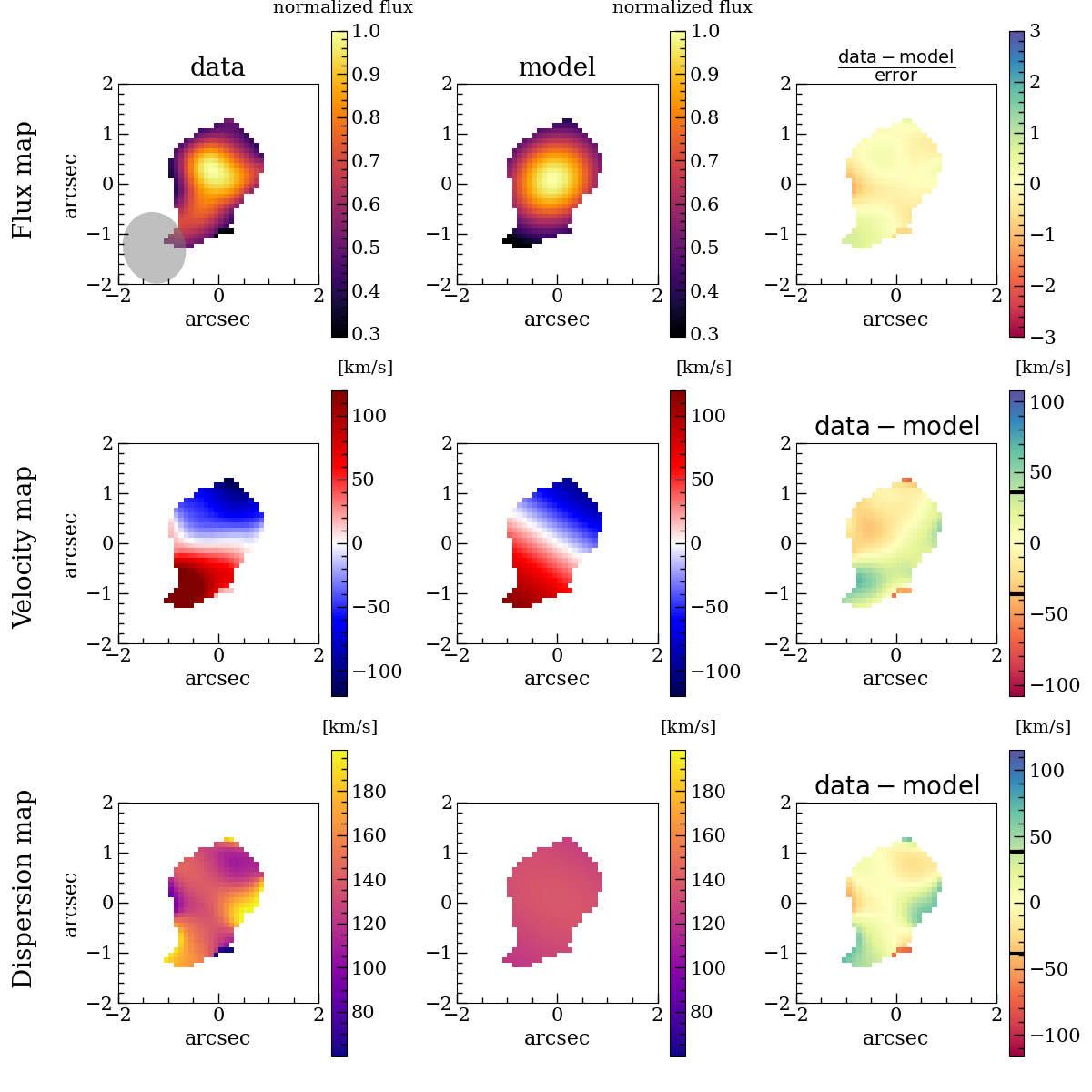}
        \includegraphics[width=\hsize]{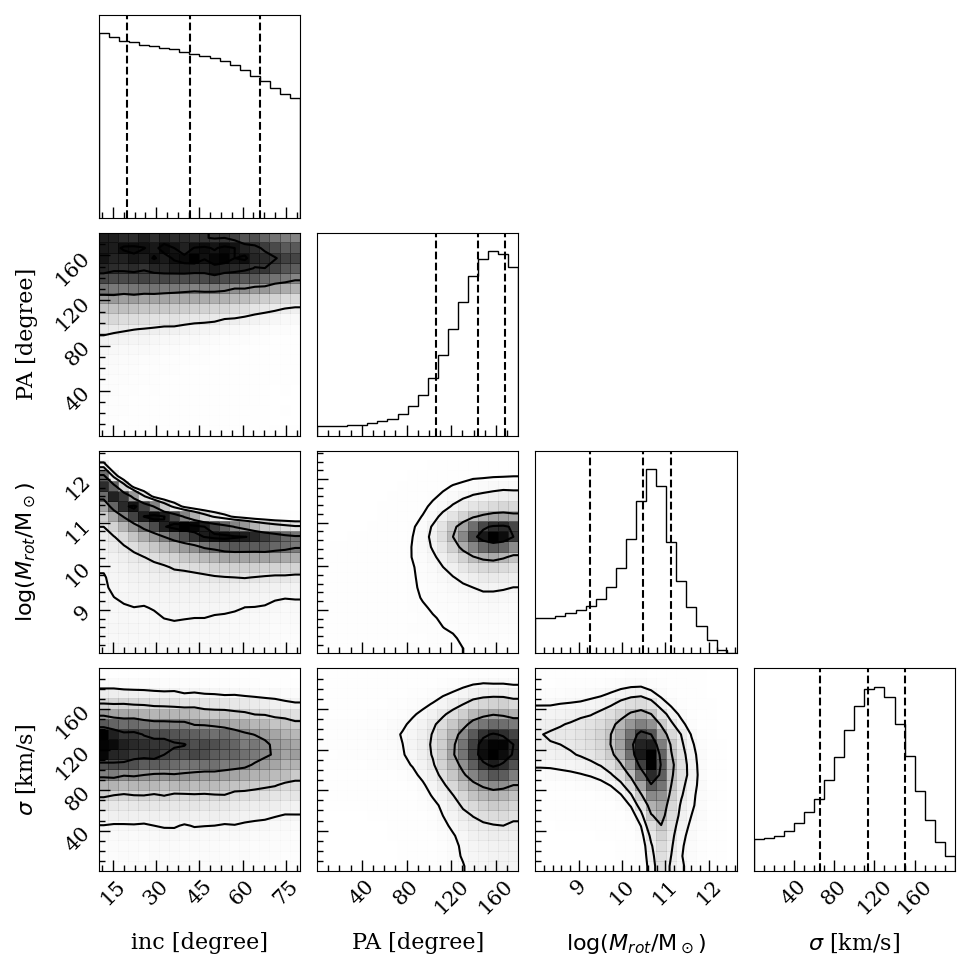}
    \includegraphics[width=\hsize]{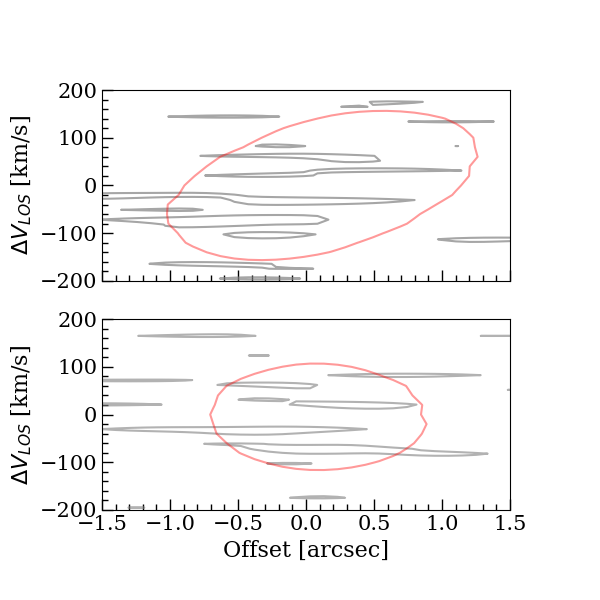}
    
    }
    \caption{Target UVISTA-Y-879 observed with the \cii emission line. See the caption of Fig. \ref{Fig:cos30}.}
    \label{Fig:z879distr}
   \end{figure*}

\begin{figure*}[p]
   \resizebox{\hsize}{!}
    {  
    \includegraphics[width=\hsize]{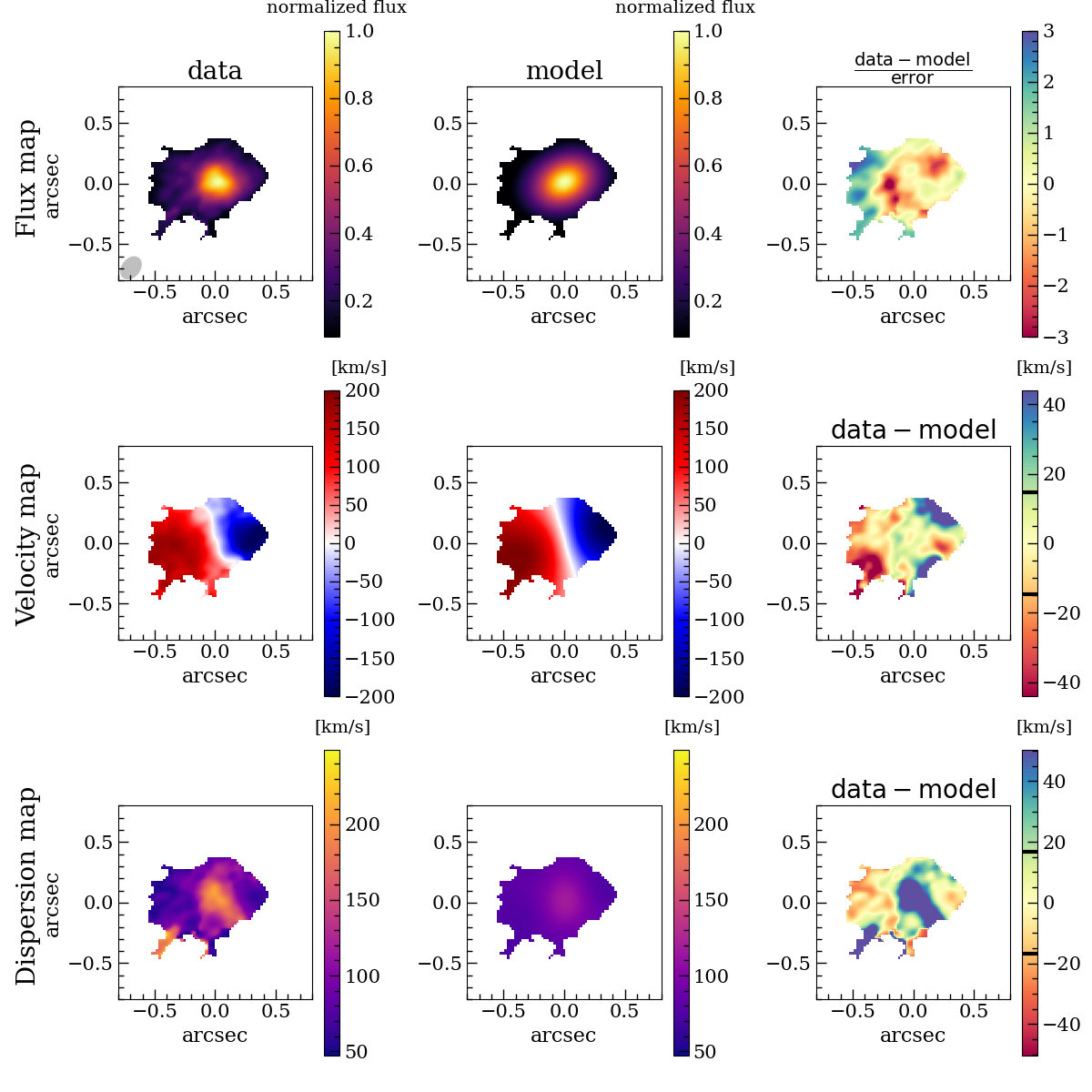}
    \includegraphics[width=\hsize]{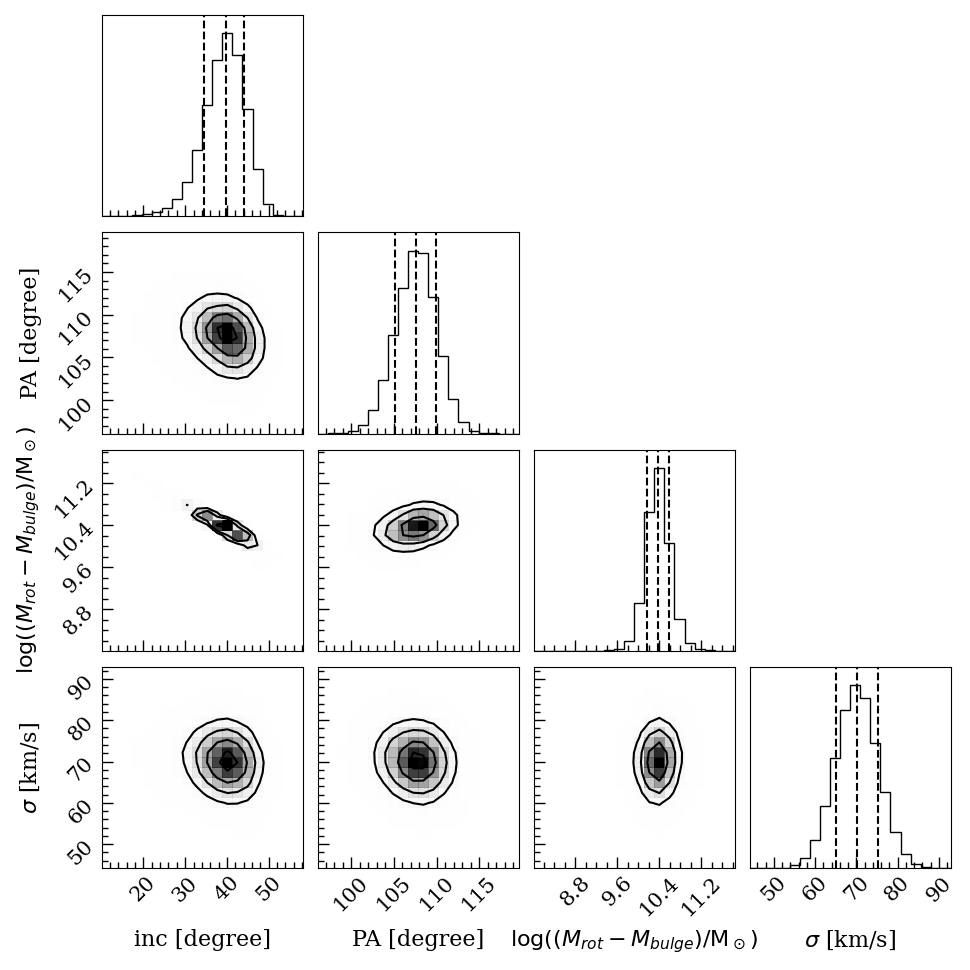}
        \includegraphics[width=\hsize]{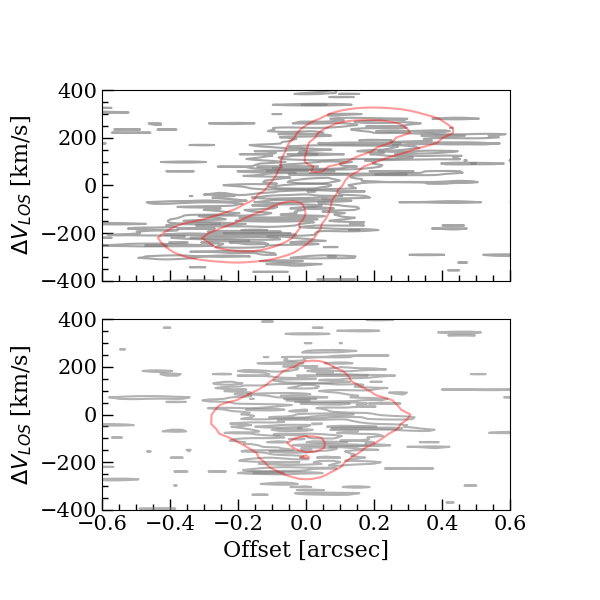}
    }
    \caption{Target DLA0817g observed with the \cii emission line. See the caption of Fig. \ref{Fig:cos30}.}
    \label{Fig:DLA}
   \end{figure*}  

\begin{figure*}[p]
   \resizebox{\hsize}{!}
    {  
    \includegraphics[width=\hsize]{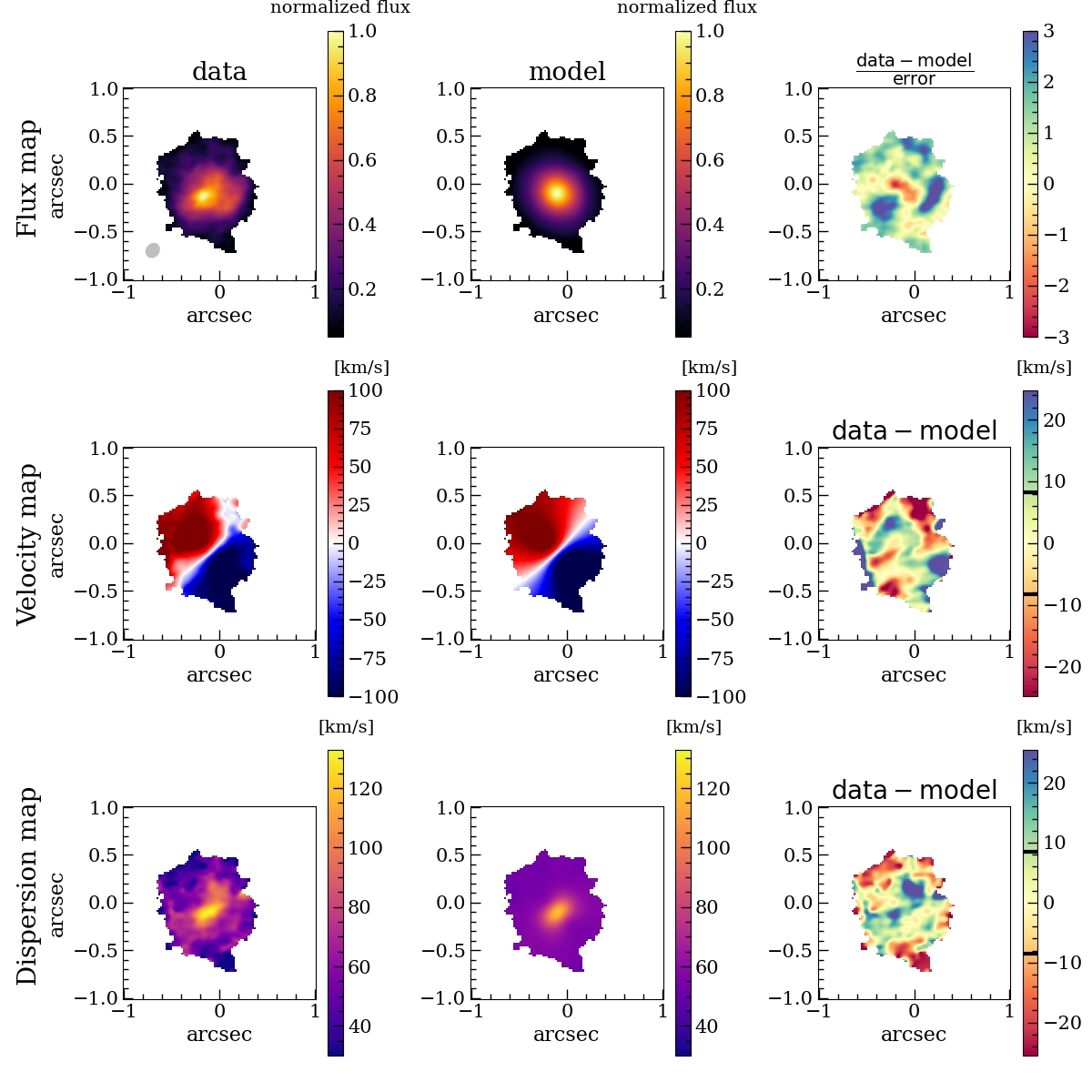}
    \includegraphics[width=\hsize]{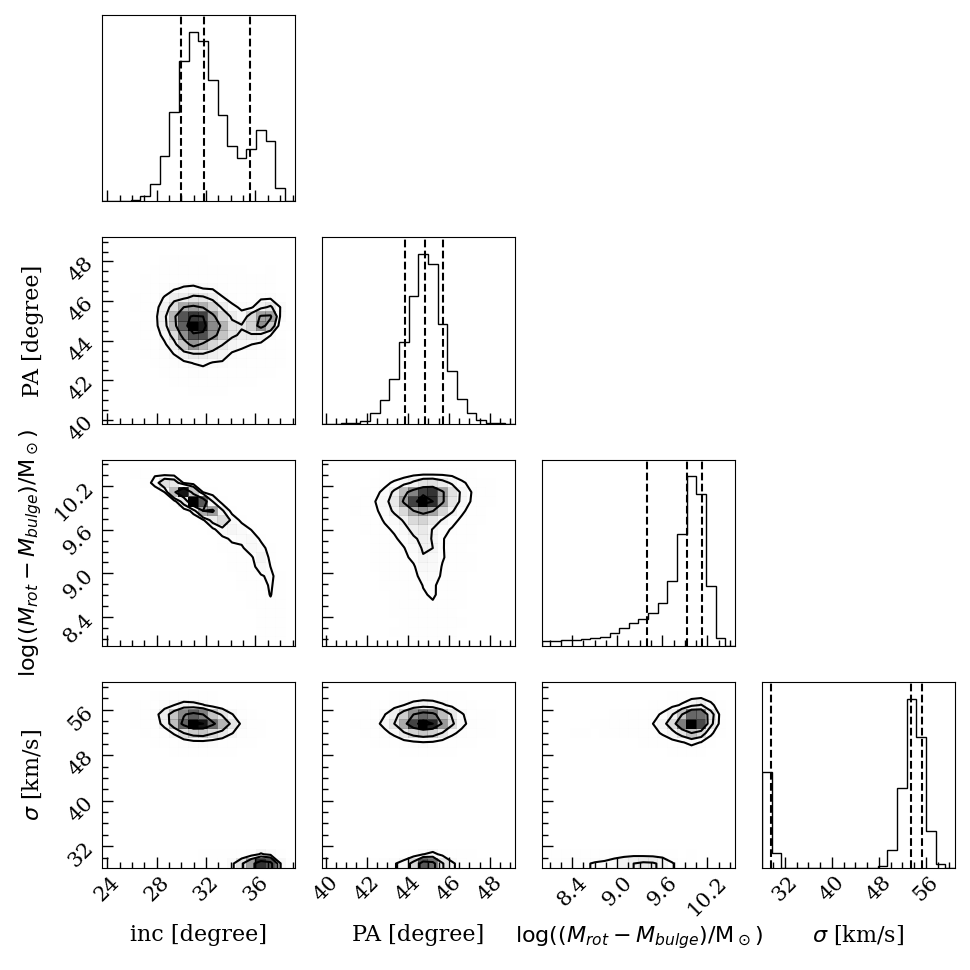}
            \includegraphics[width=\hsize]{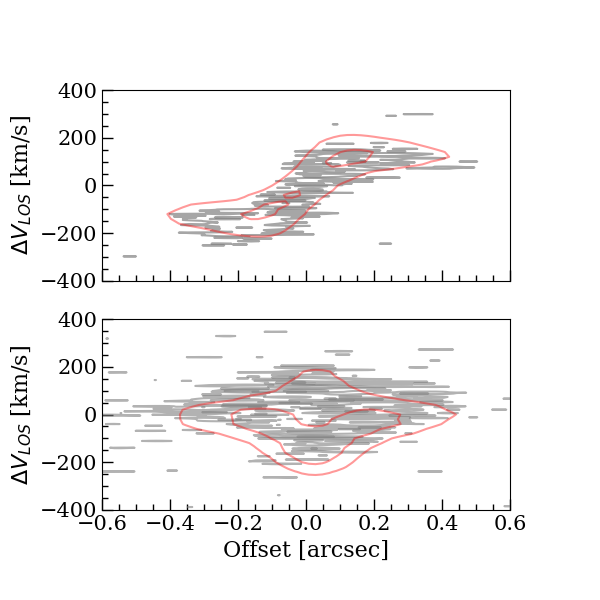}
    }
    \caption{Target ALESS 073.1 observed with the \cii emission line. See the caption of Fig. \ref{Fig:cos30}.}
    \label{Fig:aless}
   \end{figure*}

\section{1D kinematical fitting results}\label{Resultsmethod2}
In this section, we present the results of the 1D fitting procedure for galaxies with unresolved kinematics.
We recovered the radius and the galaxy center by fitting  the flux map that we present for each target in the following figures. We recovered the total baryonic mass by summing the stellar masses (estimated in UV) and the gas mass recovered from the \cii luminosity.
We then performed the fitting on the spatially integrated spectrum. We present the spectrum, the model spectrum, and the corner plot to show the distribution of the free parameters for each target in the following figures.

\centering

\begin{figure*}[p]
   \resizebox{14cm}{!}
    {   
        \includegraphics[scale=0.7]{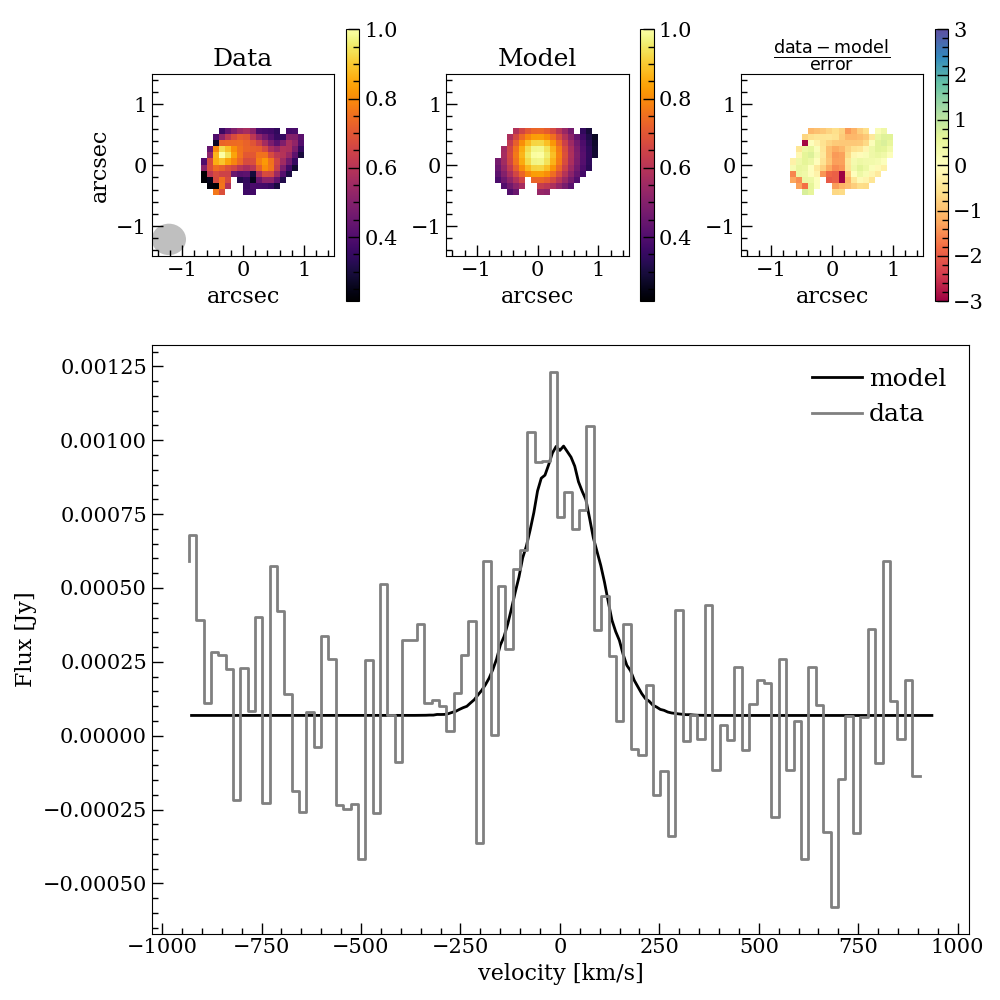}
        \includegraphics[clip, scale=0.7]{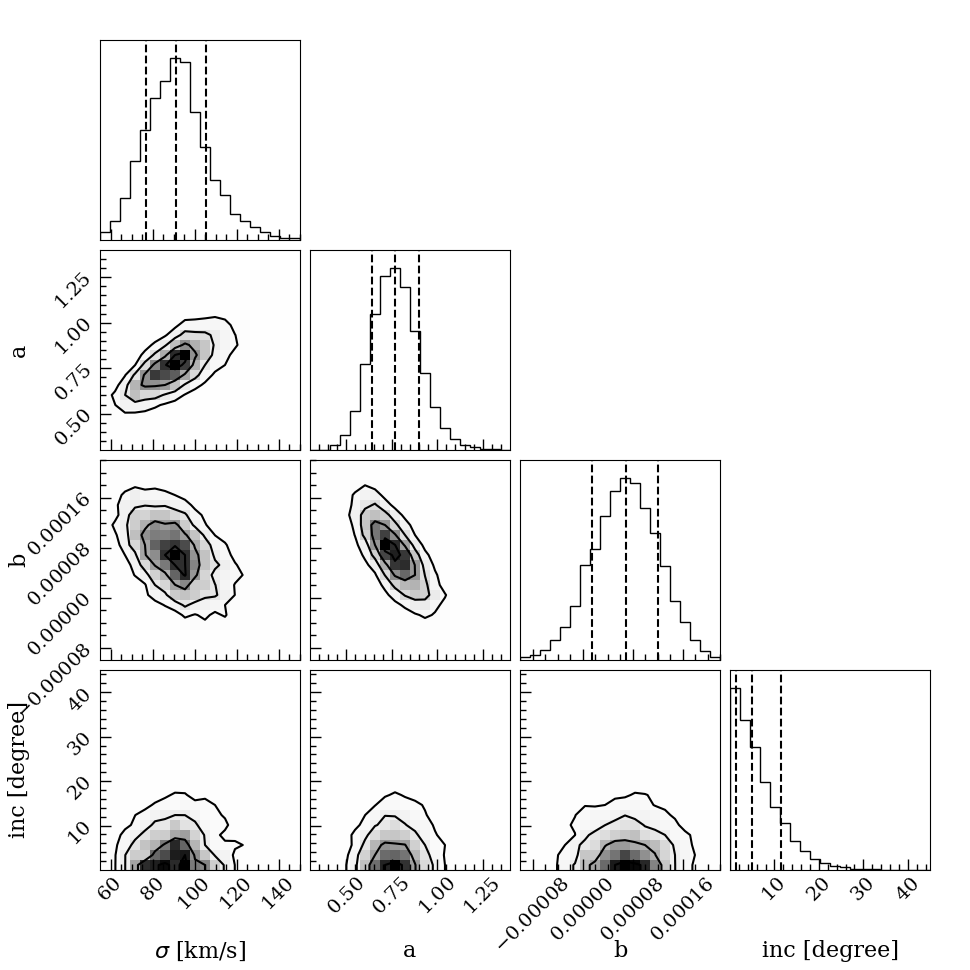}
        }
    \caption{1D fitting results for the target VR7. In the left panel, we present the flux map at the top. From left to right, we show the observed flux, the model, and the residuals. The beam size is shown as the gray ellipse. At the bottom, we show the best-fit and the observed integrated spectrum.
    On the right, we show the corner plot of the free parameter of the fitting of the integrated spectrum. 
          }

    \label{Fig:VR7}
   \end{figure*}

\begin{figure*}[htbp]
   \resizebox{14cm}{!}
    {   
    \includegraphics{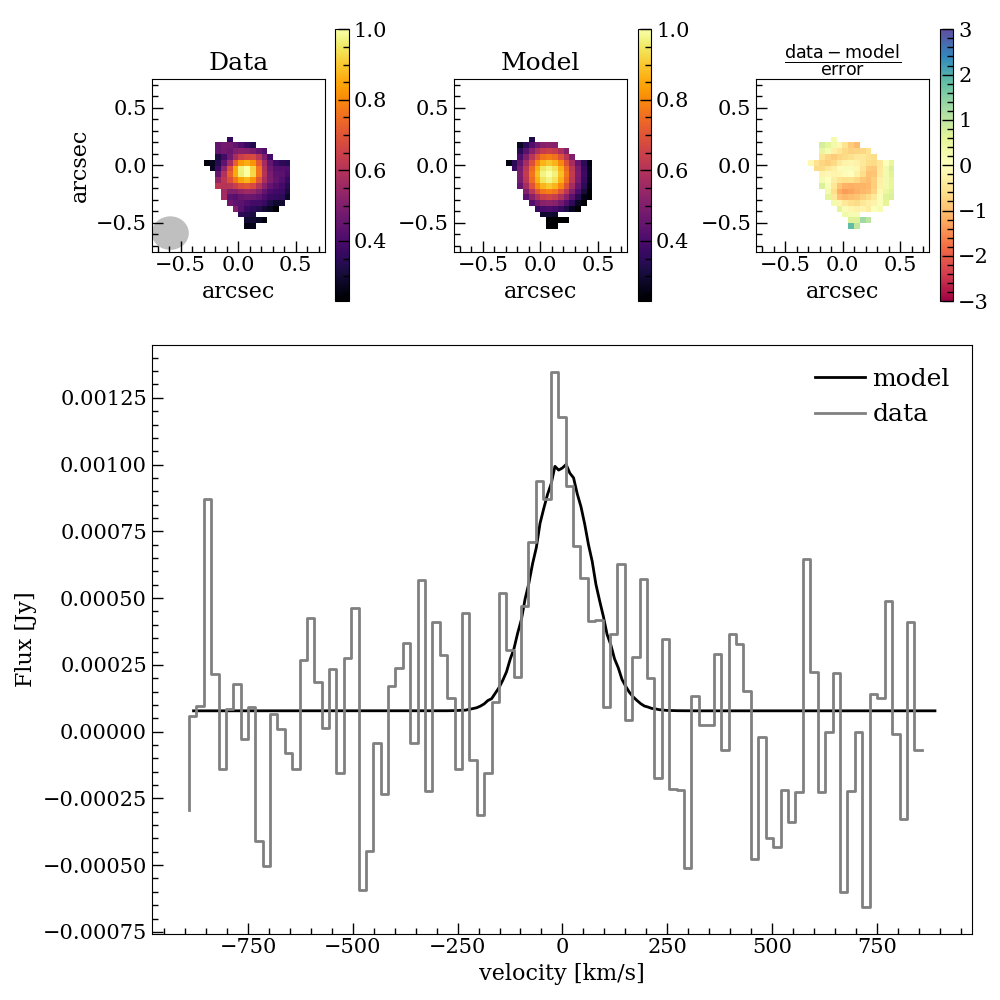}
        \includegraphics{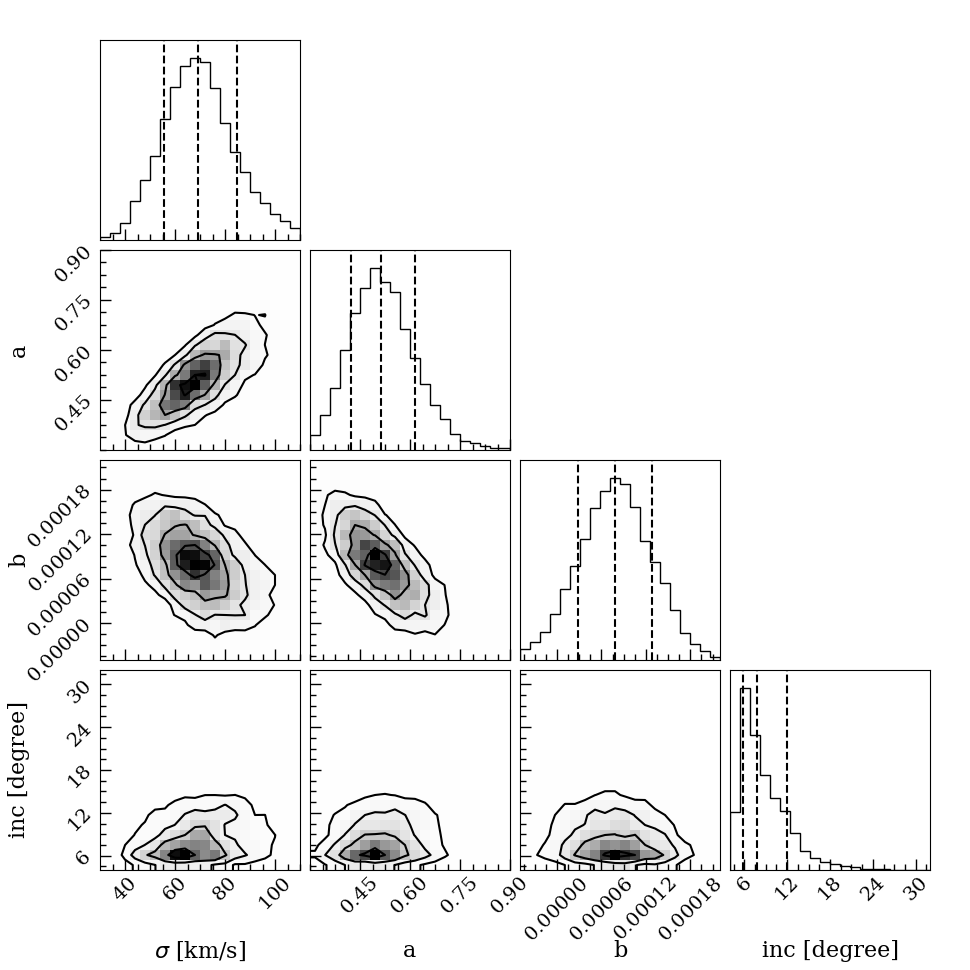}
    
        }
    \caption{Target CLM1, see the caption of \ref{Fig:VR7}.
          }

    \label{Fig:clm1}
   \end{figure*}

\begin{figure*}[htbp]
   \resizebox{14cm}{!}
    {   
        \includegraphics{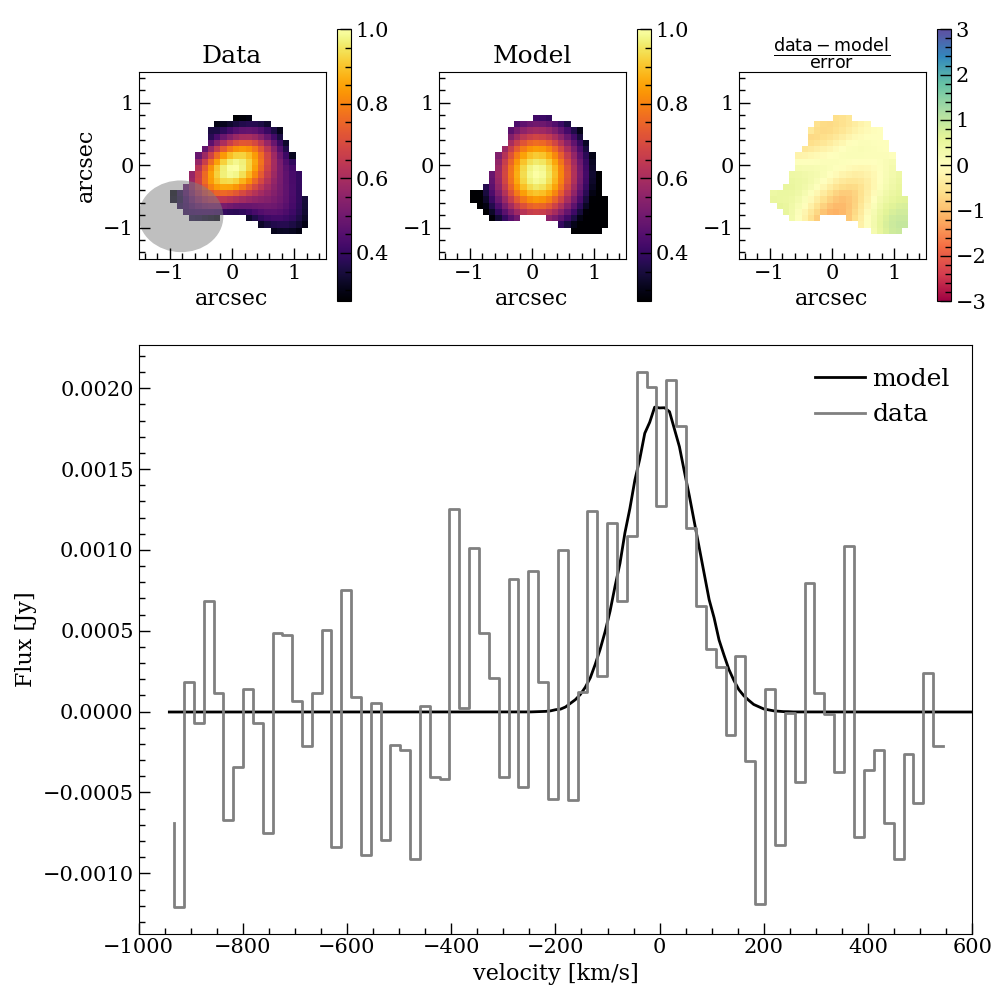}
        \includegraphics[clip]{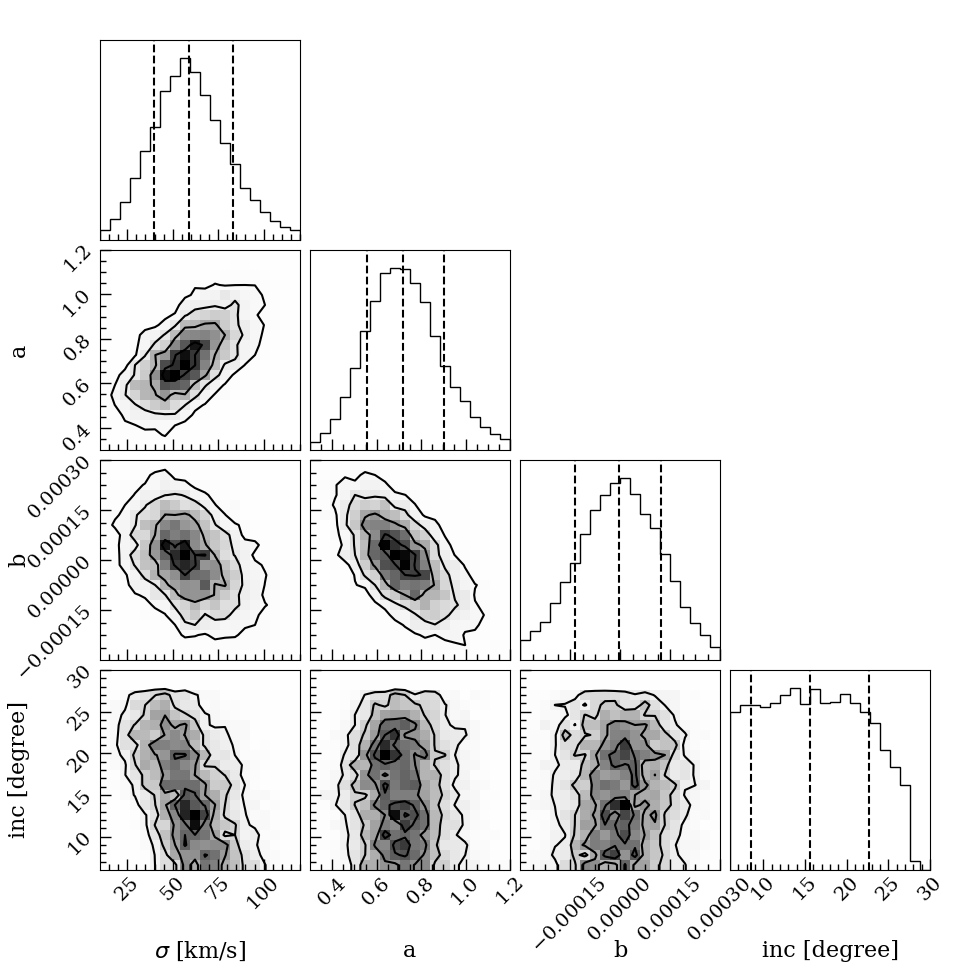}   
        }
    \caption{Target UVISTA-Z-004, see the caption of \ref{Fig:VR7}.
          }

    \label{Fig:uvista-z-004}
   \end{figure*} 

\begin{figure*}[htbp]
   \resizebox{14cm}{!}
    {   
        \includegraphics{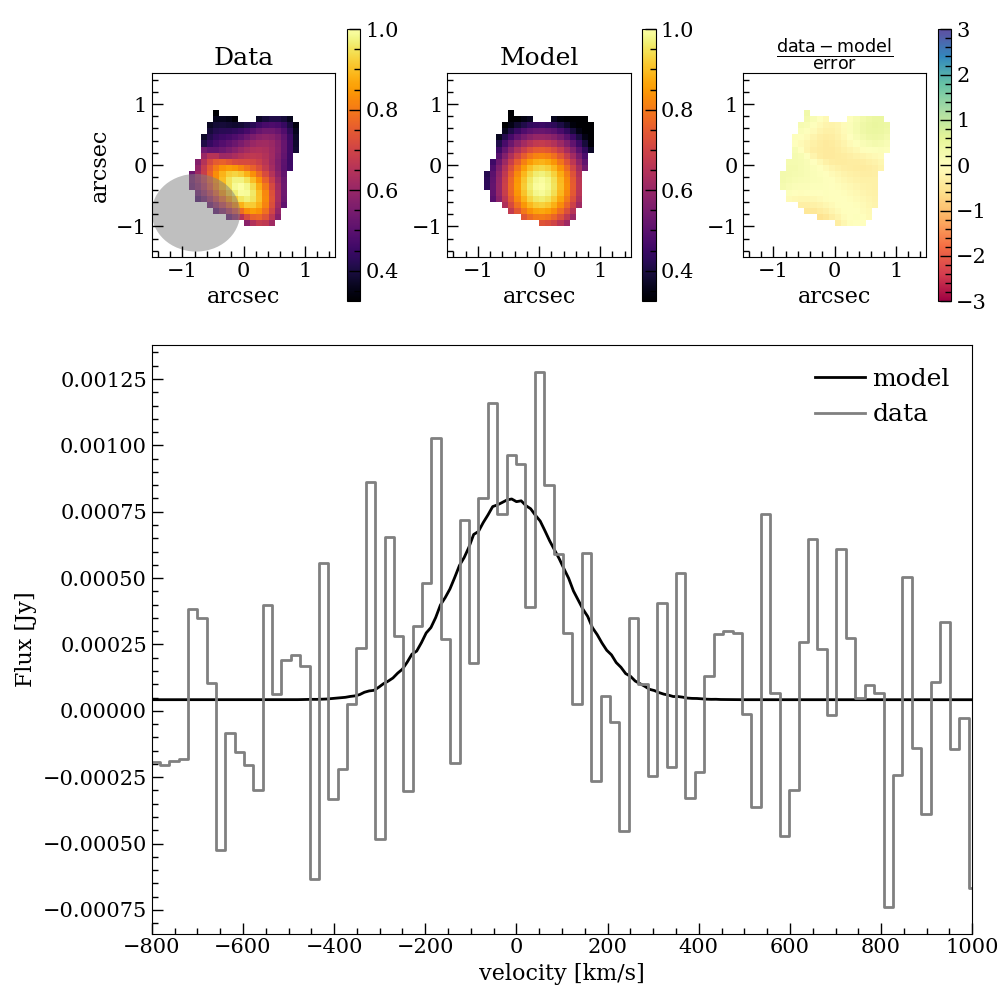}
        \includegraphics[clip]{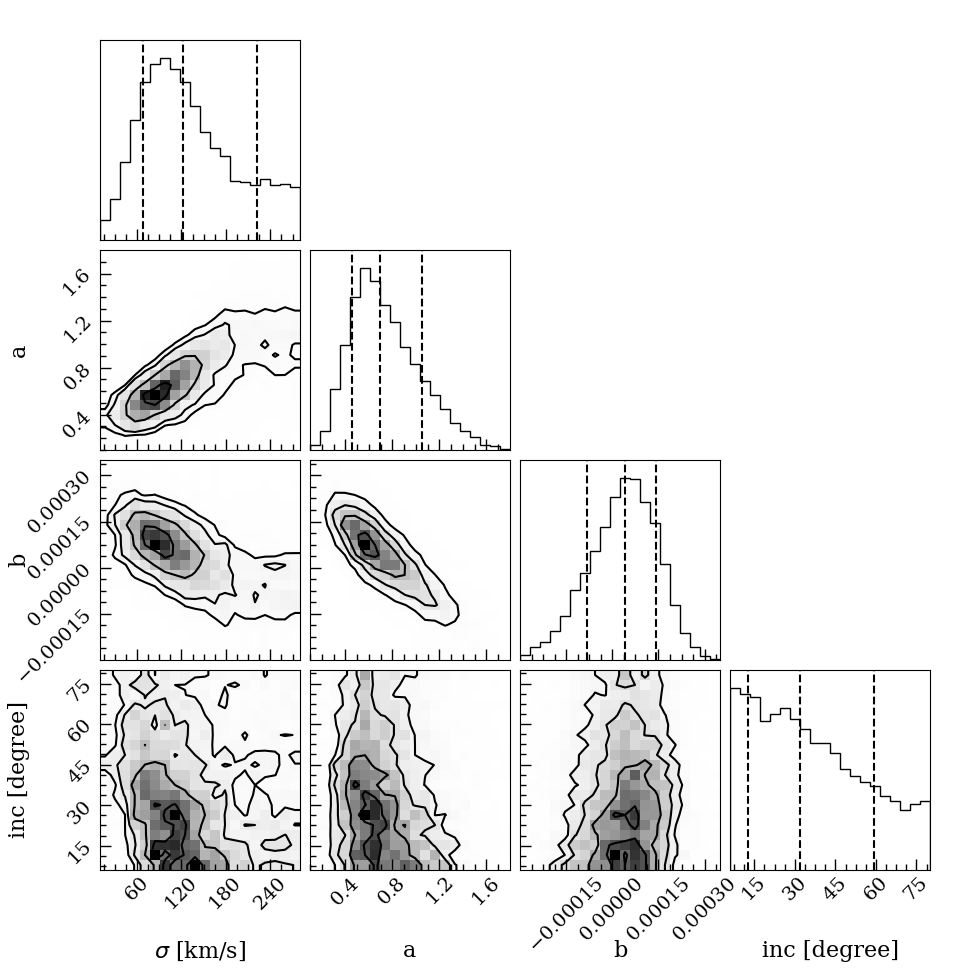}    
        }
    \caption{Target SUPER-8, see the caption of \ref{Fig:VR7}.
          }

    \label{Fig:super8}
   \end{figure*}

\begin{figure*}[htbp]
   \resizebox{14cm}{!}
    {   
        \includegraphics{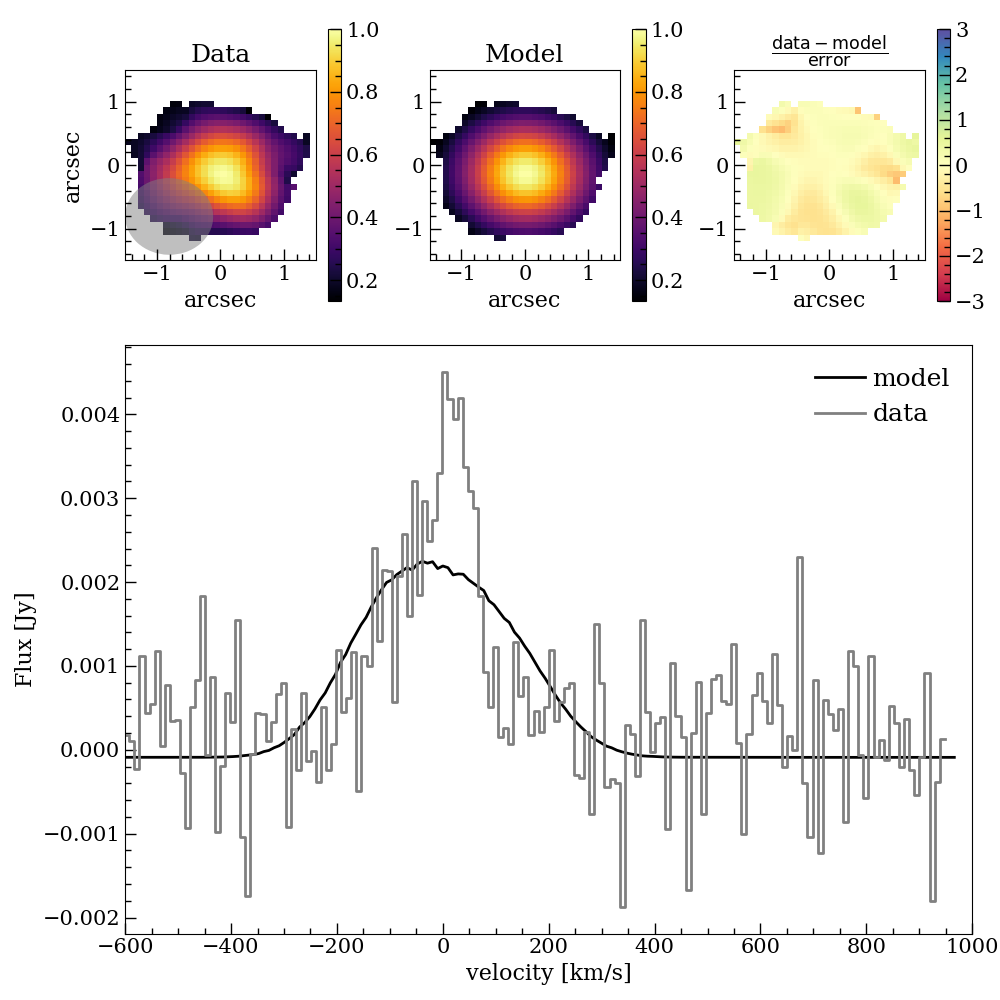}
        \includegraphics[clip]{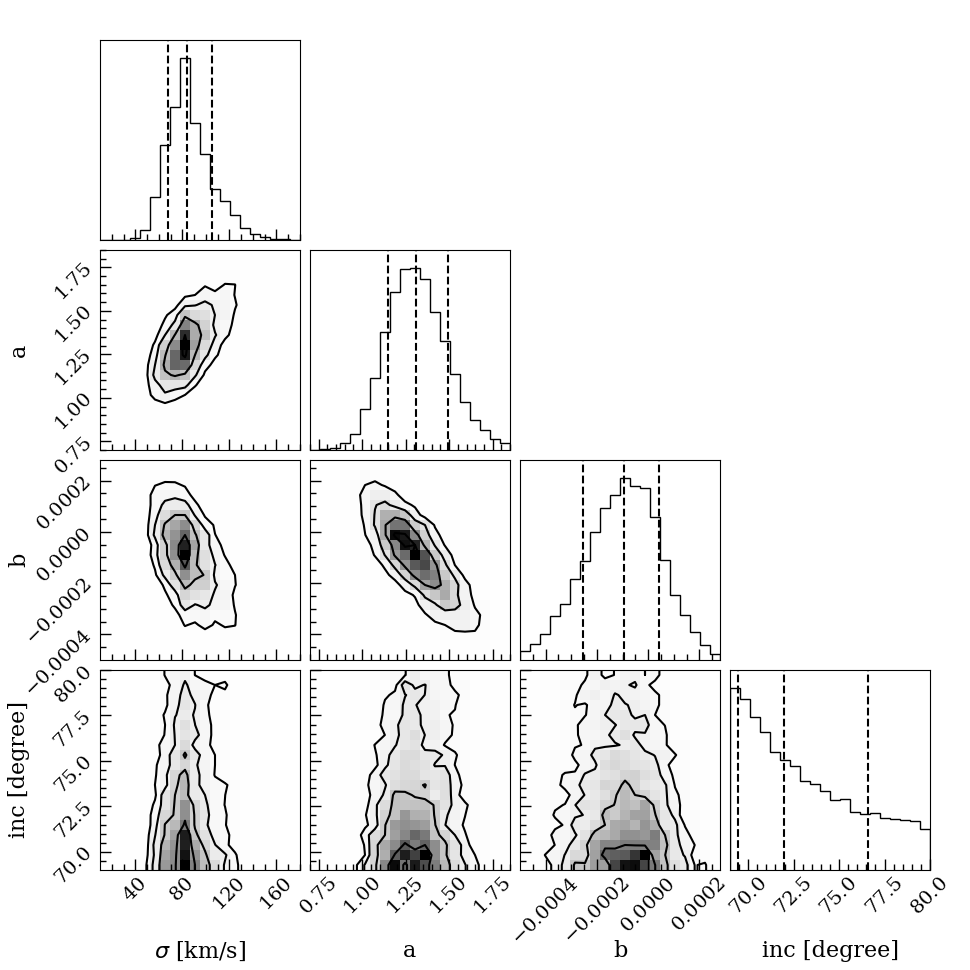}    
        }
    \caption{Target UVISTA-Z-019, see the caption of \ref{Fig:VR7}.
          }

    \label{Fig:uvista-z-019}
   \end{figure*}

\begin{figure*}[htbp]
   \resizebox{14cm}{!}
    {   
        \includegraphics{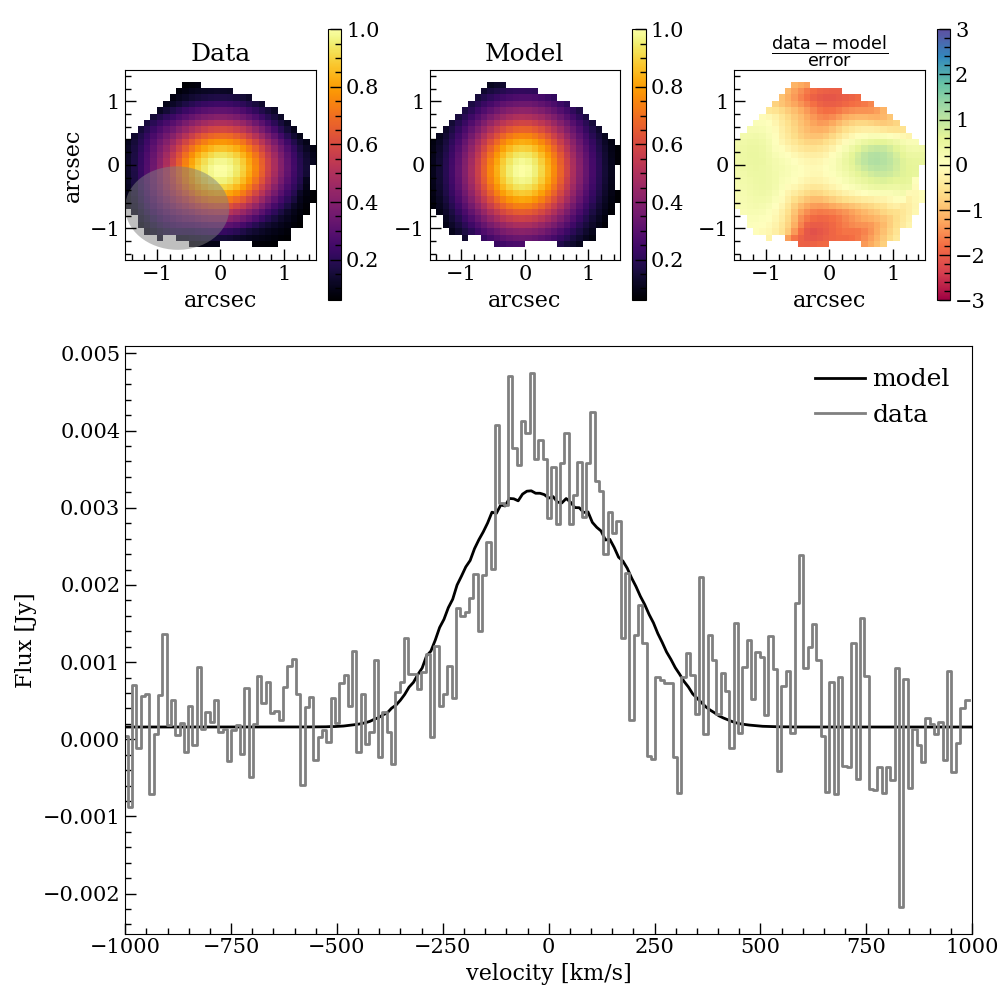}
        \includegraphics[clip]{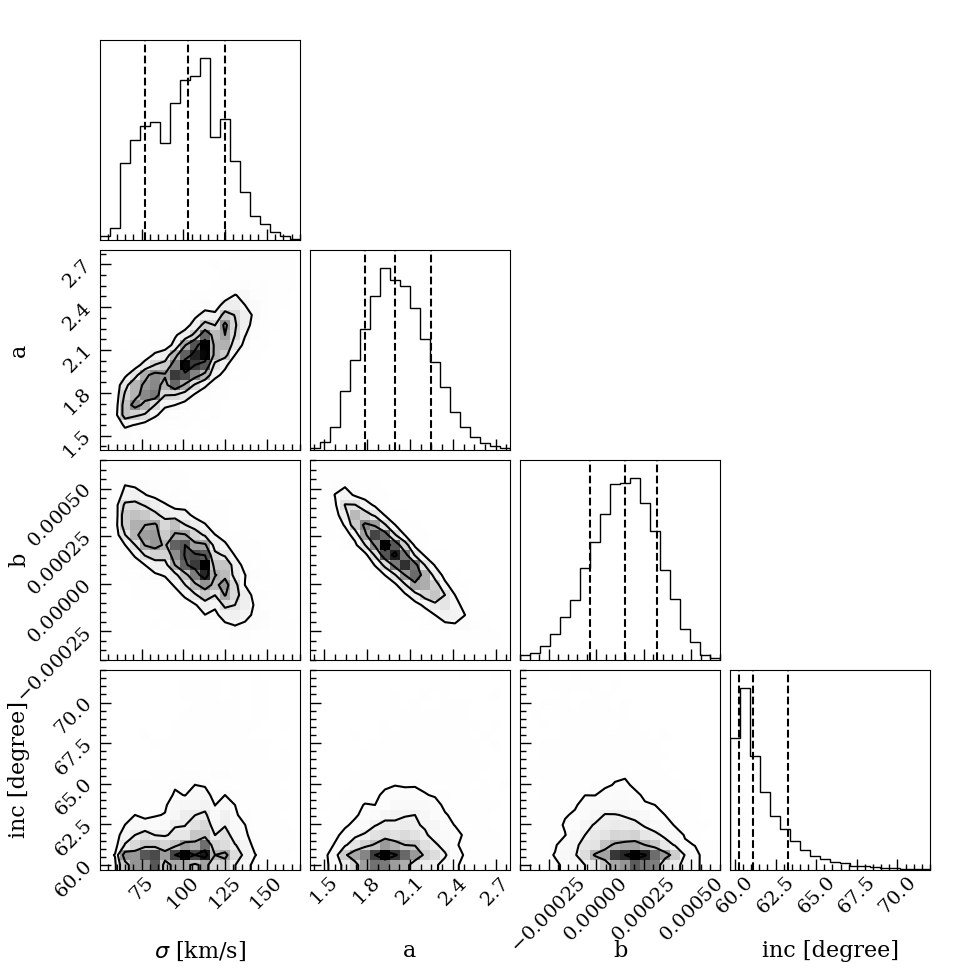}    
        }
    \caption{Target UVISTA-Y-003, see the caption of \ref{Fig:VR7}
          }

    \label{Fig:uvista-y-003}
   \end{figure*}

\begin{figure*}[htbp]
   \resizebox{14cm}{!}
    {   
        \includegraphics{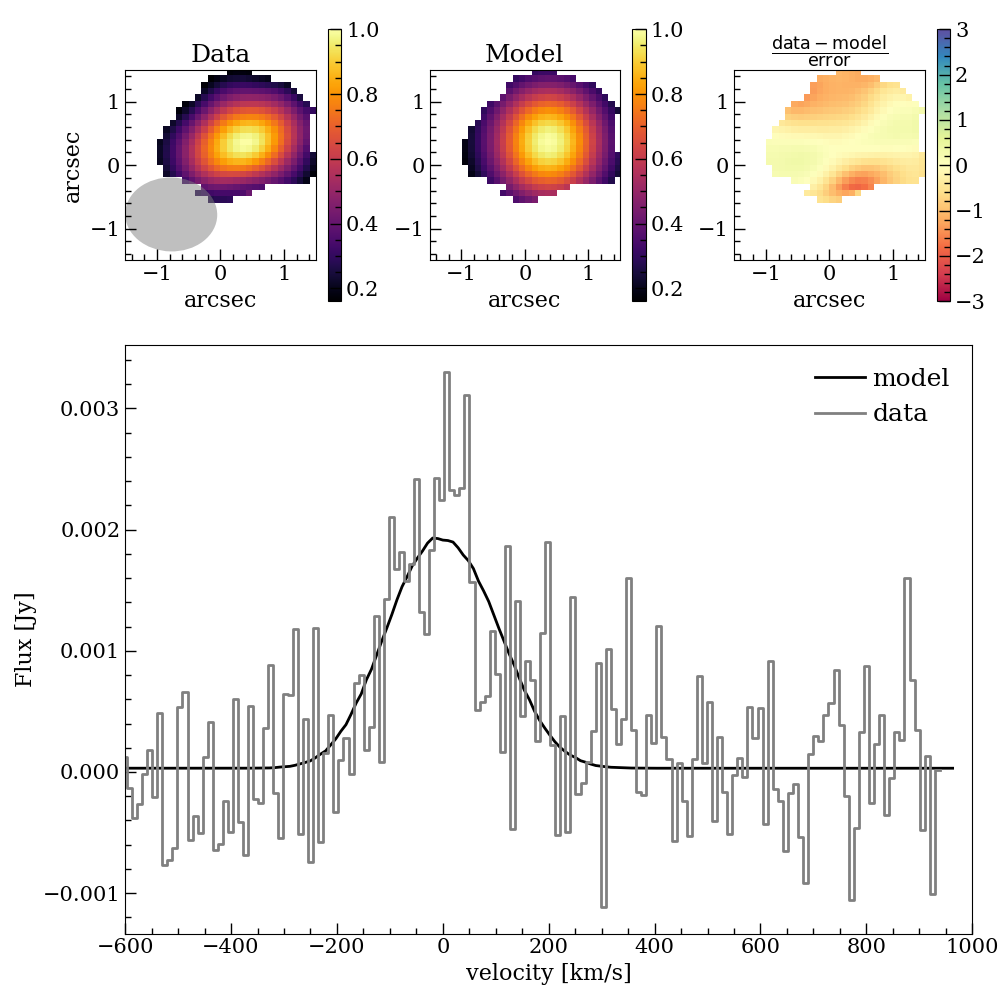}
        \includegraphics[clip]{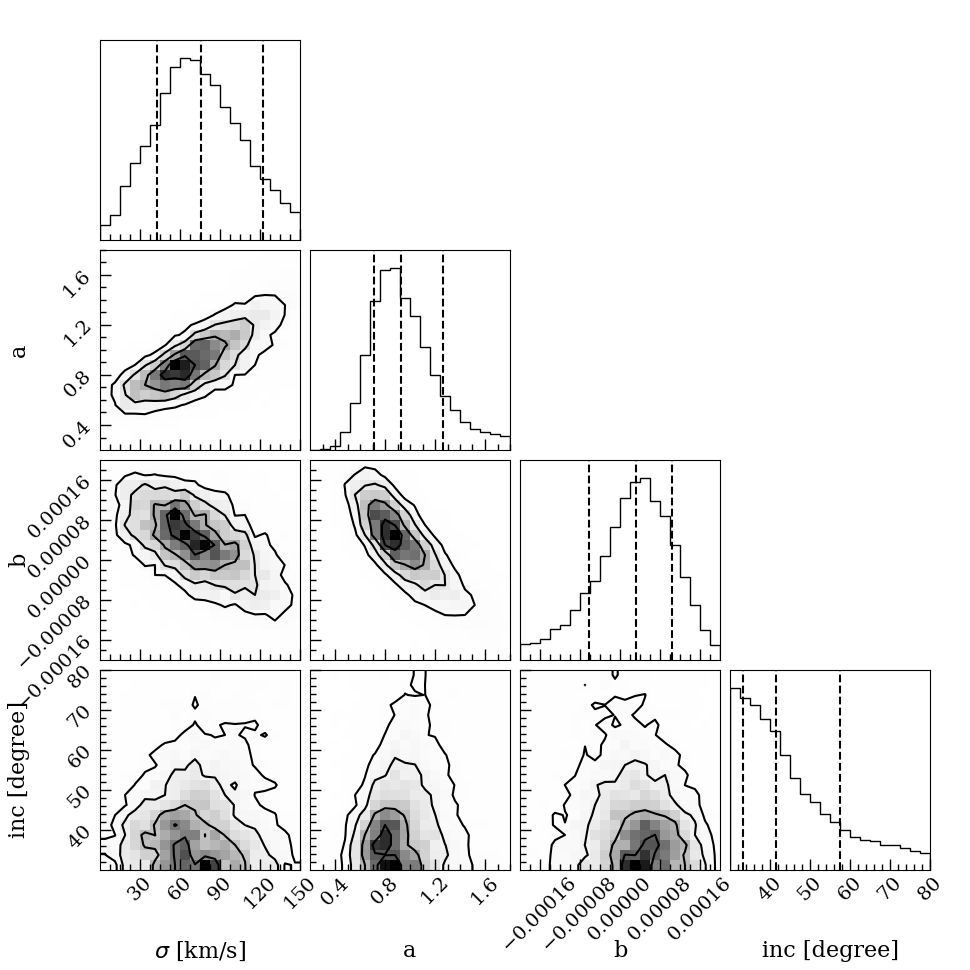}    
        }
    \caption{Target UVISTA-Z-049, see the caption of \ref{Fig:VR7}.
          }

    \label{Fig:uvista-z-049}
   \end{figure*}

\begin{figure*}[htbp]
   \resizebox{14cm}{!}
    {   
        \includegraphics{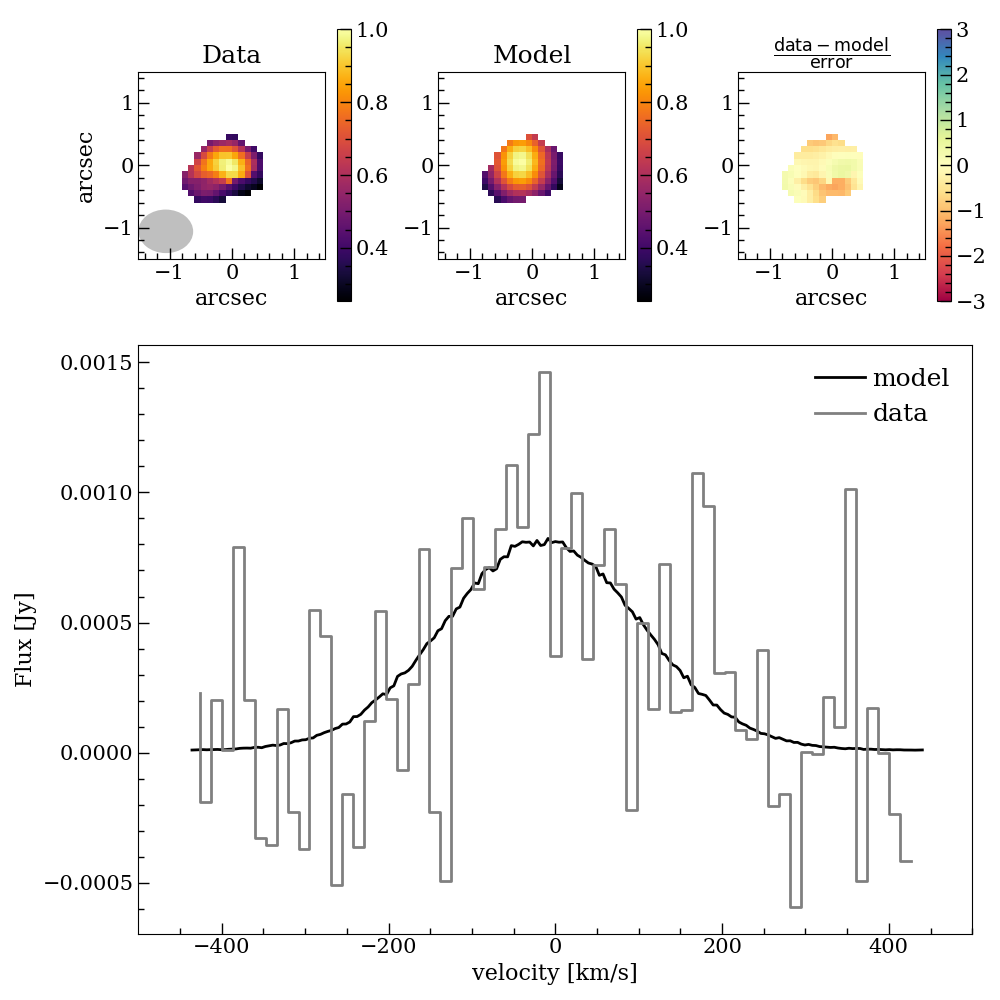}
        \includegraphics[clip]{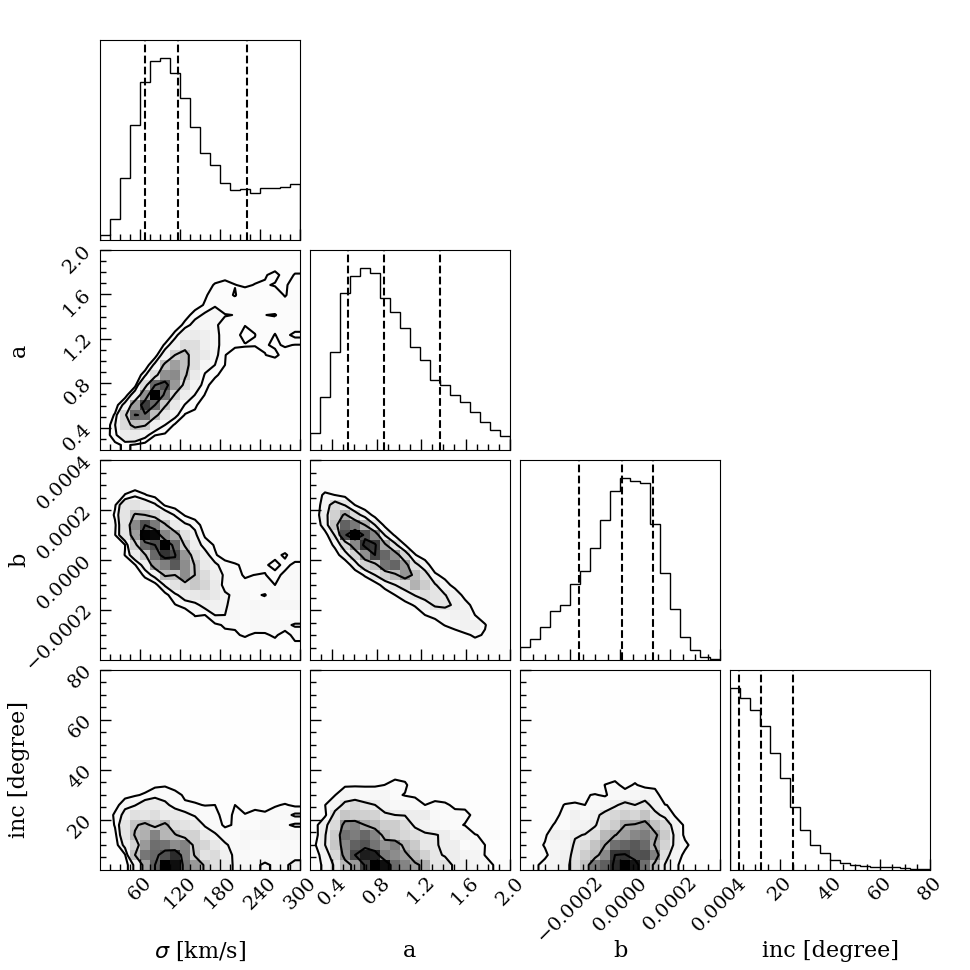}    
        }
    \caption{Target J0235 observed with the \cii emission line, see the caption of \ref{Fig:VR7}.
          }

    \label{Fig:J0235cii}
   \end{figure*} 

\begin{figure*}[htbp]
   \resizebox{14cm}{!}
    {   
        \includegraphics{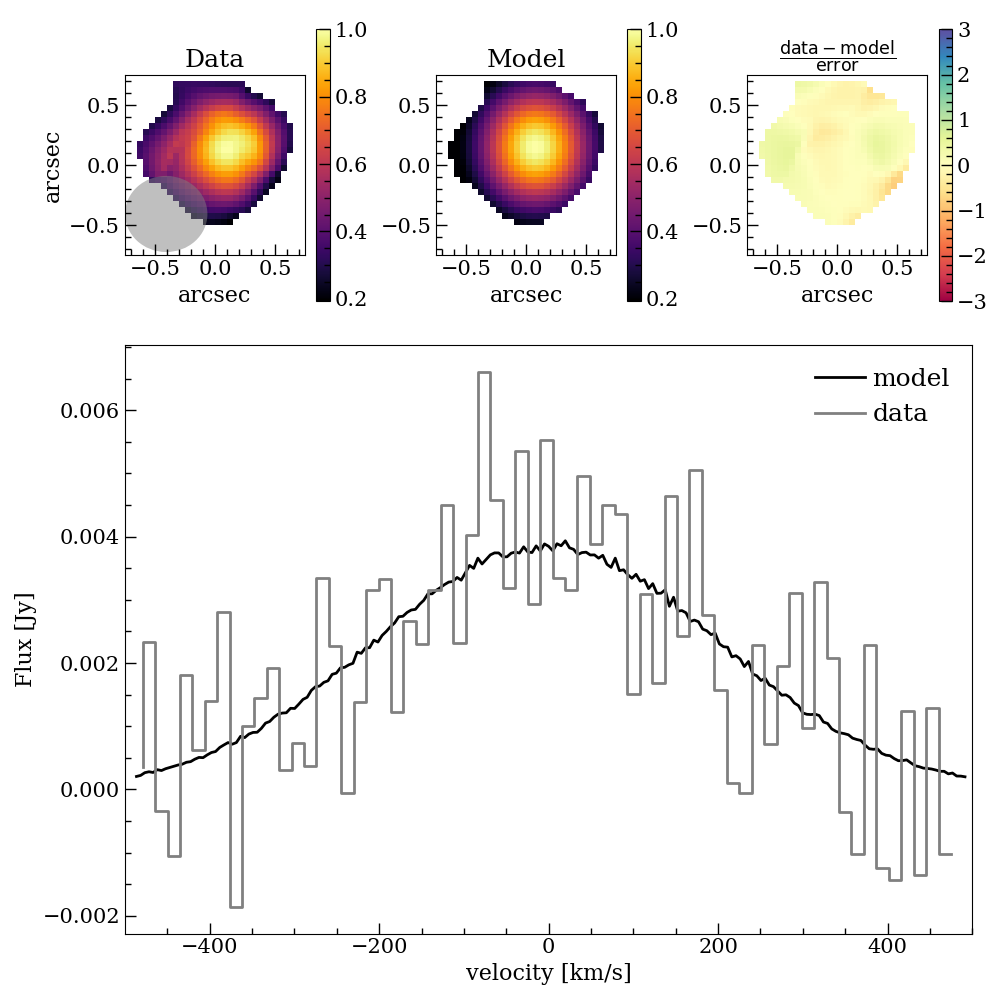}
        \includegraphics[clip]{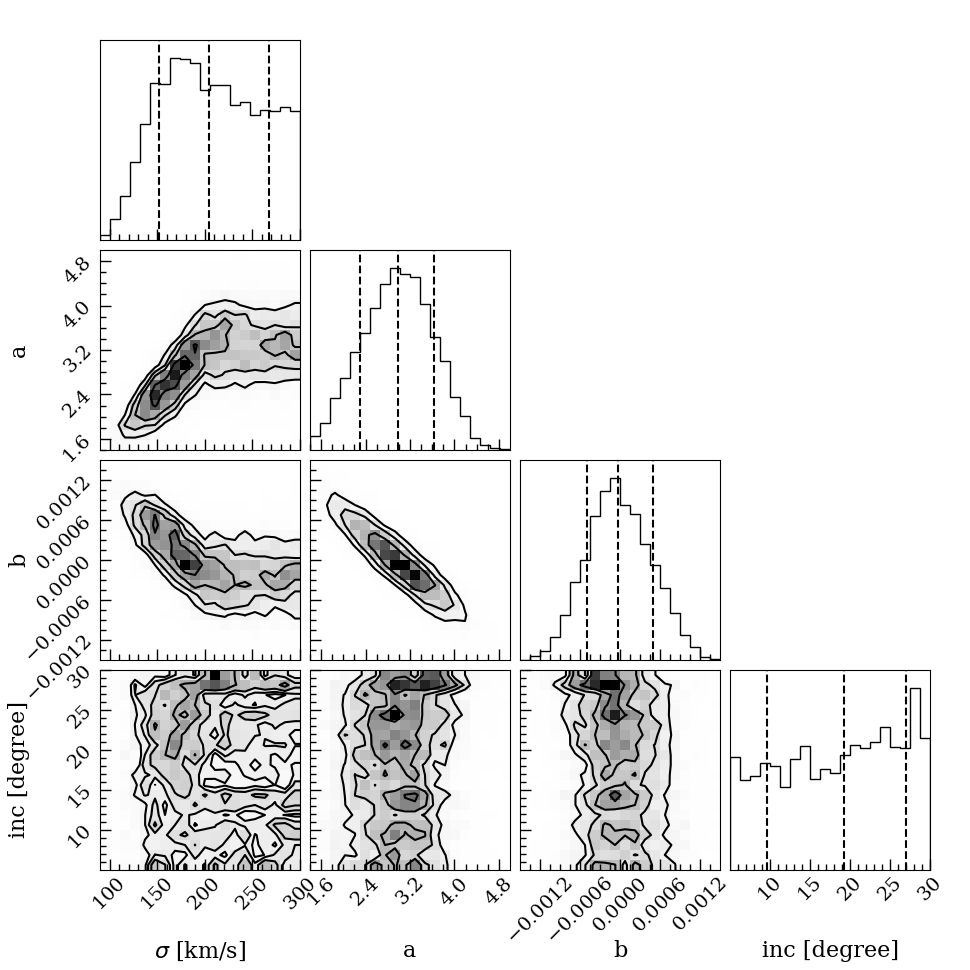}    
        }
    \caption{Target J0235 observed with the \oiii emission line, see the caption of \ref{Fig:VR7}.
          }

    \label{Fig:J0235_oiii}
   \end{figure*}

   \begin{figure*}[htbp]
   \resizebox{14cm}{!}
    {   
        \includegraphics{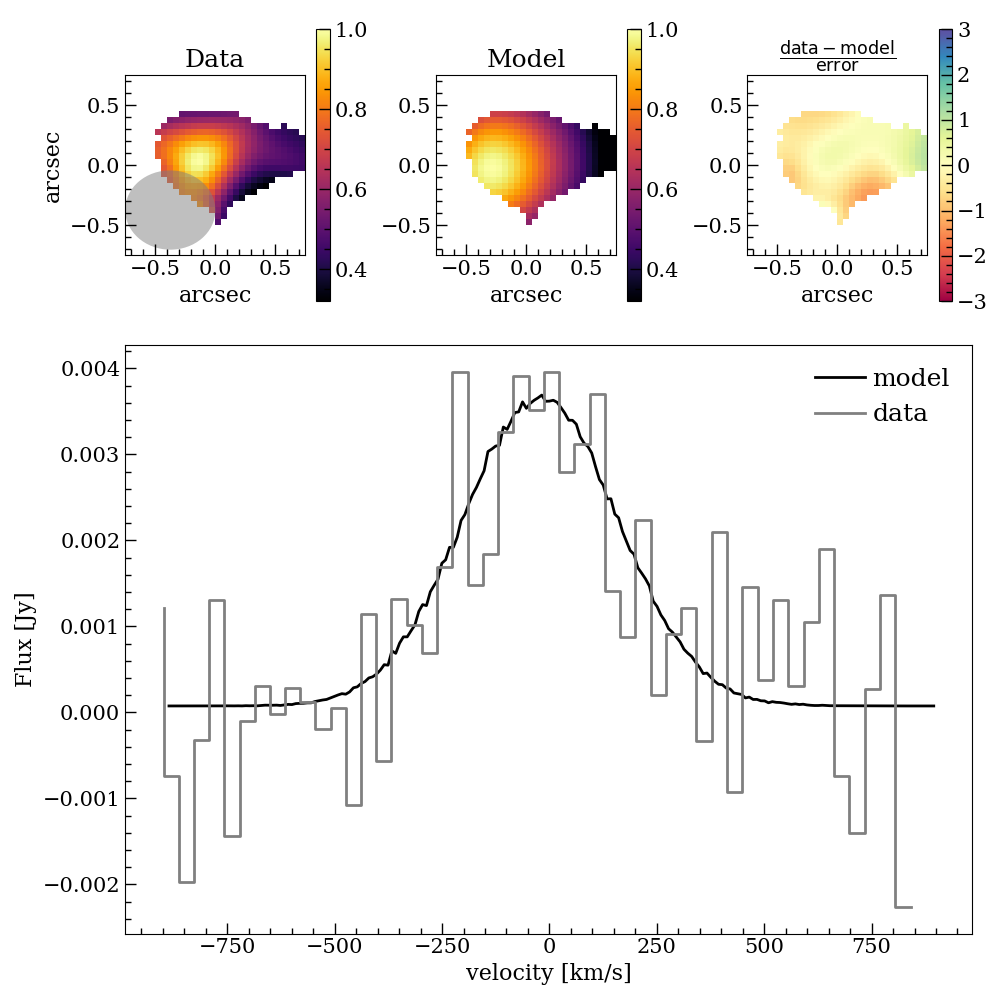}
        \includegraphics[clip]{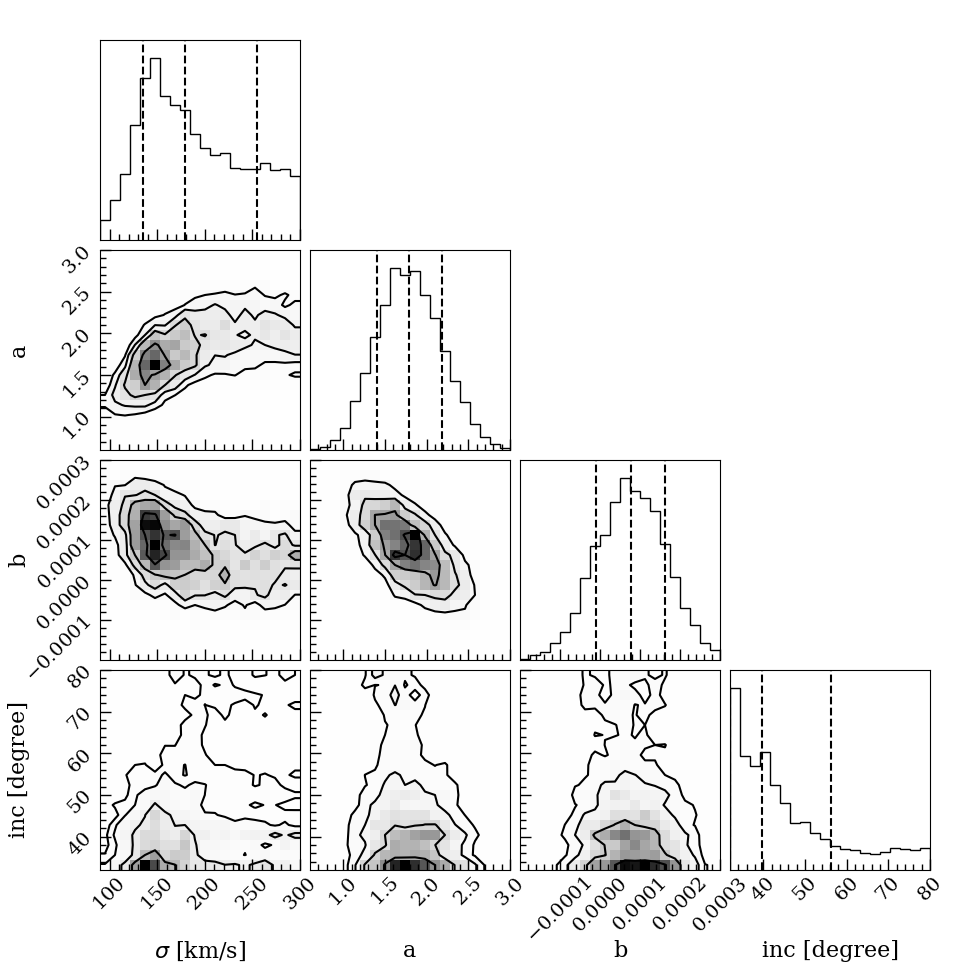}    
        }
    \caption{Target J0217 observed with the \cii emission line, see the caption of \ref{Fig:VR7}.
          }

    \label{Fig:J0217_cii}
   \end{figure*}

\begin{figure*}[htbp]
   \resizebox{14cm}{!}
    {   
        \includegraphics{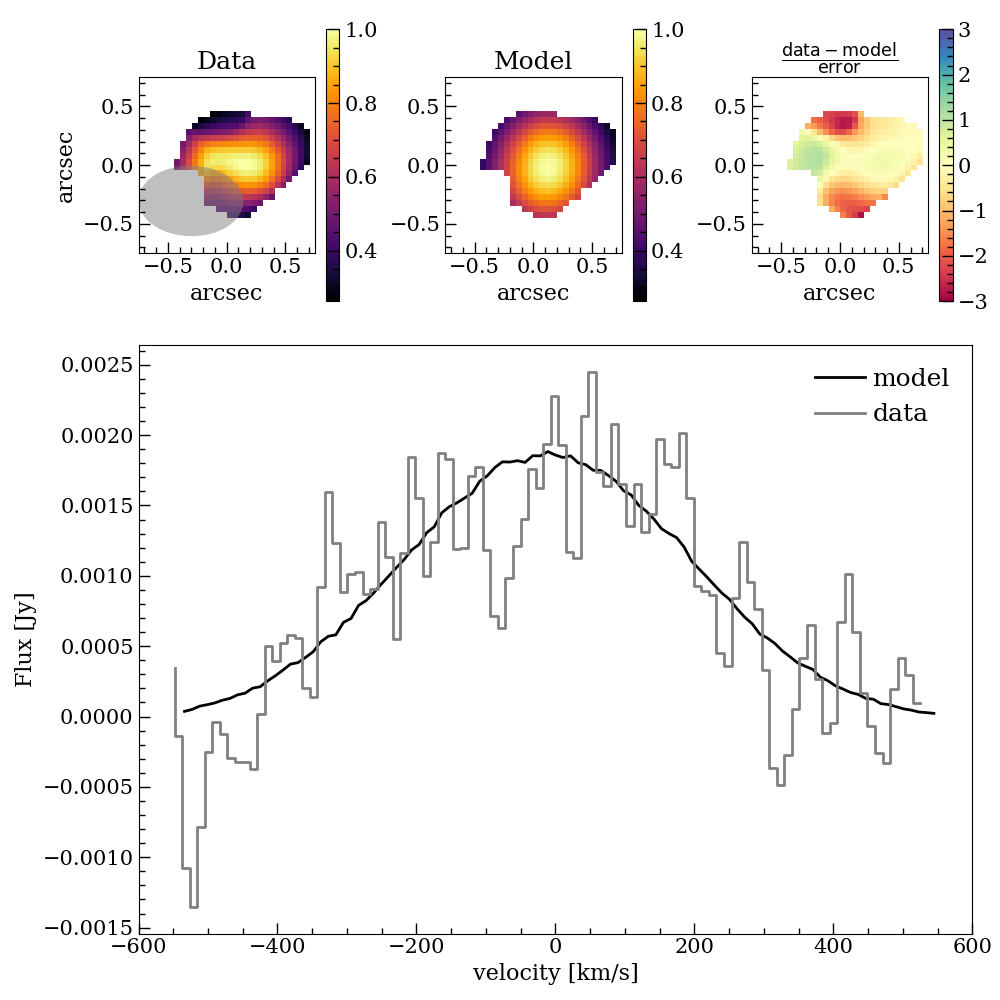}
        \includegraphics[clip]{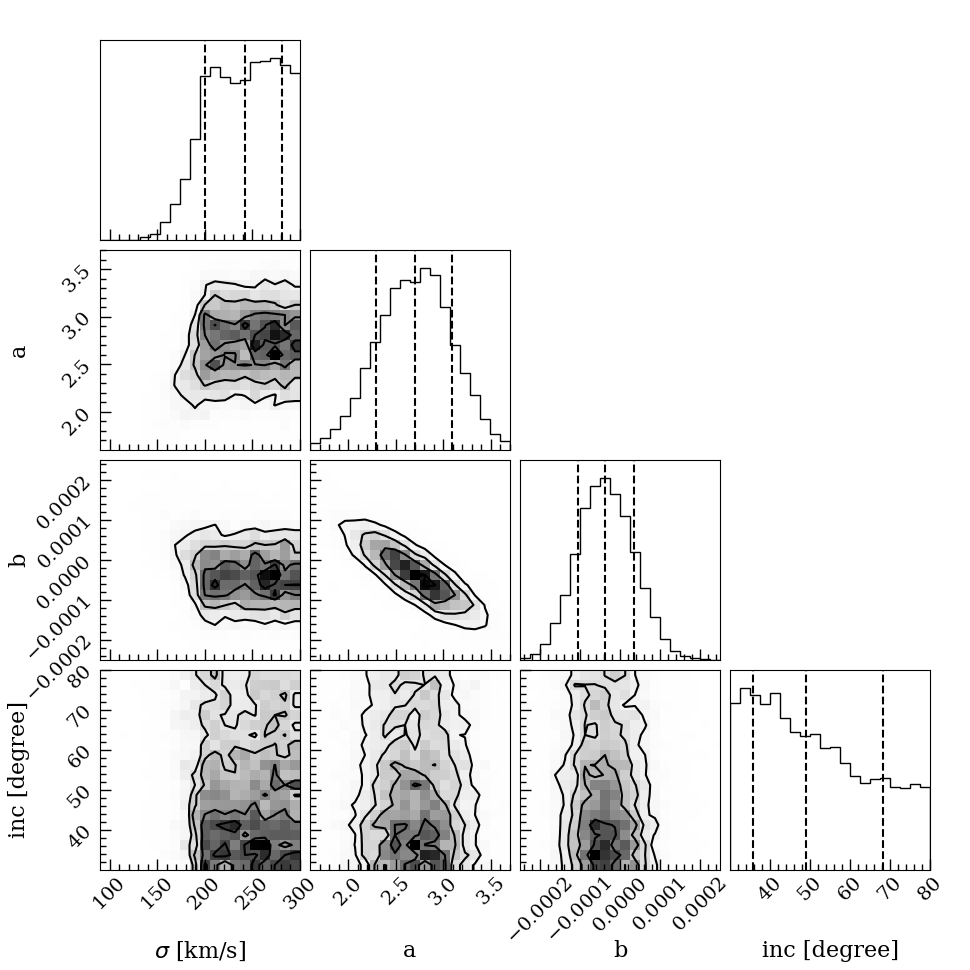}    
        }
    \caption{Target COS-30 observed with the \oiii emission line, see the caption of \ref{Fig:VR7}.
          }

    \label{Fig:cos30oiii}
   \end{figure*}

\end{appendix}
\end{document}